\documentclass{article}
\usepackage{amsfonts}
\usepackage{amssymb}
\usepackage{graphicx}
\usepackage{amsmath}
\usepackage{geometry}
\usepackage{appendix}

\input{tcilatex}
\begin{document}

\title{A Statistical Field Approach to Capital Accumulation}
\author{Pierre Gosselin\thanks{%
Pierre Gosselin : Institut Fourier, UMR 5582 CNRS-UJF, Universit\'{e}
Grenoble I, BP 74, 38402 St Martin d'H\`{e}res, France.\ E-Mail:
gosselin@ujf-grenoble.fr} \and A\"{\i}leen Lotz\thanks{%
A\"{\i}leen Lotz: Cerca Trova, BP 114, 38001 Grenoble Cedex 1, France.\
E-mail: a.lotz@erc-cercatrova.eu} \and Marc Wambst\thanks{%
Marc Wambst : IRMA, UMR 7501 CNRS, Universit\'{e} de Strasbourg, France.\
E-Mail: wambst@math.unistra.fr}}
\date{Mai 2019}
\maketitle

\begin{abstract}
This paper presents a model of capital accumulation for a large number of
heterogenous producer-consumers in an exchange space in which interactions
depend on agents' positions. Each agent is described by his production,
consumption, stock of capital, as well as the position he occupies in this
abstract space.

Each agent produces one differentiated good whose price is fixed by market
clearing conditions. Production functions are Cobb-Douglas, and capital
stocks follow the standard capital accumulation dynamic equation. Agents
consume all goods but have a preference for goods produced by their closest
neighbors.

Agents in the exchange space are subject both to attractive and repulsive
forces. Exchanges drive agents closer, but beyond a certain level of
proximity, agents will tend to crowd out more distant agents.

The present model uses a formalism based on statistical field theory
developed earlier by the authors. This approach allows the analytical
treatment of economic models with an arbitrary number of agents, while
preserving the system's interactions and complexity at the individual level.

Our results show that the dynamics of capital accumulation and agents'
position in the exchange space are correlated. Interactions in the exchange
space induce several phases of the system.

A first phase appears when attractive forces are limited. In this phase, an
initial central position in the exchange space favors capital accumulation
in average and leads to a higher level of capital, while agents far from the
center will experience a slower accumulation process. A high level of
initial capital drives agents towards a central position, i.e. improve the
terms of their exchanges: they experience a higher demand and higher prices
for their product. As usual, high capital productivity favors capital
accumulation, while higher rates of capital depreciation reduce capital
stock.

In a second phase, attractive forces are predominant. The previous results
remain, but an additional threshold effect appears. Even though no
restriction was imposed initially on the system, two types of agents emerge,
depending on their initial stock of capital.\ One type of agents will remain
above the capital threshold and occupy and benefit from a central position.
The other type will remain below the threshold, will not be able to break it
and will remain at the periphery of the exchange space. In this phase,
capital distribution is less homogenous than in the first phase.


\end{abstract}

\bigskip \newpage

\section*{Introduction}

Field theory applied to economic models is useful to overcome the pitfalls
of aggregation. This approach developed in Gosselin, Lotz and Wambst (2017,
2018) preserves the microeconomic concepts of standard economic models to
describe fully or partly rational agents, while enabling the study of the
transition from individual to collective scale given by statistical physics.
It provides an analytical treatment of a broad class of economic models with
an arbitrary number of agents, while keeping track of the system's
interactions and complexity at the individual level.

Field theory describes an environment of an infinite number of interacting
agents, from which various phases or equilibria may emerge.\ It allows to
study the agents' behaviors, the way they are influenced by and interact
with their environment. Depending on the parameters of the system, the form
of the ground state may drastically change the description at the individual
level. It is thus possible to compare the features of the macro state of a
system and those of the micro level, and their interactions. As such, it may
confirm or invalidate some aspects of the representative agent models and
shed light on up-to-now discarded micro and macro phenomena.

The present paper applies this field formalism to a model of capital
accumulation for a large number of heterogenous producer-consumers in an
exchange space in which interactions depend on agents' positions. Each agent
is described by his production, consumption, stock of capital, as well as
the position he occupies in an abstract space of exchanges. Each agent
produces one differentiated good whose price is fixed by market clearing
conditions. Production functions are Cobb-Douglas, and capital stocks follow
the standard capital accumulation dynamic equation. Agents consume all goods
but have a preference for goods produced by their closest neighbors. Thus,
demand depends not only on prices, but also on the distance between
consumers and producers in the exchange space. The closer the agents, the
higher their propensity to exchange, the higher the demand. Moreover, the
position of each agent is itself dynamic. Agents in the exchange space are
subject both to aggregation and repulsion forces. Exchanges drive agents
closer, but beyond a certain level of proximity, closest agents will crowd
out more distant agents: an increased proximity weakens more distant
exchanges.

The dynamic exchange space presented in this model allows to study the
production, exchanges, market shares and capital accumulation within a large
group of agents.\ What are the patterns of accumulation across agents? Is
there a threshold effect leading some producers to accumulate at the expense
of the others? Is there a phase in which a better wealth distribution could
be reached? All these questions can be addressed within our formalism.

Translating standard economic models into a statistical field model is a
two-step process. In a first step, the usual model of optimizing agents is
replaced by a probabilistic description of the system. In such a setting,
individual optimization problems are discarded. Each agent is described by a
time-dependent probability distribution centered around this agent's
classical optimization path. In a second step, the individual agents'
description is replaced by a more compact model of field theory that
replicates the properties of the system when $N$, the number of agents, is
large (Gosselin, Lotz and Wambst 2017, 2018, and Kleinert 1989). This
modeling, although approximate, is compact enough to allow an analytical
treatment of the system.

In our model, this formalism yields the probabilistic dynamics of individual
agents through the computation of so-called transitions functions. Given an
initial stock of capital and an initial position in the exchange space, each
individual stochastic path can be found, and depends on parameters such as
strength in exchange interactions, rate of capital depreciation and
uncertainty in economic variables. We show that, depending on the
interaction forces in the exchange space, two phases of the system appear.
One corresponds to a society in which the repulsion force is a large enough
to compensate the attractive force resulting from the exchange. We show that
in that phase capital accumulation and mobility in the exchange space depend
highly on the initial capital stock. However, this phase presents a certain
stability and allows, up to a point, some capital accumulation for most
agents. The second phase appears for an attractive force larger than the
repulsive one.

The first section details the literature review.\ Section two describes a
classical model of capital accumulation with $N$ economic agents, translates
it into a probabilistic framework and presents its associated field
formulation. In section three, we solve the dynamics of the system and
present its various phases. Section four interprets the results and section
five concludes.

\section{Literature review}

A large branch of the recent economic literature has been devoted to address
the notion of representative agent. Complex systems, Networks, Agent Based
Systems or Econophysics are among the various paths that have been explored
to remedy its pitfalls.

By several aspects, our approach is related to the Multi-Agents System
economic literature, notably Agent Based Models (see Gaffard and Napoletano
2012) and Economic Networks (Jackson 2010). Both rely on numerical
simulation of Multi-Agents System but are often concerned with different
types of model. Agent Based Models deal with general macroeconomics models,
whereas Network Models rather deal with lower scale models, such as Contract
Theory, Behavior Diffusion, Information Sharing or Learning. In both type of
settings, agents are typically defined by, and follow, various set of rules.
These rules allow for equilibria and dynamics that would otherwise remain
inaccessible to the representative agent setup.

The Agent-Based approach is similar to ours in that it does not seek to
aggregate all agents but considers the interacting system in itself. It is
however highly numerical, model-dependent, and relies on microeconomic
relations, such as ad-hoc reaction functions, that may be too simplistic. On
the contrary, Statistical Field Theory accounts for the impact of scale
changes. Macroeconomic patterns do not emerge from the sole dynamics of a
large set of agents but are grounded on particular behaviors and
interactions structures. Describing these structures in terms of Field
Theory allows to study the emergence of a phase at the macro scale, and in
turn its impact at the individual level.

Econophysics is closer to our approach (for a review, see Chakraborti, Muni
Toke, Patriarca and Abergel 2011a, b and references therein). It often
considers the set of agents as a statistical system.\ Moreover, Kleinert
(2009) has already used path integrals to model the stock prices' dynamics.
However, Econophysics does not apply the full potentiality of Field Theory
to economic systems. It focuses on empirical laws, but the lack of
micro-foundations casts some doubts on the robustness of these observed
empirical laws, that are prone, like ad-hoc macroeconomics, to the Lucas
critique (see Lucas 1976). Our approach, in contrast, keeps track of usual
microeconomics concepts such as utility functions, expectations, forward
looking behaviors. It includes these behaviors in the analytical treatment
of Multi-Agents Systems by translating the main characteristics of a system
of optimizing agents in terms of a statistical system.

Capital accumulation has been considered in several ways since the Solow
growth model (Solow 1957) and its subsequent developments (see Barro 1995
for an account). The closest approaches to this paper stems from Economic
Geography (Krugman 1999) and introduce several types of producers in a
differentiated geographical environment, "core" and "periphery". This
environment impacts the production of partly differentiated goods, such as
agricultural and manufactured goods (see Fujita and Thisse 2013 for a
review).\ There is a notion of space in these models, but the position of
agents is static, whereas in our approach, exchange position is dynamic and
interacts with capital accumulation. Moreover, although conditions for
takeoff and convergence are studied for regional industrialization or multi
countries growth models (Aghion and Durlauf 2005), it is the fixed
geographical parameters that determine the environment, notably
transportation costs and shares of immobile workers.\textbf{\ }The present
approach, on the contrary, considers an evolutive environment, that may be
endogenized to interact with capital accumulation.

Also related to our purpose, a recent combination of evolutionary theory,
complex systems and agent-based model (Gintis 2007) has allowed a detailed
study of a large number of producers' capital dynamics (Dosi and Nelson
2010; Dosi, Fagiolo and Roventini 2010; Dosi et al. 2015; Dawid et al. 2011,
2014), (Ciarli et al.\ 2010; Mandel et al. 2010; Wolf et al.\ 2013). More
refined models investigate how several production sectors interact and
compete, using neighbors' output as inputs (Mandel et al. 2016; Mandel
2012). Competition and capital accumulation are driven by random changes in
technology, and producers progressively adapt their production via
imitation, replacing parts of their inputs with newer, more efficient
technologies. In these models, the firms' interactions dynamics are
simulated numerically and track the persistence of heterogenous production
sectors and the computation of emergent macro quantities such as total
output, wages or unemployment. This focus on the evolution of independent
sectors is close to our purpose.\ However, we do not use numerical methods
and consider an exchange space determining exchanges between agents.
Moreover, we do not focus on the evolution on technology, even if it could
be included, as in Gosselin, Lotz and Wambst (2018), but rather study the
impact of the exchange dynamics on agents' capital accumulation.

\section{Description of the model}

This section describes a standard model of capital accumulation for a large
number of agents. The usual capital dynamics and production function are
maintained, but here agents interact dynamically through an exchange space.

\subsection{Setup}

There are $N$ consumer-producer agents.\ Each agent is differentiated by his
position on an exchange space, denoted $X_{i}\left( t\right) \in \left[ -1,1%
\right] $. This exchange position is a dynamic variable that interacts with
the other variables of the model. It can be seen as a geographic space, but
also as an abstract exchange space, in which the central position $X=0$
ensures higher exchanges.

\bigskip

Each agent produces a single differentiated good. Production functions are
Cobb-Douglas.\ Each agent $i$ individual capital stock is denoted $%
K_{i}\left( t\right) $. It is a fully liquid capital whose price is set to
one. Since agents are individual producers, labor\ can be discarded, and we
further assume a constant technology factor $A$. So that the individual
production function is of the form $AK_{i}^{\alpha }\left( t\right) $. The
price of each good $P_{i}\left( t\right) $ is determined by market-clearing
condition. Ultimately, the agent's income is the product of his production
and his price:%
\begin{equation}
Y_{i}\left( t\right) =P_{i}\left( t\right) AK_{i}^{\alpha }\left( t\right)
\label{Prd}
\end{equation}

\bigskip

Each agent consumes all the goods produced. We denote $C_{i}^{\left(
j\right) }\left( t\right) $\ the consumption of good $j$ by agent $i$ at
time $t$. Three factors determine the agents' consumption of each good.
First, the quantity of each good that an agent consumes is proportional to
his income, with a proportionality factor to consume $\kappa $ ($0<\kappa <1$%
).

Besides, the consumption of each good will also be a decreasing function of
the good's relative price, and of the distance between its producer and
consumer. We detail these two last conditions below.

\bigskip

Agents' consumption depends on the relative price level of goods: we assume
that agent's $i$ consumption of good $j$ is a decreasing function $g\left(
R_{i,j}\left( t\right) \right) $ with $R_{i,j}\left( t\right) =$ $%
P_{j}\left( t\right) /\bar{P}_{i}\left( t\right) $. The ratio $R_{i,j}\left(
t\right) $ is the relative price of good $j$ with respect to a general
subjective price level for agent $i$. We chose a dependency of the type $%
g\left( R_{i,j}\right) \sim \left( R_{i,j}\right) ^{-\left( 1+\gamma \right)
}$. We define $\bar{P}_{i}\left( t\right) =P_{K}^{1-\varepsilon }\left( \hat{%
P}_{i}\left( t\right) \right) ^{\varepsilon }$, where $0<\varepsilon <1$.
Thus, the price level $\bar{P}_{i}\left( t\right) $ is a combination of the
general price of capital $P_{K}$ and a subjective location-dependent
consumption price index $\hat{P}_{i}\left( t\right) $. Since the price of
capital is set to $1$, $\bar{P}_{i}\left( t\right) =\left( \hat{P}_{i}\left(
t\right) \right) ^{\varepsilon }$, and the ratio $R_{i,j}\left( t\right) $
rewrites $P_{j}\left( t\right) /\left( \hat{P}_{i}\left( t\right) \right)
^{\varepsilon }$. The index $\hat{P}_{i}\left( t\right) $ is a weighted
average of prices, where each weight is a function of the distance between
the consumer and producer. Thus, it depends on the agent's exchange position
at each moment of time, and writes:

\begin{equation*}
\hat{P}_{i}\left( t\right) =\frac{1}{d}\sum_{j}P_{j}\left( t\right) \exp
\left( -d_{ij}\left( t\right) /d\right)
\end{equation*}%
with:%
\begin{equation*}
d_{ij}\left( t\right) =\left\vert X_{i}\left( t\right) -X_{j}\left( t\right)
\right\vert
\end{equation*}%
and $d$ a constant parameter.\ The factor $\frac{1}{d}$\ inside the
exponential models the fact that agents interact in average on an interval
of length in the exchange space. The factor $\frac{1}{d}$ outside the
exponential acts as a normalization factor.

\bigskip

Agent's consumption also depends on the distance between consumer and
producers. We assume consumption to be an exponentially decreasing function
of the distance between consumer $i$ and producer $j$,$\ \left\vert
X_{i}\left( t\right) -X_{j}\left( t\right) \right\vert $. Recall that $%
X_{i}\left( t\right) $ and $X_{j}\left( t\right) \in \left[ -1,1\right] $.
In analogy with the price level, we chose a dependency of the form $\exp
\left( -d_{ij}\left( t\right) /d\right) /d$. This exponentially decreasing
factor has only a relative value. It may reflect the agents' connections%
\textbf{: }the position in the exchange space only indicates the exchanges
agents establish within the exchange space.\textbf{\ }It could also account
for transportation costs in a geographic interpretation. It also implies
that an agent at the center of the exchange space will face a higher demand
for his good.

\bigskip

Under the previous hypotheses, consumption of good $j$ by agent $i$ writes:%
\begin{equation}
C_{i}^{\left( j\right) }\left( t\right) =\frac{\kappa Y_{i}\left( t\right) }{%
\left( R_{i,j}\left( t\right) \right) ^{1+\gamma }}\frac{\exp \left(
-d_{ij}\left( t\right) /d\right) }{d}  \label{cnsptn}
\end{equation}%
For the sake of simplicity we normalize $\left( 1+\gamma \right) \varepsilon
=1$ in the sequel, but this factor could be reintroduced without impairing
the results.

The consumption function of good $j$ by agent $i$ rewrites:%
\begin{equation}
C_{i}^{\left( j\right) }\left( t\right) =\frac{\frac{\kappa }{d}Y_{i}\left(
t\right) }{P_{j}^{1+\gamma }\left( t\right) /\hat{P}_{i}\left( t\right) }%
\exp \left( -\frac{d_{ij}\left( t\right) }{d}\right)  \label{csptnpr}
\end{equation}

\bigskip

Remark that agent $i$ propensity to consume depends on his relative price
index, and thus is location. More precisely, the dependence in $X_{i}\left(
t\right) $\ of this propensity follows the pattern: 
\begin{equation}
\frac{1}{d^{2}}\sum_{j}\exp \left( -d_{ij}\left( t\right) \right) \left(
\sum_{k}\exp \left( -\frac{d_{ik}\left( t\right) }{d}\right) \right)
\label{prp}
\end{equation}%
Replacing the summation by an integral, equation (\ref{prp}) is proportional
to $\left( 1-\cosh \left( \frac{\left\vert X_{i}\left( t\right) \right\vert 
}{d}\right) \exp \left( -\frac{1}{d}\right) \right) ^{2}$. All things equal,
the agent's propensity to consume is maximal for $X_{i}\left( t\right) =0$,
minimal for $X_{i}\left( t\right) =-1$\ and $X_{i}\left( t\right) =1$. The
individual marginal propensity to consume is thus higher in the center of
the exchange space than at the periphery, reflecting the fact that exchanges
are a decreasing function of the distance between agents.

The above assumptions reflect the fact that exchanges are more frequent at
the center of the exchange space. Actually, this space can be seen as a
scale of exchanges, in which the position of the agent measures the
intensity of his exchanges.\ Moves towards the center or the periphery
depict respectively an improvement or a deterioration of his terms of
exchanges. We define the terms of exchanges of a producer as his ability to
sell his production at a given price within this exchange space -
competitors and consumers, depending on their capital, revenue, distance
from the producer, etc.

To conclude this section, note that the proportionality factor $\kappa $
should be determined by optimization of an intertemporal utility function
under the constraint of future flow of expected profits. However, assuming
some autonomous consumption proportional to the agent's revenue, $\kappa $
is constant in first approximation. Agents consume an average "minimal"
necessary level of goods, given their position in the exchange space, and
reinvest the full amount of their remaining income. In our setting, the
creation of trade relations compensates the income loss due to a high
production. This favors capital accumulation and ultimately, through a
better position in the exchange space, increases the producer price.

\subsection{Classical description of the model}

In this setting, each good's price at each point in time is determined by
market clearing conditions.\ The global demand for good $i$ at time t by all
agents $j$, $\sum_{j}C_{j}^{\left( i\right) }\left( t\right) $ matches the
production of good $i$:

\begin{equation*}
\sum_{j}C_{j}^{\left( i\right) }\left( t\right) =AK_{i}^{\alpha }\left(
t\right)
\end{equation*}%
Using (\ref{csptnpr}) and (\ref{Prd}), this equation can be rewritten as:%
\begin{equation}
\frac{\kappa }{d^{2}}\sum_{j,k}P_{j}\left( t\right) K_{j}^{\alpha }\left(
t\right) P_{k}\left( t\right) \exp \left( -\frac{d_{ij}\left( t\right)
+d_{kj}\left( t\right) }{d}\right) =P_{i}^{1+\gamma }\left( t\right)
K_{i}^{\alpha }\left( t\right)  \label{Mk}
\end{equation}%
Capital accumulation dynamics follows a standard pattern. Capital
depreciates at rate $\delta $, and capital accumulation is subject to a
shock $\epsilon _{i}\left( t\right) $. We further assume that revenue saved
is entirely reinvested in capital at a price $1$. In such a setting, the
capital dynamic equation is:%
\begin{equation}
K_{i}\left( t+1\right) =\left( 1-\delta \right) K_{i}\left( t\right)
+Y_{i}\left( t\right) -\sum_{j}P_{j}\left( t\right) C_{i}^{\left( j\right)
}\left( t\right) +\epsilon _{i}\left( t\right)  \label{KA}
\end{equation}%
Using (\ref{csptnpr}), we find:

\begin{equation}
K_{i}\left( t+1\right) \simeq \left( 1-\delta \right) K_{i}\left( t\right)
+Y_{i}\left( t\right) -\frac{\kappa }{d^{2}}Y_{i}\left( t\right) \sum_{j,k}%
\frac{P_{k}\left( t\right) }{P_{j}^{\gamma }\left( t\right) }\exp \left( -%
\frac{d_{ij}\left( t\right) +d_{ik}\left( t\right) }{d}\right) +\epsilon
_{i}\left( t\right)  \label{Kdn}
\end{equation}%
$\epsilon _{i}\left( t\right) $, $\epsilon _{i}^{\left( 1\right) }\left(
t\right) $, $\epsilon _{i}^{\left( 2\right) }\left( t\right) $ variance $%
\sigma ^{2}$.

\subsection{Probabilistic Description}

Three dynamic variables $K_{i}$, $P_{i}$\ and $X_{i}$\ describe our model.
Each agent dynamics is described by a path within the space defined by these
three variables: from an initial point in $K_{i}$, $P_{i}$\ and $X_{i}$, the
agent reaches a final point in this same space. Classically, up to some
fluctuations, an optimal path does exist for each agent.

For a large number of agents however, because of fluctuations, all possible
paths may exist with varying probabilities. So that, for a large number of
agents, we must take into account the set - or space - of all paths, and
associate to this set a \emph{probability density }centered around the
classical optimal path\footnote{%
We use the word probability density rather than probability, since due to
the infinite number of possible paths, each individual path has a null
probability to exist, just like a gaussian defining the size of a
population.\ }. This description is a good approximation of standard
descriptions and translates the fact that each agent experiments some
idiosyncratic shocks.

Once this probability density computed for an individual agent, the
probability density for the set of all agents is merely the product of the
individual probability densities. This probability density for a
configuration of $N$\ arbitrary individual paths is called the \emph{%
statistical weight of a state of the system}. By construction, it is
centered around one - or several in case of multiple equilibria -
configuration of paths that represent the classical equilibrium. Its shape
need not be known, since we are working directly with statistical weight.

In this section, we provide a probabilistic description of the model.\ To do
so we build a statistical weight for each equation of the model.\ Their
product will be the probability description of the system. The two first
variables, $K$\ and $P$, are standard economic variables, and their\ weight
will be derived from the equations of the model, as proposed in Gosselin,
Lotz, Wambst (2018). The last variable $X$\ is not a strictly standard
economic variable.\ We will depart from our methodology and ascribe to its
dynamics an ad hoc form.

\subsubsection{Probabilistic description for capital dynamics}

We associate a probability to the dynamic accumulation of capital, centered
around the average classical capital dynamics solution, for each period of
time $t$. To do so, note that (\ref{Kdn}) implies that the quantity:%
\begin{equation}
K_{i}\left( t+1\right) -\left( 1-\delta \right) K_{i}\left( t\right)
+Y_{i}\left( t\right) -\frac{\kappa }{d^{2}}Y_{i}\left( t\right) \sum_{j,k}%
\frac{P_{k}\left( t\right) }{P_{j}^{\gamma }\left( t\right) }\exp \left( -%
\frac{d_{ij}\left( t\right) +d_{ik}\left( t\right) }{d}\right)  \label{Kdnrd}
\end{equation}%
is a gaussian random variable $\epsilon _{i}\left( t\right) $ of variance $%
\sigma ^{2}$. In our context, equation (\ref{Kdn}) is replaced by the
probability density for $K_{i}\left( t\right) $:%
\begin{equation}
\exp \left( -\frac{1}{2\sigma ^{2}}\left( \dot{K}_{i}\left( t\right) +\delta
K_{i}\left( t\right) -AP_{i}\left( t\right) K_{i}^{\alpha }\left( t\right)
\left( 1-\frac{\kappa }{d^{2}}\sum_{j,k}\frac{P_{k}\left( t\right) \exp
\left( -\frac{d_{ij}\left( t\right) +d_{ik}\left( t\right) }{d}\right) }{%
P_{j}^{\gamma }\left( t\right) }\right) \right) ^{2}\right)  \label{Stw}
\end{equation}%
To account for the dynamics over the whole timespan, we sum over $t$ in the
exponential.\ This associates a density probability for a path of capital
accumulation over the whole timespan. This will account for stochastic paths 
$K_{i}\left( t\right) $\ that satisfy in average the classical dynamic
accumulation equation.

Ultimately, to associate a statistical weight to the set of paths of capital
accumulation for all agents, we sum over $i$ and $t$ in the exponential of (%
\ref{Stw}). The statistical weight associated to the capital accumulation of
the set of agents is thus:%
\begin{equation}
\exp \left( -\frac{1}{2\sigma ^{2}}\sum_{i}\int \left( \dot{K}_{i}\left(
t\right) +\delta K_{i}\left( t\right) -AP_{i}\left( t\right) K_{i}^{\alpha
}\left( t\right) \left( 1-\frac{\kappa }{d^{2}}\sum_{j,k}\frac{P_{k}\left(
t\right) \exp \left( -\frac{d_{ij}\left( t\right) +d_{ik}\left( t\right) }{d}%
\right) }{P_{j}^{\gamma }\left( t\right) }\right) \right) ^{2}dt\right)
\label{Stwk}
\end{equation}

\subsubsection{Probabilistic description for market clearing condition}

The dynamics for $P_{i}\left( t\right) $\ can be replaced, as for capital
dynamics, by a statistical weight derived from for the market clearing
condition. We assume that market clearing holds in average for a large
number of agents, but that fluctuations appear for individual agents.\ Thus (%
\ref{Mk}) only holds up to some random noise, and must be replaced by a
probability for each agent $i$\ to deviate from (\ref{Mk}).%
\begin{equation}
\exp \left( -\frac{1}{2\sigma _{1}^{2}}\left( P_{i}^{1+\gamma }\left(
t\right) K_{i}^{\alpha }\left( t\right) -\kappa \sum_{j,k}P_{j}\left(
t\right) K_{j}^{\alpha }\left( t\right) P_{k}\left( t\right) \exp \left( -%
\frac{d_{ij}\left( t\right) +d_{kj}\left( t\right) }{d}\right) \right)
^{2}\right)  \label{Stw2}
\end{equation}%
with $\sigma _{1}^{2}$ normalized to $\frac{\sigma ^{2}}{\bar{A}^{2}}$ in
the sequel where $\bar{A}^{2}$\ is a constant. We consider market clearing
as a more binding condition than capital accumulation, so that $\bar{A}%
^{2}>>1$. As for the capital, we associate a statistical weight for the
system to the market clearing condition:%
\begin{equation}
\exp \left( -\frac{\bar{A}^{2}}{2\sigma ^{2}}\sum_{i}\int \left(
P_{i}^{1+\gamma }\left( t\right) K_{i}^{\alpha }\left( t\right) -\kappa
\sum_{j,k}P_{j}\left( t\right) K_{j}^{\alpha }\left( t\right) P_{k}\left(
t\right) \exp \left( -\frac{d_{ij}\left( t\right) +d_{kj}\left( t\right) }{d}%
\right) \right) ^{2}dt\right)  \label{Stwm}
\end{equation}

\subsubsection{Probabilistic description of the exchange space dynamics}

The exchange position $X$ does not correspond to a usual economic variable.
We could postulate its dynamic equations, along with its interactions with
the other economic variables, and from there, deduce its probabilistic
description for an arbitrary number of agents.\textbf{\ }However choosing an
ad hoc form is equivalent, simpler and faster.

We postulate three types of forces governing agents' dynamics within the
exchange space and directly write the associated statistical weights.

A first force applies to all agents whatever their position and attracts
them towards the center of the exchange space. Without this force the system
would not exist or would tend to disintegrate. This force could be a
political or social structure assuring the cohesion and exchanges of the
group. More broadly, it can also represent the set of all factors insuring a
minimal level of exchanges for each good. In the following, we will refer to
this force as the "cohesion force".

We postulate a second force induced by the exchanges existing between
agents. We suppose that exchanging agents create connections that will
smooth their further exchanges, and get closer within the exchange space.

Finally, we postulate a third force that counterweight the second force.\ We
suppose that a small group of agents, i.e. close and exchanging, tend to
repell potential new entrants. This force can model exclusive connections
such as clientelism or various degrees of market openness.

Considering these three assumptions, we chose the following statistical
weight for the all set of agents:

\begin{equation}
\exp \left( -\sum_{i}\int \left( \frac{\left( \dot{X}_{i}\left( t\right)
\right) ^{2}}{\sigma _{X}^{2}}+V_{0}\left( X_{i}\left( t\right) \right)
+\sum_{j}V_{1}\left( d_{ij}\left( t\right) \right) +\sum_{j,k}V_{2}\left(
d_{ij}\left( t\right) ,d_{ik}\left( t\right) ,d_{jk}\left( t\right) \right)
\right) dt\right)  \label{STx}
\end{equation}

where $\sigma _{X}^{2}$ is a constant parameter measuring the inertia of $%
X_{i}$. For $\sigma _{X}^{2}<<1$, the variable $X_{i}$ presents a strong
inertia, whereas, for $\sigma _{X}^{2}>>1$ the variable $X_{i}$ adjusts
freely.

This statistical weight describes a random dynamic for the variables $%
X_{i}\left( t\right) $. The first term $\frac{\left( \dot{X}_{i}\left(
t\right) \right) ^{2}}{\sigma _{X}^{2}}$ represents the inertia of the
variable $X_{i}\left( t\right) $. The variation of $X_{i}\left( t\right) $\
over one period, measured by $\dot{X}_{i}\left( t\right) $, is in average%
\textbf{\ }of order $\sigma _{X}$: the value of $X_{i}\left( t\right) $
cannot be changed instantaneously.

The three other terms represent the forces acting on each individual agent,
but also on groups of various size.

The term $V_{1}$\ represents an attraction force between agents exchanging
at the individual level.

\begin{equation}
V_{1}\left( d_{ij}\left( t\right) \right) =-\frac{\kappa _{1}}{4}\frac{%
K_{i}\left( t\right) K_{j}\left( t\right) }{\left\langle K\right\rangle
_{X_{i}\left( t\right) }\left\langle K\right\rangle _{X_{j}\left( t\right) }}%
\exp \left( -\chi _{1}d_{ij}\left( t\right) \right)  \label{V1}
\end{equation}%
Two exchanging agents have a tendency to get closer in an exchange space. $%
\chi _{1}$ is a parameter and $\left\langle K\right\rangle _{X_{i}\left(
t\right) }$ and $\left\langle K\right\rangle _{X_{j}\left( t\right) }$ are
the average capital stock of agents at position $X_{i}\left( t\right) $ and $%
X_{j}\left( t\right) $. As (\ref{V1}) shows, $V_{1}\left( d_{ij}\left(
t\right) \right) $ is proportional to $K_{i}\left( t\right) K_{j}\left(
t\right) /\left\langle K\right\rangle _{X_{i}\left( t\right) }\left\langle
K\right\rangle _{X_{j}\left( t\right) }$: the attraction force is
proportional to agents' exchanges, revenues, and consequently capital stocks.

The term $V_{2}$\ describes some repulsive forces that occur in the
interaction of small groups.\ 

\begin{equation*}
V_{2}\left( d_{ij}\left( t\right) ,d_{ik}\left( t\right) ,d_{jk}\left(
t\right) \right) =\frac{\kappa _{2}}{6}\exp \left( -\chi _{2}\left(
d_{ij}\left( t\right) +d_{ik}\left( t\right) +d_{jk}\left( t\right) \right)
\right)
\end{equation*}%
Here, we have chosen interactions within a group of three agents, but this
could be generalized to $k$ agents, with $k<<N$, where $N$ is the total
number of agents. The idea behind this force is that when several - more
than two - agents interact, the interactions they already have deter
additional interaction.

The term $\chi _{2}$ is a parameter, and the last term, $V_{0}$ is chosen to
be:

\begin{equation*}
V_{0}\left( X_{i}\left( t\right) \right) =\frac{\kappa _{0}}{2\sigma _{X}^{2}%
}\left( X_{i}\left( t\right) -\left\langle X_{i}\left( t\right)
\right\rangle \right) ^{2}
\end{equation*}%
with $\kappa _{0}<<1$\ . This describes a "weak" global force that tends to
regroup all agents towards the center of the exchange space, a cohesion
force.

\subsubsection{Probabilistic description of the system}

Gathering the contributions (\ref{Stwk}), (\ref{Stwm}) and (\ref{STx})\
leads to the probabilistic weight:%
\begin{eqnarray}
&&\exp \left( -\frac{1}{2\sigma ^{2}}\sum_{i}\int dt\left( \dot{K}_{i}\left(
t\right) +\delta K_{i}\left( t\right) \right. \right.  \label{ST} \\
&&\left. \left. -AP_{i}\left( t\right) K_{i}^{\alpha }\left( t\right) \left(
1-\frac{\kappa }{d^{2}}\sum_{j,k}\frac{P_{k}\left( t\right) \exp \left( -%
\frac{d_{ij}\left( t\right) +d_{ik}\left( t\right) }{d}\right) }{%
P_{j}^{\gamma }\left( t\right) }\right) \right) ^{2}\right)  \notag \\
&&\times \exp \left( -\frac{\bar{A}^{2}}{2\sigma ^{2}}\sum_{i}\int \left(
P_{i}^{1+\gamma }\left( t\right) K_{i}^{\alpha }\left( t\right) -\kappa
\sum_{j,k}P_{j}\left( t\right) K_{j}^{\alpha }\left( t\right) P_{k}\left(
t\right) \exp \left( -\frac{d_{ij}\left( t\right) +d_{kj}\left( t\right) }{d}%
\right) \right) ^{2}dt\right)  \notag \\
&&\times \exp \left( -\sum_{i}\int \left( \frac{\left( \dot{X}_{i}\left(
t\right) \right) ^{2}}{\sigma _{X}^{2}}+V_{0}\left( X_{i}\left( t\right)
\right) +\sum_{j}V_{1}\left( d_{ij}\left( t\right) \right)
+\sum_{j,k}V_{2}\left( d_{ij}\left( t\right) ,d_{ik}\left( t\right)
,d_{jk}\left( t\right) \right) \right) dt\right)  \notag
\end{eqnarray}%
Appendix 0 rewrites this functional in an alternative form , more suitable
to switch to the field theory formalism.

\subsection{Statistical fields description}

Describing the whole system by a statistical weight that accounts for a
large number of stochastic paths has allowed us to introduce the various
interactions between agents directly inside the statistical weight.

Formally, this statistical weight could be used to study the evolution of
the system. Equation (\ref{ST}) would allow to compute the transition
probability from an initial state for $N$ agents to a final state. Yet this
approach, possible in some cases, is intractable for a large number of
agents: it would imply to keep track of the $N$\ agents' probability
transitions.

It is however a necessary first step (see Gosselin, Lotz and Wambst 2017,
2018) to turn to the more compact \emph{field formalism}. This formalism
replaces the probabilistic description of $N$\ agents (\ref{ST}) by a
collective field formalism.\ Rather than tracking $N$\ variables, we define
an abstract function, the field, that depends only on one single set of
variable, here $K$, $P$, $X$.

These variables are no more indexed by $i$, the label of individual agents,
because field formalism, rather than considering $N$\ copies of the same
variables and defining a probability for the set of these $N$\ copies,
defines a probability density on the space of complex valued functions of
the variables $K$, $P$, $X$.\textbf{\ }

These complex valued functions $\Psi \left( K,P,X\right) $ replace the set
of paths used in the previous paragraph. Field formalism reduces the model
described by (\ref{ST}) but preserves the essential information contained in
the probabilistic description (\ref{ST}) of the system with $N$\ agents.
Switching to field formalism is a change of perspective: rather than keeping
track of agents $1,2,...,N$, this formalism describes dynamics and
interactions as a collective thread of all possible anonymous paths. This
thread is the environment that conditions the dynamics of individual agents
from one state to another.

To each function $\Psi \left( K,P,X\right) $, we associate a statistical
weight $\exp \left( -S\left( \Psi \right) \right) $\ computing the density
of probability associated to the a particular configuration $\Psi \left(
K,P,X\right) $. The functional $S\left( \Psi \right) $\ is called the field
action. The form of $S\left( \Psi \right) $\ is directly derived from the
probabilistic description of our model (\ref{ST}). Technical details about
the derivation of the field action $S\left( \Psi \right) $\ are given in
Gosselin, Lotz and Wambst (2017) and a detailed abstract can be found in
Gosselin, Lotz and Wambst (2018). In the following, we briefly recall the
main steps and Appendix 1 provides some extensions adapted to our purposes.

\subsubsection{Field theoretic formulation}

The set of variables $\left( K_{i}\left( t\right) ,P_{i}\left( t\right)
,X_{i}\left( t\right) \right) _{i=1,...,N}$ is replaced by a function $\Psi
\left( K,P,X,\theta \right) $, where $K,P,X$ represent all the possible
values of capital, price and exchange position for a non labelled agent. The
parameter $\theta $ is a counting variable and plays the role of time.

The statistical weight is replaced by a weight for this function $\Psi
\left( K,P,X,\theta \right) $, and describes the probability of a set of
configurations $\left( K,P,X,\theta \right) $ for all agents. We will define 
$Z=\left( K,P,X,\theta \right) $ and $D_{ij}=\left\vert
X_{i}-X_{j}\right\vert $. By convention, we will consider that, unless
otherwise mentioned, the sign of integration $\int $ refers to all variables
involved. The integrations ranges for $K,P$ are $%
\mathbb{R}
^{+}$ and for $X$, the interval $\left[ -1,1\right] $. The transformation
from the probabilistic description (\ref{fllwgt}) to a field theoretic model
is described in Gosselin, Lotz and Wambst 2017, 2018, and detailed in
Appendix 1. The derivation of the results of this section is given in
Appendix 2.

We show that the $\left( K,P\right) $ part of the field action is:

\begin{eqnarray}
S_{1}\left( \bar{\Psi}\right) &=&\int \bar{\Psi}^{\dag }\left( Z,\theta
\right) \left( -\sigma ^{2}\nabla _{K}^{2}-\sigma _{X}^{2}\nabla
_{X}^{2}-\vartheta ^{2}\nabla _{\theta }^{2}+\frac{1}{\vartheta ^{2}}+\alpha
\right) \bar{\Psi}\left( Z,\theta \right)  \label{S1} \\
&&+\int \left( \frac{\left( \delta K-APK^{\alpha }\left( 1-\hat{U}%
_{1}\right) \right) ^{2}}{\sigma ^{2}}+\frac{\bar{A}^{2}\left( P^{1+\gamma
}K^{\alpha }+U_{2}\right) ^{2}}{\sigma ^{2}}\right) \left\vert \bar{\Psi}%
\left( Z,\theta \right) \right\vert ^{2}  \notag
\end{eqnarray}%
with:%
\begin{eqnarray*}
\hat{U}_{1} &=&\kappa \int \frac{P_{3}\exp \left( -\left( \frac{D_{12}+D_{13}%
}{d}\right) \right) }{P_{2}^{\gamma }}\left\vert \bar{\Psi}\left(
Z_{2},\theta \right) \right\vert ^{2}\left\vert \bar{\Psi}\left(
Z_{3},\theta \right) \right\vert ^{2} \\
U_{2} &=&-\kappa \int P_{2}\left( K_{2}\right) ^{\alpha }P_{3}\exp \left(
-\left( \frac{D_{12}+D_{23}}{d}\right) \right) \left\vert \Psi \left(
Z_{2},\theta \right) \right\vert ^{2}\left\vert \Psi \left( Z_{3},\theta
\right) \right\vert ^{2}
\end{eqnarray*}%
In the sequel, we simplify the notation by replacing $\bar{\Psi}$ by $\Psi $%
, but the change of variable will be accounted for while computing the
transition functions of the model.

The $X$ part of the weight yields the field contribution: 
\begin{eqnarray}
S_{2}\left( \Psi \right) &=&\sigma _{X}^{2}\int \left\vert \nabla \Psi
\left( Z,\theta \right) \right\vert ^{2}+\int V_{0}\left( X\right)
\left\vert \Psi \left( Z,\theta \right) \right\vert ^{2}+\int V_{1}\left(
D_{12}\right) \left\vert \Psi \left( Z_{1},\theta \right) \right\vert
^{2}\left\vert \Psi \left( Z_{2},\theta \right) \right\vert ^{2}  \label{S2}
\\
&&+\int V_{2}\left( D_{12},D_{13},D_{23}\right) \left\vert \Psi \left(
Z_{1},\theta \right) \right\vert ^{2}\left\vert \Psi \left( Z_{2},\theta
\right) \right\vert ^{2}\left\vert \Psi \left( Z_{3},\theta \right)
\right\vert ^{2}  \notag
\end{eqnarray}%
where:%
\begin{eqnarray*}
V_{0}\left( X\right) &=&\frac{\kappa _{0}}{2\sigma _{X}^{2}}\left(
X-\left\langle X\right\rangle \right) ^{2} \\
V_{1}\left( \left\vert X-Y\right\vert \right) &=&-\frac{\kappa _{1}}{4}\frac{%
KK^{\prime }\exp \left( -\chi _{1}\left\vert X-Y\right\vert \right) }{%
\left\langle K\right\rangle ^{2}} \\
V_{2}\left( \left\vert X-Y\right\vert ,\left\vert X-Z\right\vert ,\left\vert
Y-Z\right\vert \right) &=&\frac{\kappa _{2}}{6}\exp \left( -\chi _{2}\left(
\left\vert X-Y\right\vert +\left\vert X-Z\right\vert +\left\vert
Y-Z\right\vert \right) \right)
\end{eqnarray*}%
and the full following field action $S\left( \Psi \right) =S_{1}\left( \Psi
\right) +S_{2}\left( \Psi \right) $ becomes:%
\begin{eqnarray}
S\left( \Psi \right) &=&\int \Psi ^{\dag }\left( Z,\theta \right) \left(
-\sigma ^{2}\nabla _{K}^{2}-\sigma _{X}^{2}\nabla _{X}^{2}-\vartheta
^{2}\nabla _{\theta }^{2}+\frac{1}{\vartheta ^{2}}+\alpha \right) \Psi
\left( Z,\theta \right)  \label{fllctn} \\
&&+\int \left( V_{0}\left( X\right) +\frac{\left( \delta K-APK^{\alpha
}\left( 1-\hat{U}_{1}\right) \right) ^{2}}{\sigma ^{2}}+\frac{\bar{A}%
^{2}\left( P^{1+\gamma }K^{\alpha }+U_{2}\right) ^{2}}{\sigma ^{2}}\right)
\left\vert \Psi \left( Z,\theta \right) \right\vert ^{2}  \notag \\
&&+\int V_{1}\left( D_{12}\right) \left\vert \Psi \left( Z_{1},\theta
\right) \right\vert ^{2}\left\vert \Psi \left( Z_{2},\theta \right)
\right\vert ^{2}+\int V_{2}\left( D_{12},D_{13},D_{23}\right) \left\vert
\Psi \left( Z_{1},\theta \right) \right\vert ^{2}\left\vert \Psi \left(
Z_{2},\theta \right) \right\vert ^{2}\left\vert \Psi \left( Z_{3},\theta
\right) \right\vert ^{2}  \notag
\end{eqnarray}

\section{Resolution}

Several results about the model can be derived from the field action $%
S\left( \Psi \right) $\ and its statistical weight $\exp \left( -S\left(
\Psi \right) \right) $: the phases of the system, that describe the
collective background of the system, and the transition functions of the
system.

The \emph{phases of the system} are defined by the field(s) $\Psi _{0}\left(
Z,\theta \right) $\ that maximize(s) the statistical weight $\exp \left(
-S\left( \Psi \right) \right) $, i.e. minimize(s) $S\left( \Psi \right) $.
The field $\Psi _{0}\left( Z,\theta \right) $\ is the most likely
configuration: it represents some collective background field(s) that
condition(s) the dynamics of individual agents.

The existence of a minimum for $S\left( \Psi \right) $\ (\ref{fllctn})
depends on the parameters of the system. For some values of the parameters,
only the trivial phase $\Psi _{0}\left( Z,\theta \right) =0$\ exists, and
amounts to a system linearized around a static equilibrium. When non-trivial
phases $\Psi _{0}\left( Z,\theta \right) \neq 0$\ exist, they reveal other
types of equilibria. The configuration $\Psi _{0}\left( Z,\theta \right) $\
is the background state in which probability transitions and some average
values can be computed. Thus, the form of $\Psi _{0}$ has direct
implications on the system's dynamics.

The \emph{transition functions, or Green functions, }for $k$ agents compute,
for a given phase, the probability for $k$ agents to evolve from one initial
state - the given values of capital, price and exchange position for each of
the $k$ agents - to a final state in a certain time span. Thus, the field
formalism recovers the random description of one or any number of agents. We
will see below that, in a given phase, the $k$ agent transitions functions
can be recovered from the single agent transition probability.

To find the transition functions, we use the fact that $\exp \left( -S\left(
\Psi \right) \right) $\ represents itself a statistical weight for the
system.\ In fact, the second order expansion of $S\left( \Psi \right) $\
around $\Psi _{0}\left( Z,\theta \right) $\ yields a quadratic action that
determines directly the Green functions of the system. The derivation of
this point can be found in (Gosselin, Lotz, Wambst 2018).

In this section, we study the conditions for the appearance of a non-trivial
phase for the system described by (\ref{fllctn}). We find that, depending on
the parameters of the system, two potential phases arise. We then compute
for each possible phase an "effective" quadratic action. This is a
simplified version of (\ref{fllctn}) in which the price variable can be
replaced and the specificities of a given phase are taken into account. This
quadratic action is then used to compute the transition functions in each
phase.

\subsection{Possibility of several phase}

In this paragraph, inspecting the configurations $\Psi _{0}\left( Z,\theta
\right) $ that minimize the action, we find the conditions of appearance of
a non-trivial phase for the system.

In the absence of any dynamics for $X$, since the potential for $K,P$ is
positive, the minimal configuration is null. Thus, the possibility of
non-trivial configuration depends on the $X$ part of the action (\ref{S2}).
Using a first approximation for $S_{2}\left( \bar{\Psi}\right) $ the
minimization of (\ref{S2}) yields: 
\begin{eqnarray*}
0 &=&-\sigma _{X}^{2}\nabla _{X}^{2}\Psi \left( Z,\theta \right)
+V_{0}\left( X\right) \Psi \left( Z,\theta \right) +\left( \int V_{1}\left(
X-X_{2}\right) \left\vert \Psi \left( Z_{2},\theta \right) \right\vert
^{2}\right) \Psi \left( Z,\theta \right) \\
&&+\left( \int V_{2}\left( X-X_{2},X-X_{3},D_{23}\right) \left\vert \Psi
\left( Z_{2},\theta \right) \right\vert ^{2}\left\vert \Psi \left(
Z_{3},\theta \right) \right\vert ^{2}\right) \Psi \left( Z,\theta \right)
\end{eqnarray*}%
and we replace some quantities by their average:%
\begin{eqnarray*}
\left\langle X\right\rangle &=&0 \\
\left\vert X-X_{i}\right\vert &\simeq &\sqrt{\left\langle \left(
X-X_{i}\right) ^{2}\right\rangle }\simeq \sqrt{2\left\langle
X^{2}\right\rangle }\simeq \sqrt{2}\kappa _{0}^{-\frac{1}{4}}\sigma _{X}
\end{eqnarray*}%
\begin{eqnarray*}
V_{1}\left( \left\vert X-X_{2}\right\vert \right) &=&-\frac{\kappa _{1}}{4}%
\frac{KK^{\prime }\exp \left( -\chi _{1}\left\vert X-X_{2}\right\vert
\right) }{\left\langle K\right\rangle ^{2}}\exp \left( -\chi _{1}\left\vert
X-X_{2}\right\vert \right) \\
&\simeq &-\frac{\kappa _{1}}{4}\frac{KK^{\prime }\exp \left( -\chi
_{1}\left\vert X-X_{2}\right\vert \right) }{\left\langle K\right\rangle ^{2}}%
\exp \left( -\chi _{1}\sqrt{2}\kappa _{0}^{-\frac{1}{4}}\sigma _{X}\right)
\equiv -\frac{\bar{\kappa}_{1}}{4}
\end{eqnarray*}%
and similarly:%
\begin{equation*}
V_{2}\left( X-X_{2},X-X_{3},D_{23}\right) \simeq \frac{\kappa _{2}}{6}\exp
\left( -\chi _{2}3\sqrt{2}\kappa _{0}^{-\frac{1}{4}}\sigma _{X}\right)
\equiv \frac{\bar{\kappa}_{2}}{6}
\end{equation*}%
In the sequel, the parameters $\chi _{1}$, $\chi _{2}$ and $\kappa _{0}$
will be considered relatively small.\ Since we are only concerned with
finding approximate conditions for the existence of a non-trivial phase, we
can approximate $\bar{\kappa}_{1}\simeq \kappa _{1}$, $\bar{\kappa}%
_{2}\simeq \kappa _{2}$ . We are thus left with an approximated equation:%
\begin{eqnarray}
&&0=-\sigma _{X}^{2}\nabla _{X}^{2}\Psi \left( Z,\theta \right) +\frac{%
\kappa _{0}}{\sigma _{X}^{2}}X^{2}\Psi \left( Z,\theta \right)
\label{prtfndm} \\
&&+\kappa _{1}\left( \int \left\vert \Psi \left( Z_{2},\theta \right)
\right\vert ^{2}\right) \Psi \left( Z,\theta \right) +\kappa _{2}\left( \int
\left\vert \Psi \left( Z_{2},\theta \right) \right\vert ^{2}\left\vert \Psi
\left( Z_{3},\theta \right) \right\vert ^{2}\right) \Psi \left( Z,\theta
\right)  \notag
\end{eqnarray}%
We define $\rho ^{2}=\int \left\vert \Psi \left( Z,\theta \right)
\right\vert ^{2}$, and assume a fundamental of the form $\Psi _{0}\left(
Z,\theta \right) =\Psi _{0}\left( K\right) \Psi _{0}\left( P\right) \Psi
_{0}\left( X\right) \Psi _{0}\left( \theta \right) $, that we will justify
later.\ Equation (\ref{prtfndm}) rewrites:%
\begin{equation*}
0=-\sigma _{X}^{2}\nabla _{X}^{2}\Psi \left( Z,\theta \right) +\frac{\kappa
_{0}}{\sigma _{X}^{2}}X^{2}\Psi \left( Z,\theta \right) +\kappa _{1}\rho
^{2}\Psi \left( Z,\theta \right) +\kappa _{2}\rho ^{4}\Psi \left( Z,\theta
\right)
\end{equation*}%
The associated fundamental eigenvalue satisfies:%
\begin{equation*}
\frac{1}{2}\kappa _{0}^{\frac{1}{2}}-\kappa _{1}\rho ^{2}+\kappa _{2}\rho
^{4}=0
\end{equation*}%
Thus for $\kappa _{1}^{2}-2\kappa _{0}^{\frac{1}{2}}\kappa _{2}<0$ the
system has only one phase, the trivial phase $\rho =0$, i.e. $\Psi \left(
Z,\theta \right) =0$.

On the contrary, for $\kappa _{1}^{2}-2\kappa _{0}^{\frac{1}{2}}\kappa
_{2}>0 $, there is a possibility of non-trivial phase with: 
\begin{equation*}
\rho \simeq \frac{\kappa _{1}+\sqrt{\kappa _{1}^{2}-2\kappa _{0}^{\frac{1}{2}%
}\kappa _{2}}}{2\kappa _{2}}
\end{equation*}%
The value of $\rho $ will be refined below.

As a consequence, the possibility of a non-trivial phase depends on the
relative strength of the repulsive force over the attractive one. A
non-trivial phase is possible only for a strong enough repulsive force.

\subsection{Derivation of the effective action}

For each phase, we derive a simpler form of the "effective action" (\ref%
{fllctn}). Due to the binding market clearing condition, the price $P$ in (%
\ref{fllctn})\ is not a dynamic variable: there are no derivatives in $P$,
i.e. $\nabla _{P}$. We will show that this variable can consequently be
expressed in terms of $K$, $X$ and the field. It is this simplification that
allows to find a quadratic effective action for a field depending on two
variables $K$ and $X$.

\subsubsection{Phase 1: $\protect\rho =0$}

We first inspect the case $\rho =0$. Recall that we assume relatively small
individual fluctuations around the market clearing condition, so that $\bar{A%
}>>A$. We also assume that $\sigma ^{2}<1$. The ratio $\frac{A}{\bar{A}}$\
will thus measure the effect of these fluctuations on the system.

As a consequence, in the statistical weight $\exp \left( -\left( S\left(
\Psi \right) \right) \right) $\ with $S\left( \Psi \right) $\ defined by (%
\ref{fllctn}), the potential term:%
\begin{equation}
\int \left( \frac{\left( \delta K-APK^{\alpha }\left( 1-\hat{U}_{1}\right)
\right) ^{2}}{\sigma ^{2}}+\frac{\bar{A}^{2}\left( P^{1+\gamma }K^{\alpha
}+U_{2}\right) ^{2}}{\sigma ^{2}}\right) \left\vert \Psi \left( Z,\theta
\right) \right\vert ^{2}  \label{Ptn}
\end{equation}

is predominant. It implies that $\exp \left( -\left( S\left( \Psi \right)
\right) \right) $ is peaked around fields minimizing (\ref{Ptn}). Since the
dependence of $\Psi $\ in $P$ is static, i.e. no gradient in $P$ appears in $%
S\left( \Psi \right) $, we can consider that the fields minimizing (\ref{Ptn}%
) have the form:%
\begin{equation}
\Psi \left( Z,\theta \right) \rightarrow \delta \left( P-F\left( K,X\right)
\right) \Psi \left( K,X,\theta \right)  \label{Prct}
\end{equation}%
In other words, the most likely fields are those of the form (\ref{Prct}).
Classically, the interpretation is straightforward: due to the market
clearing condition, the price is a function $F$ of $\left( K,X\right) $.

Inserting (\ref{Prct}) into the potential terms (\ref{Ptn}), the
minimization equation of (\ref{Ptn}) becomes:%
\begin{equation}
\Psi \left( K,P,X,\theta \right) \left( -UAK^{\alpha }\left( \delta
K-APK^{\alpha }U\right) +\left( 1+\gamma \right) P^{\gamma }\bar{A}%
^{2}K^{\alpha }\left( P^{1+\gamma }K^{\alpha }-V\right) \right) \Psi ^{\dag
}\left( K,P,X,\theta \right) =0  \label{Pmnz}
\end{equation}%
where $U=1-\hat{U}_{1}$. Appendix $2$ shows that inserting (\ref{Prct}) in (%
\ref{Pmnz}) leads to identify: 
\begin{equation}
P=F\left( K,X\right) =K^{-\frac{\alpha }{1+\gamma }}f\left( X\right)
\label{Pr}
\end{equation}%
where $f$ is a trial function that must be identified in (\ref{Pmnz}).\ We
thus arrive to the following equation for $f\left( X\right) $: 
\begin{equation}
\left( f\left( X_{1}\right) \right) ^{1+\gamma }=\kappa \int \left(
K_{2}\right) ^{\frac{\alpha \gamma }{1+\gamma }}\left( K_{3}\right) ^{-\frac{%
\alpha }{1+\gamma }}f\left( X_{3}\right) f\left( X_{2}\right) \exp \left(
-\left( \frac{D_{12}+D_{23}}{d}\right) \right) \left\vert \Psi \left(
K_{3},X_{3}\right) \right\vert ^{2}\left\vert \Psi \left( K_{2},X_{2}\right)
\right\vert ^{2}  \label{eqtn1}
\end{equation}%
Appendix $2$ shows that we obtain in first approximation: 
\begin{equation*}
f\left( X\right) =D\exp \left( -\frac{\left\vert X\right\vert }{d\left(
1+\gamma \right) }\right)
\end{equation*}%
with:%
\begin{eqnarray*}
D &=&\left( \frac{A}{\delta }\left( 1-h\exp \left( -\frac{1}{2d}\right)
\left( 1-\frac{\exp \left( -\frac{1}{2d}\right) }{2}\right) \right) \right)
^{\frac{\alpha }{\left( 1-\alpha \right) \left( \gamma +1\right) }} \\
&&\times \left( \frac{1+\gamma \left( 1-\alpha \right) }{\left( 2-\alpha
\right) \bar{\kappa}}\left( \left( \frac{2-\alpha }{1+\gamma \left( 1-\alpha
\right) }\right) ^{2}-1\right) \right) ^{\frac{1+\gamma \left( 1-\alpha
\right) }{\left( 1-\gamma ^{2}\right) \left( 1-\alpha \right) }}
\end{eqnarray*}%
Appendix $2$ also shows that the average capital stock $\left\langle
K\right\rangle _{X}$ for agents at position $X$ in the exchange space
rewrites:%
\begin{equation}
\left\langle K\right\rangle _{X}=\left( \frac{A}{\delta }f\left( X\right)
\left( 1-h\exp \left( -\frac{\left\vert X\right\vert }{d}\right) \left( 1-%
\frac{\cosh \frac{X}{d}}{\exp \left( \frac{1}{d}\right) }\right) \right)
\right) ^{\frac{1+\gamma }{1+\gamma \left( 1-\alpha \right) }}  \label{KavX}
\end{equation}%
where $h$ and $\bar{\kappa}$\ are defined by: 
\begin{eqnarray}
\bar{\kappa} &=&\kappa \left( 1-h\frac{A^{2}}{\bar{A}^{2}}\left( 1-\exp
\left( -\frac{1}{d}\right) \right) \right)  \label{qnkp} \\
h &=&\frac{\bar{h}}{\left( 1-\frac{A^{2}}{\bar{A}^{2}}\bar{h}\left( 1-\exp
\left( -\frac{1}{d}\right) \right) \right) }  \label{qnh} \\
\bar{h} &=&\left( 1+\gamma \left( 1-\alpha \right) \right) \left( \frac{%
1+\gamma \left( 1-\alpha \right) }{2\left( 2-\alpha \right) }\left( \left( 
\frac{2-\alpha }{1+\gamma \left( 1-\alpha \right) }\right) ^{2}-1\right)
\right) \left( \frac{3}{2}-\frac{e^{-\frac{1}{d}}}{2}\right)  \label{qnhb}
\end{eqnarray}%
Moreover, we also show in Appendix $2$ that $P$, as a function of $K$ and $X$%
, writes:%
\begin{equation}
P=\frac{\left( \frac{1+\gamma \left( 1-\alpha \right) }{\left( 2-\alpha
\right) \bar{\kappa}}\left( \left( \frac{2-\alpha }{1+\gamma \left( 1-\alpha
\right) }\right) ^{2}-1\right) \right) ^{\frac{1}{1-\gamma }}\exp \left( -%
\frac{\left( 1-\alpha \right) \left\vert X\right\vert }{d\left( 1+\gamma
\left( 1-\alpha \right) \right) }\right) }{\left( \frac{K}{\left\langle
K\right\rangle _{X}}\right) ^{\frac{\alpha }{1+\gamma }}\left( \left( \frac{%
1-h\exp \left( -\frac{\left\vert X\right\vert }{d}\right) \left( 1-\frac{%
\cosh \frac{X}{d}}{\exp \left( \frac{1}{d}\right) }\right) }{1-h\exp \left( -%
\frac{1}{2d}\right) \left( 1-\frac{\exp \left( -\frac{1}{2d}\right) }{2}%
\right) }\right) \right) ^{\frac{\alpha }{1+\gamma \left( 1-\alpha \right) }}%
}  \label{PX}
\end{equation}%
Once the price level associated to $K$ and $X$ is found, the potential (\ref%
{Ptn}) becomes:

\begin{equation}
\frac{\omega ^{2}}{\sigma ^{2}}\int \Psi ^{\dag }\left( K,X,\theta \right)
\left( K-\left\langle K\right\rangle _{X}\right) ^{2}\Psi \left( K,X,\theta
\right)  \label{Ufpr}
\end{equation}%
with:%
\begin{equation*}
\omega =\sigma \sqrt{\delta ^{2}+\frac{\bar{A}^{2}A^{2}}{\left( A^{2}\left(
1-h\left( 1-\exp \left( -\frac{1}{d}\right) \right) \right) ^{2}+\bar{A}%
^{2}\right) ^{2}}}
\end{equation*}%
and the effective action, is now equal to: 
\begin{eqnarray}
S\left( \Psi \right) &=&\int \Psi ^{\dag }\left( K,X,\theta \right) \left(
-\sigma ^{2}\nabla _{K}^{2}-\sigma _{X}^{2}\nabla _{X}^{2}-\vartheta
^{2}\nabla _{\theta }^{2}+\frac{\omega ^{2}}{\sigma ^{2}}\left(
K-\left\langle K\right\rangle _{X}\right) ^{2}+\frac{1}{\vartheta ^{2}}%
+\alpha \right) \Psi \left( K,X,\theta \right)  \label{fllctn1} \\
&&+\int \left( V_{0}\left( X\right) \right) \left\vert \Psi \left(
K,X,\theta \right) \right\vert ^{2}+\int V_{1}\left( D_{12}\right)
\left\vert \Psi \left( K_{1},X_{1},\theta \right) \right\vert ^{2}\left\vert
\Psi \left( K_{2},X_{2},\theta \right) \right\vert ^{2}  \notag \\
&&+\int V_{2}\left( D_{12},D_{13},D_{23}\right) \left\vert \Psi \left(
K_{1},X_{1},\theta \right) \right\vert ^{2}\left\vert \Psi \left(
K_{2},X_{2},\theta \right) \right\vert ^{2}\left\vert \Psi \left(
K_{3},X_{3},\theta \right) \right\vert ^{2}  \notag
\end{eqnarray}%
This simplified form will be useful to compute the transition functions of
the agents in the system.

We can now interpret these first set of results. From the above, we can see
that the average level of capital is a function of the variable $X$ and the
parameter $d$. The average level of capital of an agent (\ref{KavX})\
decreases exponentially as a function of his distance from the center $X$.
Indeed, an agent at the center has more opportunities to exchange.\ He faces
a higher demand, has a higher price, a higher income, and accumulates more
capital. Our results show that capital accumulation is an increasing
function of parameter $d$, the average distance of interactions with other
agents.\ Recall that agents exchange with all other agents, but that due to
the exponential form of consumption an agent exchanges mainly with the
agents on an interval of length $d$, centered around his position. The more
the agent can exchange - the better the infrastructures for instance - the
higher capital accumulation. Besides, inspection of (\ref{qnh}) shows that
the fluctuations of prices around the market clearing condition deteriorate
the average level of capital stock.

Finally, and as expected, equation (\ref{KavX}) shows that high capital
productivity favors capital accumulation while higher rates of capital
depreciation reduce capital stock.

Moreover, the above results show that the level of prices is a function of
the variables $K$, $X$, and the parameter $d$. The price (\ref{PX}) of a
good produced by an agent in a specific position $X$ is a decreasing
function of the ratio between the level of capital and the level of average
capital in $X$. We retrieve the result that the higher the level of capital,
the more agents produce and sell at lower prices. However, this dependence
to the level of capital is relative.\ Rather, it is the ratio of the agent's
level of capital to the average level of capital that is determinant. The
price (\ref{PX}) is also an exponentially decreasing function of position $X$%
.\ Given a constant level of capital, an agent at the periphery faces a
lower demand and has lower prices than at the center of the exchange space.
Ultimately, the price of a good produced is an increasing function of the
parameter $d$.\ The higher $d$, the higher the volume of echanges, and the
higher the prices.

\subsubsection{Phase 2: $\protect\rho >0$}

In phase 2, computations are similar to phase 1 but expectations are
computed for a field $\Psi _{0}\left( Z,\theta \right) +\Psi \left( Z,\theta
\right) $. We thus compute the corrections to phase 1 due to the non-trivial 
$\Psi _{0}\left( Z,\theta \right) $.

Appendix $3$ shows that the previous equations defining the potential (\ref%
{Ptn}), $P$ (\ref{Pmnz}) and $\left\langle K\right\rangle _{X}$ are still
valid: 
\begin{equation}
P=F\left( K,X\right) =K^{-\frac{\alpha }{1+\gamma }}f\left( X\right)
\label{Pr2}
\end{equation}%
However, now the trial function $f$ satisfies: 
\begin{equation}
\left( f\left( X_{1}\right) \right) ^{1+\gamma }=\kappa \int \left(
K_{2}\right) ^{\frac{\alpha \gamma }{1+\gamma }}\left( K_{3}\right) ^{-\frac{%
\alpha }{1+\gamma }}f\left( X_{3}\right) f\left( X_{2}\right) \exp \left(
-\left( \frac{D_{12}+D_{23}}{d}\right) \right) \left\vert \bar{\Psi}\left(
K_{3},X_{3}\right) \right\vert ^{2}\left\vert \bar{\Psi}\left(
K_{2},X_{2}\right) \right\vert ^{2}  \label{eqtn2}
\end{equation}%
with $\bar{\Psi}\left( K,X,\theta \right) =\Psi _{0}\left( K,X,\theta
\right) +\Psi \left( K,X,\theta \right) $. Equation (\ref{eqtn2}) is
actually equation (\ref{eqtn1}), but evaluated now for a "translated field",
since the properties of $f\left( X_{1}\right) $ are modified by the
non-trivial fundamental state. As explained above, this reflects the
particular features of the non-trivial phase and the collective contribution
of the environment created by the set of all agents.

Appendix $3$ computes the function $f$ in two steps corresponding to the
decomposition $\bar{\Psi}\left( K,X,\theta \right) =\Psi _{0}\left(
K,X,\theta \right) +\Psi \left( K,X,\theta \right) $. We find:

\begin{equation*}
f\left( X\right) =D\exp \left( -\frac{\left\vert X\right\vert }{1+\gamma }%
\right)
\end{equation*}%
with:%
\begin{eqnarray*}
D &=&\left( \frac{A}{\delta }\left( 1-h\exp \left( -\frac{1}{2d}\right)
\left( 1-\frac{\exp \left( -\frac{1}{2d}\right) }{2}\right) \right) \right)
^{\frac{\alpha }{\left( 1-\alpha \right) \left( \gamma +1\right) }} \\
&&\times \left( \frac{1+\gamma \left( 1-\alpha \right) }{\left( 2-\alpha
\right) \bar{\kappa}\rho ^{4}}\left( \left( \frac{2-\alpha }{1+\gamma \left(
1-\alpha \right) }\right) ^{2}-1\right) \right) ^{\frac{1+\gamma \left(
1-\alpha \right) }{\left( 1-\gamma ^{2}\right) \left( 1-\alpha \right) }}
\end{eqnarray*}%
and where $h$ is defined as in (\ref{qnh}). The associated average capital
stock is given by:%
\begin{equation}
\left\langle K\right\rangle _{X}=\left( \frac{A}{\delta }f\left( X\right)
\left( 1-h\exp \left( -\frac{\left\vert X\right\vert }{d}\right) \left( 1-%
\frac{\cosh \frac{X}{d}}{\exp \left( \frac{1}{d}\right) }\right) \right)
\right) ^{\frac{1+\gamma }{1+\gamma \left( 1-\alpha \right) }}  \label{KavX2}
\end{equation}%
with $h$ is defined as in (\ref{qnh}). Moreover, we also show in Appendix $3$
that $P$, as a function of $K$ and $X$, writes: 
\begin{equation}
P=\frac{\left( \frac{1+\gamma \left( 1-\alpha \right) }{\left( 2-\alpha
\right) \bar{\kappa}\rho ^{4}}\left( \left( \frac{2-\alpha }{1+\gamma \left(
1-\alpha \right) }\right) ^{2}-1\right) \right) ^{\frac{1}{1-\gamma }}\exp
\left( -\frac{\left( 1-\alpha \right) \left\vert X\right\vert }{d\left(
1+\gamma \left( 1-\alpha \right) \right) }\right) }{\left( \frac{K}{%
\left\langle K\right\rangle _{X}}\right) ^{\frac{\alpha }{1+\gamma }}\left(
\left( \frac{1-h\exp \left( -\frac{\left\vert X\right\vert }{d}\right)
\left( 1-\frac{\cosh \frac{X}{d}}{\exp \left( \frac{1}{d}\right) }\right) }{%
1-h\exp \left( -\frac{1}{2d}\right) \left( 1-\frac{\exp \left( -\frac{1}{2d}%
\right) }{2}\right) }\right) \right) ^{\frac{\alpha }{1+\gamma \left(
1-\alpha \right) }}}  \label{PX2}
\end{equation}%
Having found $\left\langle K\right\rangle _{X}$, appendix $3$ shows that the
potential part involving $K$ has the same form as in the first phase (\ref%
{Ufpr}): 
\begin{equation}
\frac{\omega ^{2}}{\sigma ^{2}}\int \Psi ^{\dag }\left( K,X,\theta \right)
\left( K-\left\langle K\right\rangle _{X}\right) ^{2}\Psi \left( K,X,\theta
\right)  \label{Ufpr2}
\end{equation}%
These results are similar to those of phase 1. The patterns of (\ref{PX2})
and (\ref{KavX2}) as a function of $K$, $X$ and $d$, are the same as those
of (\ref{PX}) et (\ref{KavX}).\ Only their magnitudes differ. This is due to 
$\rho ^{2}$, the norm of the fundamental state $\Psi _{0}$, a parameter
specific to phase 2. The two phases will be compared in our interpretation
of the results. The formula for the effective action in phase 2 is also the
same as (\ref{fllctn1}). Yet again, in phase $2$, these formulas depend
explicitly on the value of $\rho ^{2}$. We previously found an approximative
value for $\rho $. With the results now at hand, the value of $\rho $ can be
found more precisely by writing the equation for the state $\Psi _{0}\left(
K,X,\theta \right) $. Appendix $3$ shows that:%
\begin{equation*}
\rho ^{2}=\frac{\kappa _{1}+\sqrt{\kappa _{1}^{2}-2\kappa _{2}\left( 2\alpha
+\sqrt{\kappa _{0}}+\sigma \delta \right) }}{2\kappa _{2}}
\end{equation*}%
and the precise form of $\Psi _{0}\left( K,X,\theta \right) $ is given by:%
\begin{equation*}
\Psi _{0}\left( K,X,\theta \right) =\rho N\Psi _{0}\left( K\right) \Psi
_{0}\left( X\right)
\end{equation*}%
where:%
\begin{eqnarray*}
N &=&\frac{1}{2\pi \sqrt{\kappa _{0}^{-\frac{1}{2}}\sigma \sqrt{\delta ^{2}+%
\frac{\bar{A}^{2}A^{2}}{\left( A^{2}\left( 1-h\left( 1-\exp \left( -\frac{1}{%
d}\right) \right) \right) ^{2}+\bar{A}^{2}\right) ^{2}}}}} \\
\rho ^{2} &=&\frac{\kappa _{1}+\sqrt{\kappa _{1}^{2}-2\kappa _{2}\left(
2\alpha +\frac{2}{\vartheta ^{2}}+\sqrt{\kappa _{0}}+\sigma \delta \right) }%
}{2\kappa _{2}} \\
\Psi _{0}\left( K\right) &=&\exp \left( -\frac{\sqrt{\delta ^{2}+\frac{\bar{A%
}^{2}A^{2}}{\left( A^{2}\left( 1-h\left( 1-\exp \left( -\frac{1}{d}\right)
\right) \right) ^{2}+\bar{A}^{2}\right) ^{2}}}\left( K-\left\langle
K\right\rangle _{X}\right) ^{2}}{2\sigma }\right) \\
\Psi _{0}\left( X\right) &=&\exp \left( -\frac{\kappa _{0}^{-\frac{1}{2}%
}\left( X+\delta X\right) ^{2}}{2}\right) H\left( x\right) +\exp \left( -%
\frac{\kappa _{0}^{-\frac{1}{2}}\left( X-\delta X\right) ^{2}}{2}\right)
H\left( -x\right) \\
\delta X &=&\left( \chi _{1}\kappa _{1}\frac{K}{2\left\langle K\right\rangle 
}\rho ^{2}-\chi _{2}\kappa _{2}\rho ^{4}\right)
\end{eqnarray*}

\subsection{Transition probabilities}

In the previous paragraph, replacing $P$ as a function of $K$, $X$ and the
field has reduced the system to a field theory of the two variables $K$ and $%
X$ with quadratic potentials (\ref{Ufpr}) and (\ref{Ufpr2}). This
simplification allows to compute the transition probabilities for the system
in both phases, i.e. the probabilities for agents to evolve from one initial
state of $K$, $X$, and $P$ towards a final state during a certain time-span,
given the environment created by the set of $N$ agents (see Gosselin, Lotz,
Wambst 2018). To do so, we first simplify in each phase the part of the
action dependent on $X$.

\subsubsection{Phase 1: $\protect\rho =0$}

Using the effective action (\ref{fllctn1}), we can isolate the part of the
action dependent on $X$: 
\begin{eqnarray}
S_{2}\left( \Psi \right) &=&\sigma _{X}^{2}\int \left\vert \nabla \Psi
\left( Z,\theta \right) \right\vert ^{2}+\int V_{0}\left( X\right)
\left\vert \Psi \left( Z,\theta \right) \right\vert ^{2}+\int V_{1}\left(
D_{12}\right) \left\vert \Psi \left( Z_{1},\theta \right) \right\vert
^{2}\left\vert \Psi \left( Z_{2},\theta \right) \right\vert ^{2}
\label{S2pt} \\
&&+\int V_{2}\left( D_{12},D_{13},D_{23}\right) \left\vert \Psi \left(
Z_{1},\theta \right) \right\vert ^{2}\left\vert \Psi \left( Z_{2},\theta
\right) \right\vert ^{2}\left\vert \Psi \left( Z_{3},\theta \right)
\right\vert ^{2}  \notag
\end{eqnarray}%
This action $S_{2}\left( \Psi \right) $ can be evaluated replacing the
interaction potentials by their average in phase 1:

\begin{eqnarray*}
&&-\int \frac{\kappa _{1}}{2}\left\vert \Psi \left( K,X,\theta \right)
\right\vert ^{2}\frac{KK^{\prime }\exp \left( -\chi _{1}\left\vert
x-y\right\vert \right) }{\left\langle K\right\rangle ^{2}}\left\vert \Psi
\left( K^{\prime },Y,\theta \right) \right\vert ^{2} \\
&\rightarrow &-\kappa _{1}\int \left\langle \frac{KK^{\prime }\exp \left(
-\chi _{1}\left\vert x-y\right\vert \right) }{\left\langle K\right\rangle
^{2}}\left\vert \Psi \left( K^{\prime },Y,\theta \right) \right\vert
^{2}\right\rangle \left\vert \Psi \left( K,X,\theta \right) \right\vert ^{2}
\end{eqnarray*}%
and:%
\begin{eqnarray*}
&&\int \frac{\kappa _{2}}{3}\left\vert \Psi \left( K,X,\theta \right)
\right\vert ^{2}\exp \left( -\chi _{2}\left\vert X-Y\right\vert -\chi
_{2}\left\vert X-Z\right\vert -\chi _{2}\left\vert Y-Z\right\vert \right)
\left\vert \Psi \left( K^{\prime },Y,\theta \right) \right\vert
^{2}\left\vert \Psi \left( K^{^{\prime \prime }},Z,\theta \right)
\right\vert ^{2} \\
&\rightarrow &\kappa _{2}\int \left\vert \Psi \left( K,X,\theta \right)
\right\vert ^{2}\left\langle \exp \left( -\chi _{2}\left\vert X-Y\right\vert
-\chi _{2}\left\vert X-Z\right\vert -\chi _{2}\left\vert Y-Z\right\vert
\right) \left\vert \Psi \left( K^{\prime },Y,\theta \right) \right\vert
^{2}\left\vert \Psi \left( K^{^{\prime \prime }},Z,\theta \right)
\right\vert ^{2}\right\rangle
\end{eqnarray*}%
The computations, performed in Appendix 4, lead to an approximation:%
\begin{equation*}
S_{2}\left( \Psi \right) =-\sigma _{X}^{2}\Psi ^{\dag }\left( x\right)
\nabla _{X}^{2}\Psi \left( x\right) +\omega _{X}\Psi ^{\dag }\left( x\right)
X^{2}\Psi \left( x\right) +\alpha _{X}\Psi ^{\dag }\left( x\right) \Psi
\left( x\right)
\end{equation*}%
with:%
\begin{eqnarray}
\omega _{X} &=&\kappa _{0}+\kappa _{1}\frac{K}{\left\langle K\right\rangle }%
\exp \left( -\chi _{1}\right) \chi _{1}-\kappa _{2}\exp \left( -\chi
_{2}\right) \chi _{2}  \label{Omega X} \\
\alpha _{X} &=&\alpha -\kappa _{1}\frac{K}{\left\langle K\right\rangle }%
\frac{2\left( 1-\exp \left( -\chi _{1}\right) \right) }{\chi _{1}}+\kappa
_{2}\frac{2\left( 1-\exp \left( -\chi _{2}\right) \right) }{\chi _{2}} 
\notag
\end{eqnarray}%
Gathering the part of the effective action depending on $K$ and our
simplification for $S_{2}\left( \Psi \right) $, the system is thus described
by an overall action:%
\begin{equation}
S\left( \Psi \right) =\Psi ^{\dag }\left( K,X,\theta \right) \left( -\sigma
^{2}\nabla _{K}^{2}-\vartheta ^{2}\nabla _{\theta }^{2}-\sigma
_{X}^{2}\nabla _{X}^{2}+\frac{\omega ^{2}}{\sigma ^{2}}\left( K-\left\langle
K\right\rangle _{X}\right) ^{2}+\omega _{X}X^{2}+\frac{1}{\vartheta ^{2}}%
+\alpha _{X}\right) \Psi \left( K,X,\theta \right)  \label{SpsiPh1}
\end{equation}%
The equation (\ref{SpsiPh1}) describes a system that looks like a
three-variable system with quadratic potential. An important difference
however is that each variable of the system influences the other
non-linearly.

To get a better insight on the dynamics, we will simplify the mechanism at
hand, and consider that when an agent moves from $K,X$ towards $K^{\prime
},X^{^{\prime }}$, we will average along the path the influence of one
variable on the other. To do so, we first set for two values $X$ and $%
X^{\prime }$: 
\begin{equation}
\left\langle K\right\rangle =\frac{\left\langle K\right\rangle
_{X}+\left\langle K\right\rangle _{X^{\prime }}}{2}  \label{Kav}
\end{equation}

where $\left\langle K\right\rangle $ is an average of capital for agents
whose trajectory in the exchange space starts from $X$ to reach $X^{\prime }$%
. This represents the average influence of X on the capital stock within the
dynamics.

Similarly, we define average values for $\omega _{X}$ and $\alpha _{X}$
along a path of capital from $K$ to $K^{\prime }$. 
\begin{eqnarray}
\bar{\omega}_{X} &=&\kappa _{0}+\kappa _{1}\frac{\kappa _{1}}{2}\left( \frac{%
K}{\left\langle K\right\rangle _{X}}+\frac{K^{\prime }}{\left\langle
K\right\rangle _{Y}}\right) \exp \left( -\chi _{1}\right) \chi _{1}-\kappa
_{2}\exp \left( -\chi _{2}\right) \chi _{2}  \label{inertX} \\
\bar{\alpha}_{X} &=&\alpha -\kappa _{1}\frac{\kappa _{1}}{2}\left( \frac{K}{%
\left\langle K\right\rangle _{X}}+\frac{K^{\prime }}{\left\langle
K\right\rangle _{Y}}\right) \frac{2\left( 1-\exp \left( -\chi _{1}\right)
\right) }{\chi _{1}}+\kappa _{2}\frac{2\left( 1-\exp \left( -\chi
_{2}\right) \right) }{\chi _{2}}  \notag
\end{eqnarray}

Once (\ref{SpsiPh1}) found and given (\ref{Kav}) and (\ref{inertX}), the
transition probability for one agent in the first phase can be computed.\
Recall that this is the probability for an agent to evolve from an initial
state to a final state during a certain timespan. Appendix 4 shows that:

\begin{eqnarray}
&&G\left( K,K^{\prime },P,P^{\prime },X,X^{\prime },\theta ,\theta ^{\prime
}\right)  \notag \\
&=&\exp \left( -\left[ \frac{\left( K-\left\langle K\right\rangle
_{X}\right) ^{2}}{2\sigma ^{2}}_{\left( K,X\right) }^{\left( K^{\prime
},X^{\prime }\right) }\right] _{\left( K,X\right) }^{\left( K^{\prime
},X^{\prime }\right) }\right)  \label{Trsntn1} \\
&&\times \sqrt{\frac{\bar{\omega}_{X}}{2\pi \sinh \left( \bar{\omega}%
_{X}t\right) }}\exp \left( -\frac{\bar{\omega}_{X}\left( \left( X^{2}+\left(
X^{\prime }\right) ^{2}\right) \cosh \left( \bar{\omega}_{X}t\right)
-2XX^{\prime }\right) }{2\sinh \left( \bar{\omega}_{X}t\right) }\right) 
\notag \\
&&\times \sqrt{\frac{\omega }{2\pi \sigma ^{2}\sinh \left( \omega t\right) }}%
\exp \left( -\frac{\omega \left( \left( \left( K-\left\langle K\right\rangle
\right) ^{2}+\left( K^{\prime }-\left\langle K\right\rangle \right)
^{2}\right) \cosh \left( \omega t\right) -2\left( K-\left\langle
K\right\rangle \right) \left( K^{\prime }-\left\langle K\right\rangle
\right) \right) }{2\sigma ^{2}\sinh \left( \omega t\right) }\right)  \notag
\\
&&\times \delta \left( P-\frac{\left( \frac{1+\gamma \left( 1-\alpha \right) 
}{\left( 2-\alpha \right) \kappa \left( 1-h\frac{A^{2}}{\bar{A}^{2}}\left(
1-\exp \left( -\frac{1}{d}\right) \right) \right) d}\left( \left( \frac{%
2-\alpha }{1+\gamma \left( 1-\alpha \right) }\right) ^{2}-1\right) \right) ^{%
\frac{1}{1-\gamma }}\exp \left( -\frac{\left( 1-\alpha \right) \left\vert
X\right\vert }{1+\gamma \left( 1-\alpha \right) }\right) }{\left( \frac{K}{%
\left\langle K\right\rangle _{X}}\right) ^{\frac{\alpha }{1+\gamma }}\left(
1-h\exp \left( -\left\vert X\right\vert \right) \left( 1-\frac{\cosh X}{e}%
\right) \right) ^{\frac{\alpha }{1+\gamma }}}\right)  \notag \\
&&\times \delta \left( P^{\prime }-\frac{\left( \frac{1+\gamma \left(
1-\alpha \right) }{\left( 2-\alpha \right) \kappa \left( 1-h\frac{A^{2}}{%
\bar{A}^{2}}\left( 1-\exp \left( -\frac{1}{d}\right) \right) \right) d}%
\left( \left( \frac{2-\alpha }{1+\gamma \left( 1-\alpha \right) }\right)
^{2}-1\right) \right) ^{\frac{1}{1-\gamma }}\exp \left( -\frac{\left(
1-\alpha \right) \left\vert X^{\prime }\right\vert }{1+\gamma \left(
1-\alpha \right) }\right) }{\left( \frac{K^{\prime }}{\left\langle
K\right\rangle _{X^{\prime }}}\right) ^{\frac{\alpha }{1+\gamma }}\left(
1-h\exp \left( -\left\vert X^{\prime }\right\vert \right) \left( 1-\frac{%
\cosh X^{\prime }}{e}\right) \right) ^{\frac{\alpha }{1+\gamma }}}\right)
H\left( \theta -\theta ^{\prime }\right)  \notag
\end{eqnarray}

where $\delta \left( u\right) $\ is the Dirac function.\ A Dirac function is
a function which is null for all $u\neq 0$ and peaked at $u=0$\textbf{. }

Equation (\ref{Trsntn1}) computes the probability for an agent starting at $%
\left( K,P,X,\theta \right) $\ to reach $\left( K^{\prime },P^{\prime
},X^{\prime },\theta ^{\prime }\right) $ during a timespan $\theta -\theta
^{\prime }$.\ 

This formula also allows to compute the probability for $k$ agents with
initial state $\left( K_{i},P_{i},X_{i},\theta \right) _{i=1,...,N}$ to
reach the state $\left( K_{i}^{\prime },P_{i}^{\prime },X_{i}^{\prime
},\theta ^{\prime }\right) _{i=1,...,N}$. Formula (\ref{Trsntn1}) is the
product of Green functions:%
\begin{equation*}
\prod\limits_{i=1}^{k}G\left( K_{i},K_{i}^{\prime },P_{i},P_{i}^{\prime
},X_{i},X_{i}^{\prime },\theta ,\theta ^{\prime }\right)
\end{equation*}%
This probability is a product of independent probabilities. Indeed, recall
that in this set up interactions have been absorbed in the Green function
parameters.\ This illustrates the fact that the transition functions of
agents are shaped by the global environment, and not by some specific
interactions.

\bigskip

Formula (\ref{Trsntn1}) allows to understand the main features of individual
agents' dynamics.

Consider an agent starting from $K^{\prime }$, $X^{\prime }$\ and arriving
at a final point $K$, $X$\ after some time. This final point is partly
random, but due to the parameters of the system some values of $K$, $X$\ are
more likely, and some dynamic patterns appear.

We will in turn examine each of the terms of this equation (\ref{Trsntn1}).

The first exponential in (\ref{Trsntn1}) is trivial: it merely translates
the fact that the capital stock of agents in position $X$ is bounded around
the average stock $\left\langle K\right\rangle _{X}$.

The second exponential in Formula (\ref{Trsntn1}) describes the dynamics on $%
X$.\ It has the form of a stochastic harmonic oscillator, with one important
difference: the frequency $\bar{\omega}_{X}$\ is not constant and satisfies (%
\ref{inertX}).\ This exponential can be rewritten: 
\begin{equation}
\exp \left( -\frac{\bar{\omega}_{X}\left( \left( \left( X-X^{\prime }\right)
^{2}\right) +\left( \cosh \left( \bar{\omega}_{X}t\right) -1\right) \left(
X^{2}+\left( X^{\prime }\right) ^{2}\right) \right) }{2\sinh \left( \bar{%
\omega}_{X}t\right) }\right)  \label{prw}
\end{equation}%
The first term in the exponential $\left( X-X^{\prime }\right) ^{2}$
represents a Brownian random walk: any agent starting at $X^{\prime }$ could
move randomly with a standard deviation of $\sqrt{\frac{\sinh \left( \bar{%
\omega}_{X}t\right) }{\bar{\omega}_{X}}}$.

The additional term however, $\left( \cosh \left( \bar{\omega}_{X}t\right)
-1\right) \left( X^{2}+\left( X^{\prime }\right) ^{2}\right) $, favors a
final $X$ closer from the center $X=0$: the cohesion force drives agents
toward the center of the exchange space.

The parameter $\bar{\omega}_{X}$ is a rough estimator of the speed of this
move towards the center.

For $\bar{\omega}_{X}<<1$, the weight (\ref{prw}) is dampened for all final
value of $X$, which means that all final points $X$ is equally probable. The
driving force is weak, and the convergence very slow.

Inspection of (\ref{inertX}) shows that this speed $\bar{\omega}_{X}$
depends on the agent's initial level of capital. As expected, this speed
also depends positively on $\kappa _{0}$, the magnitude of the cohesion
force, and negatively on $\kappa _{2}$, the magnitude of the repulsive force.

Most importantly $\kappa _{1}$, the magnitude of the attractive force is
dampened by the ratio of the agent's current capital stock to $\left\langle
K\right\rangle _{X}$. Recall that $\left\langle K\right\rangle _{X}$ is the
average capital stock of agent's current - transitory - exchange position.
Since $\left\langle K\right\rangle _{X}$ increases as the agent moves
towards the center of the exchange space ( $X\rightarrow 0$), the closer the
agent gets to the center, the more he decelerates. A strong enough repulsive
force and a relatively low initial level of capital may prevent the agent to
ever reach the center of the exchange space. The attractive force towards
the center is thus inequal among agents. It favors those endowed with a
higher initial capital.

\bigskip

The third exponential in (\ref{Trsntn1}) determines the dynamics on $K$. It
can be rewritten:%
\begin{equation}
\exp \left( -\frac{\omega \left( \left( \left( K-K^{\prime }\right)
^{2}\right) +\left( \cosh \left( \omega t\right) -1\right) \left( \left(
K-\left\langle K\right\rangle \right) ^{2}+\left( K^{\prime }-\left\langle
K\right\rangle \right) ^{2}\right) \right) }{2\sigma ^{2}\sinh \left( \omega
t\right) }\right)  \label{expK}
\end{equation}%
where $\left\langle K\right\rangle $ is defined by (\ref{Kav}). For a
relatively stable $X$, an agent's capital stock is in average driven towards
the average stock of his exchange position. The quadratic term $\left(
K-K^{\prime }\right) ^{2}$ represents a random Brownian walk around this
trend.

Equation (\ref{KavX}) shows that as long as an agent moves towards the
center his average capital stock will increase in average.

As a matter of fact, the two dynamics of $X$ and $K$ interact and can create
a virtuous circle. A move towards the center tends to increase capital, and
an increase of capital tends to accelerate the move towards the center. This
positive interaction will only stop when the repulsive force will overcome
the attractive forces. Given (\ref{inertX}), this occurs when initial
capital is insufficient. The attractive force depends on the ratio of
initial over average capital. It may also occur if an adverse shock reduces
the stock of capital.\ The agent can then be driven back towards the
periphery.

\bigskip

The fourth term in equation (\ref{Trsntn1}) describes the market clearing
conditions given $X$ and $K$. The Dirac function in the last line of (\ref%
{Trsntn1}) implements (\ref{PX}). For a given level of capital $K$, prices
are higher towards the center of the exchange space, and for a given
position in the exchange space $X$, prices are a decreasing function of $K$.

\subsubsection{Phase 2: $\protect\rho >0$}

As in the non-trivial phase, we isolate in the effective action (\ref%
{fllctn1}) the part dependent on $X$.\ Appendix 4 shows that in phase 2 its
second order approximation is:

\begin{eqnarray}
&&S_{2}\left( \Psi \right) =\Psi ^{\dag }\left( K,X,\theta \right) \left(
-\nabla _{X}^{2}+\kappa _{0}\left( 1-\frac{\chi _{1}^{2}\kappa _{1}\rho ^{2}%
}{2}+2\chi _{2}\kappa _{2}\rho ^{4}\right) \left( X+sgn\left( X\right)
\delta X\right) ^{2}\right) \Psi \left( K,X,\theta \right)  \label{Ph2S2} \\
&&+\left( \rho ^{4}\kappa _{2}-\rho ^{2}\kappa _{1}-\left( \delta X\right)
^{2}\right) \left\vert \Psi \left( K,X,\theta \right) \right\vert ^{2} 
\notag
\end{eqnarray}%
with:%
\begin{equation}
\rho ^{2}=\frac{\kappa _{1}+\sqrt{\kappa _{1}^{2}-2\kappa _{2}\left( 2\alpha
+\sqrt{\kappa _{0}}+\sigma \delta \right) }}{2\kappa _{2}}  \label{r2}
\end{equation}%
and:%
\begin{equation}
\delta X=\chi _{1}\kappa _{1}\frac{K}{2\left\langle K\right\rangle }\rho
^{2}-\chi _{2}\kappa _{2}\rho ^{4}  \label{deltX}
\end{equation}%
Thus, the full action for the system is:%
\begin{eqnarray}
&&S\left( \Psi \right) =\int \Psi ^{\dag }\left( K,X,\theta \right) \left(
-\sigma ^{2}\nabla _{K}^{2}-\vartheta ^{2}\nabla _{\theta }^{2}-\sigma
_{X}^{2}\nabla _{X}^{2}+\frac{\omega ^{2}}{\sigma ^{2}}\left( K-\left\langle
K\right\rangle \right) ^{2}\right) \Psi \left( K,X,\theta \right)
\label{Ph2S} \\
&&+\kappa _{0}\int \left( \left( X+\chi sgn\left( X\right) \left( \kappa _{1}%
\frac{K}{2\left\langle K\right\rangle }\rho ^{2}-\kappa _{2}\rho ^{4}\right)
\right) ^{2}+\alpha _{X}\right) \left\vert \Psi \left( K,X,\theta \right)
\right\vert ^{2}  \notag \\
&&+\left( 2\rho ^{2}\kappa _{2}-\kappa _{1}\right) \left\vert \int \Psi
^{\dag }\left( K,X,\theta \right) \Psi _{0}\left( K,X,\theta \right)
\right\vert ^{2}  \notag
\end{eqnarray}%
where:%
\begin{eqnarray*}
\omega &=&\sigma \sqrt{\delta ^{2}+\frac{\bar{A}^{2}A^{2}}{\left( A^{2}U^{2}+%
\bar{A}^{2}\right) ^{2}}} \\
\alpha _{X} &=&-\frac{1}{2}\kappa _{0}^{\frac{1}{2}}\sqrt{1-\frac{\chi
_{1}^{2}\kappa _{1}\rho ^{2}}{2}+2\chi _{2}^{2}\kappa _{2}\rho ^{4}}-\frac{%
\sigma \sqrt{\delta ^{2}+\frac{\bar{A}^{2}A^{2}}{\left( A^{2}U^{2}+\bar{A}%
^{2}\right) ^{2}}}}{2}
\end{eqnarray*}%
Given (\ref{Ph2S}), the transition function in the phase $\rho \neq 0$:%
\begin{eqnarray}
&&G\left( K,K^{\prime },P,P^{\prime },X,X^{\prime },\theta ,\theta ^{\prime
}\right)  \notag \\
&=&\exp \left( -\left[ \frac{\left( K-\left\langle K\right\rangle
_{X}\right) ^{2}}{2\sigma ^{2}}_{\left( K,X\right) }^{\left( K^{\prime
},X^{\prime }\right) }\right] _{\left( K,X\right) }^{\left( K^{\prime
},X^{\prime }\right) }\right) \times G_{\frac{K+K^{\prime }}{2}}\left(
X,X^{\prime }\right)  \label{Trsntn2} \\
&&\times \sqrt{\frac{\bar{\omega}/2\pi \sigma ^{2}}{\sinh \left( \bar{\omega}%
\left( \theta -\theta ^{\prime }\right) \right) }}\exp \left( -\frac{\bar{%
\omega}\left( \left( \left( K-\left\langle K\right\rangle \right)
^{2}+\left( K^{\prime }-\left\langle K\right\rangle \right) ^{2}\right)
\cosh \left( \bar{\omega}\left( \theta -\theta ^{\prime }\right) \right)
-2\left( K-\left\langle K\right\rangle \right) \left( K^{\prime
}-\left\langle K\right\rangle \right) \right) }{2\sigma ^{2}\sinh \left( 
\bar{\omega}\left( \theta -\theta ^{\prime }\right) \right) }\right)  \notag
\\
&&\times \delta \left( P-\frac{\left( \frac{1+\gamma \left( 1-\alpha \right) 
}{\left( 2-\alpha \right) \kappa \left( 1-h\frac{A^{2}}{\bar{A}^{2}}\left(
1-\exp \left( -\frac{1}{d}\right) \right) \right) \rho ^{4}d}\left( \left( 
\frac{2-\alpha }{1+\gamma \left( 1-\alpha \right) }\right) ^{2}-1\right)
\right) ^{\frac{1}{1-\gamma }}\exp \left( -\frac{\left( 1-\alpha \right)
\left\vert X\right\vert }{1+\gamma \left( 1-\alpha \right) }\right) }{\left( 
\frac{K}{\left\langle K\right\rangle _{X}}\right) ^{\frac{\alpha }{1+\gamma }%
}\left( 1-h\exp \left( -\left\vert X\right\vert \right) \left( 1-\frac{\cosh
X}{e}\right) \right) ^{\frac{\alpha }{1+\gamma }}}\right)  \notag \\
&&\times \delta \left( P^{\prime }-\frac{\left( \frac{1+\gamma \left(
1-\alpha \right) }{\left( 2-\alpha \right) \kappa \left( 1-h\frac{A^{2}}{%
\bar{A}^{2}}\left( 1-\exp \left( -\frac{1}{d}\right) \right) \right) \rho
^{4}d}\left( \left( \frac{2-\alpha }{1+\gamma \left( 1-\alpha \right) }%
\right) ^{2}-1\right) \right) ^{\frac{1}{1-\gamma }}\exp \left( -\frac{%
\left( 1-\alpha \right) \left\vert X^{\prime }\right\vert }{1+\gamma \left(
1-\alpha \right) }\right) }{\left( \frac{K^{\prime }}{\left\langle
K\right\rangle _{X^{\prime }}}\right) ^{\frac{\alpha }{1+\gamma }}\left(
1-h\exp \left( -\left\vert X^{\prime }\right\vert \right) \left( 1-\frac{%
\cosh X^{\prime }}{e}\right) \right) ^{\frac{\alpha }{1+\gamma }}}\right)
H\left( \theta -\theta ^{\prime }\right)  \notag
\end{eqnarray}%
where:%
\begin{eqnarray}
G_{K}\left( X,X^{\prime }\right) &=&\bar{G}\left( X+\delta X,X^{\prime
}+\delta X\right) H\left( X\right) H\left( X^{\prime }\right) +\bar{G}\left(
X-\delta X,X^{\prime }-\delta X\right) H\left( -X\right) H\left( -X^{\prime
}\right)  \label{PH2G} \\
&&+\bar{G}\left( X+\delta X,X^{\prime }-\delta X\right) H\left( X\right)
H\left( -X^{\prime }\right) +\bar{G}\left( X-\delta X,X^{\prime }+\delta
X\right) H\left( -X\right) H\left( X^{\prime }\right)  \notag
\end{eqnarray}%
with:%
\begin{equation}
\bar{G}\left( X,X^{\prime }\right) =\sqrt{\frac{\kappa _{0}}{2\pi \sigma
_{X}^{2}\sinh \left( \kappa _{0}t\right) }}\exp \left( \left( -\frac{\kappa
_{0}}{2\sigma _{X}^{2}\sinh \left( \kappa _{0}t\right) }\right) \left(
\left( X-X^{\prime }\right) ^{2}+\left( \cosh \left( \kappa _{0}t\right)
-1\right) \left( X^{2}+\left( X^{\prime }\right) ^{2}\right) \right) \right)
\label{PH2GPP}
\end{equation}%
where $H\left( X\right) $ is the Heaviside function defined by $H\left(
X\right) =1$ for $X>0$ and $0$ otherwise. As in phase 1, and for the same
reasons, the definition of $\left\langle K\right\rangle $ is given by (\ref%
{Kav}). The quantity $\delta X$\ is defined in (\ref{deltX}).

\bigskip

Here again, the first exponential in (\ref{Trsntn2}) is trivial: it merely
translates the fact that the capital stock of agents in position $X$ is
bounded to be around the average stock $\left\langle K\right\rangle _{X}$.

\bigskip

The second exponential in formula (\ref{Trsntn2}) describes the dynamics on $%
X$.\ Given the formulas (\ref{Trsntn2}) and (\ref{PH2G}), this dynamic is
different from that of phase 1. Equation (\ref{Ph2S}) shows that an
individual variable attraction point appears for each agent, along with $X=0$%
. This point depends both on the agent's initial position and capital. \ 

The Green function $G_{\frac{K+K^{\prime }}{2}}\left( X,X^{\prime }\right) $
in (\ref{PH2G}) commands the dynamics of the exchange position. Assume $X$
and $X^{\prime }$\ are positive.\ Using (\ref{PH2GPP}), one has: 
\begin{eqnarray*}
G_{\frac{K+K^{\prime }}{2}}\left( X,X^{\prime }\right) &=&\sqrt{\frac{\kappa
_{0}}{2\pi \sigma _{X}^{2}\sinh \left( \kappa _{0}t\right) }} \\
&&\times \exp \left( \left( -\frac{\kappa _{0}}{2\sigma _{X}^{2}\sinh \left(
\kappa _{0}t\right) }\right) \left( \left( X-X^{\prime }\right) ^{2}+\left(
\cosh \left( \kappa _{0}t\right) -1\right) \left( \left( X+\delta X\right)
^{2}+\left( X^{\prime }+\delta X\right) ^{2}\right) \right) \right)
\end{eqnarray*}%
As in phase 1, the quadratic term $\left( X-X^{\prime }\right) ^{2}$\
represents a random Brownian walk around the starting point $X^{\prime }$.
The second quadratic term: 
\begin{equation*}
\left( \cosh \left( \kappa _{0}t\right) -1\right) \left( \left( X+\delta
X\right) ^{2}+\left( X^{\prime }+\delta X\right) ^{2}\right)
\end{equation*}%
drives the position $X$ towards the position $-\delta X$. Thus, the
attraction point is shifted from $0$ to $-\delta X$.

For a large level of capital, equation (\ref{deltX}) shows that the
attraction point appears on the agent's opposite side of the center $X=0$.\
This variable attraction point quickly drives the agent towards $X=0$. When $%
X=0$ is reached, the agent does not cross this point in average. Were he to
cross it randomly, he would be driven back towards $X=0$.\ Actually, when $%
X^{\prime }<0$ and $X<0$, i.e. when an agent has crossed the center $X=0$,
equations (\ref{PH2G}) and (\ref{PH2GPP}) show that the agent is driven back
toward the variable attraction point $\delta X>0$, which in turn drives back
the agent toward $0$. The whole oscillatory process reveals that $X=0$ is
indeed an attractive point.

For a low level of initial capital, the dynamics is different. The variable
attraction point appears on the initial position side.\ The agent is driven
towards this point, even if he was closer to the center $X=0$. This implies
an eviction mechanism from the center for agents with low capital.

At first sight, the dynamics of capital accumulation should be identical to
phase 1, since the third exponential in (\ref{Trsntn2}) is similar to (\ref%
{Trsntn1}). Up to some random fluctuations, agents are driven to the average
level of capital of their current exchange position. However, since capital
accumulation and exchange position are interacting, the features of the
dynamics of $X$ in phase 2 impact capital accumulation.

In phase 1, a positive shock in capital could drive the agent towards the
center, in turn increasing his capital stock.

Here, due to the presence of the variable attraction point, a small positive
capital shock may not initiate a move towards the center. An agent with a
low level of capital will be driven back towards his variable attraction
point, possibly farther from the center.

\bigskip

The fourth term in equation (\ref{Trsntn2}) describes the market clearing
conditions given $X$ and $K$. Actually, the Dirac function in the last line
of (\ref{Trsntn2}) implements (\ref{PX2}). For a given level of capital $K$,
prices are higher towards the center of the exchange space, and for a given
position in the exchange space $X$, prices are a decreasing function of $K$.

A last effect which is a direct consequence of the non-trivial phase cannot
be read from the Green function (\ref{Trsntn2}) but can be tracked back to
the last term of (\ref{Ph2S}):%
\begin{equation*}
\left( 2\rho ^{2}\kappa _{2}-\kappa _{1}\right) \left\vert \int \Psi ^{\dag
}\left( K,X,\theta \right) \Psi _{0}\left( K,X,\theta \right) \right\vert
^{2}
\end{equation*}%
The effective action in phase 2 includes a positive term, since now $\rho
^{2}$ is greater than $\frac{\kappa _{1}}{2\kappa _{2}}$. Appendix 5 shows
that this term induces a barrier between agents in position $X$. Agents with
an initial capital stock below $\left\langle K\right\rangle _{X}$ will find
it harder to accumulate above $\left\langle K\right\rangle _{X}$ and thus
improve their exchange position while agents with $K>\left\langle
K\right\rangle _{X}$ will more likely remain above this threshold.
Therefore, in phase 2, a threshold effect appears at each point on the
exchange space. Agents with high initial capital will more likely overcome
barriers and accumulate, while the others will be evicted from positions in
the exchange space for which their initial stock of capital is too low.

\section{Synthesis and discussion}

To interpret the previous results, recall that this model considers loosely
defined variables. The exchange space for instance can be seen both as a
geographical space, such as national markets or individual sectors, and as
scale of terms of exchanges.

Three forces characterize this exchange space. The cohesion force is the
only force that attracts all agents towards the center of the exchange
space.\ The attractive and repulsive forces bring agents together or apart
on the $X$ axis, respectively. Since exchanges are proportional to the
distance between agents, the attractive forces facilitate exchanges between
agents, whereas the repulsive force reduces competition, by preventing new
entrants or more distant agents to establish or strengthen exchanges with
consumers. These two forces, attractive and repulsive, always exist within a
market, yet our work stresses that it is $r=\kappa _{2}/\kappa _{1}$, the
relative magnitude of the repulsive force with respect to the attractive
force, that matters. This ratio commands the occurrence of the two phases of
the dynamics.

In phase $1$, $r$ is high and free exchange is limited. Markets are
relatively protected, or exchanges may be prevented for other reasons, such
as distance, lack of infrastructure, regulation, etc. The cohesion force
improves in average the terms of exchange across all agents and contributes
to homogenizing the agents. This tendency impacts all agents, even though
agents with high initial capital are favored in the process. On the
contrary, in phase $2$, the repulsive force is relatively weaker, and $r$ is
low.\ The agent will experience higher mobility within the exchange space.

The two phases present similarities. In both, the price of the producer and
the average capital stock of agents decrease exponentially with the distance
to the center. Agents at the periphery produce less and sell at a lower
price than central agents. Yet dynamically the two phases are different.
Phase $1$ displays stability in exchanges, lower competition, a broad based
although slow improvement in the terms of exchanges. In this phase, random
shocks can redistribute capital and initiate a virtuous circle of capital
accumulation.

In phase $2$, the weakness of the repulsive force induces greater
instability in trade relations, but also greater competition among agents.
Increased mobility favors capital accumulation for producers with high
initial level of capital.\ Producers with a low level of initial capital
will experience eviction from the exchange space and a deterioration in
their prices and revenues. This is due to an indirect effect: increased
exchanges induce greater competition and increased inequality in wealth
distribution. Actually, removing the repulsive force opens market to agents
with higher level of capital.\ Selling their products at a relatively low
price, these agents gain market shares over their competitors. A producer
that has penetrated a market experiences a higher demand, higher prices, a
higher revenue, and in turn higher capital. This same mechanism, repeated
over time, may drive him along the exchange space to a relatively dominant
position.

On the contrary producers with low capital on an existing market will be
progressively evicted. To them, market liberalization is an increase in
neighboring competitors. They will face a lower demand, lose market shares
and experience a decrease in income that will push them towards the
periphery, and exclude them from their market. Such mechanisms are at play
within sectors where liberalization favors agents with high concentration of
capital. The arrival of big agricultural producers in Africa evicted local
producers, that ended up selling their work force to their previous
competitors, and in some cases, lead them to migrate.

Phase $2$ displays a seemingly counter-intuitive result: except under mild
market liberalization, i.e. when $r$ is in some range, average capital is
lower in phase $2$ than in phase $1$. This result is mainly due to our
assumption of constant technology.\ Because of this assumption, our model is
only valid within a relatively short period of time. Therefore, our
conclusions do not contradict the standard accounts of industrial
revolution. The short periods of liberalization during the technological
boost of the nineteenth century did not lead decisively to higher
accumulation, and freedom of exchanges has been at times contemporaneous of
a crisis. Thus, our model sheds a nuanced light on free exchange dynamics.
In our context, a high disparity in capital and thus in revenue does not
imply an average improvement of wealth. Disparities in capital accumulation
are directly related to a collapse in global demand and a lower average
wealth.

Our model has a last implication. In our setting, capital must be renewed to
produce: when an agent cannot do so, he disappears as a producer within his
market. This mechanism repeated over time may leave the agent with the sole
capital he may renew, namely his labor force. Thus, phase $2$\ of our model
also describes the evolution from a society with a large number of small
producers towards a society with few capitalistic producers and a large
number of workers: it accounts for de-homogenization of society.

\section{Conclusion}

In this paper, we have developed a model of capital accumulation with a
large number of heterogeneous agents.\ This model keeps some features of the
classical economic models - a standard equation of capital accumulation,
production function, and market clearing condition. However, it includes an
exchange space that dynamically interacts with capital accumulation.\
Besides, the classical description of the model is replaced by field theory
techniques. Our results show that, depending on the parameters of the model,
the system displays various phases, each describing different accumulation
processes. Depending on the parameters of the model, capital accumulation is
not necessarily favored by greater market liberalization. Besides, capital
accumulation highly depends on each agent's initial conditions, and shows
that the capital dynamics cannot be reduced to simple aggregates.\ Various
dynamic patterns appear depending on agents. Yet our model enlightens the
fact that by its dynamic divergences, a society of atomic producers can
divide between groups of accumulating producers, and agents progressively
losing their capital. Our results show a dynamic that can produce a society
divided in two classes, the first one accumulating capital, the other one
losing its capital, up to the point where its only capital left will be its
labor. This model shows the advantages of field theory modeling in
economics. It leads to a finer description of the agents' dynamics, both at
a global level and individual level, the two descriptions interacting
permanently. Our future work will continue to explore this approach to other
fields of economics.

\bigskip

\pagebreak

\section{Appendices}

\subsection{Appendix 0}

It will be useful for the description in term of field theory to rewrite the
weight (\ref{ST}) in a different form. We introduce a variable $\tilde{X}%
_{i}\left( t\right) $ so that (\ref{ST}) is equal to: 
\begin{eqnarray*}
&&\exp \left( -\sum_{i}\left( \int \frac{1}{2\sigma ^{2}}\left( \dot{K}%
_{i}\left( t\right) +\delta K_{i}\left( t\right) -AP_{i}\left( t\right)
K_{i}^{\alpha }\left( t\right) +\tilde{X}_{i}\left( t\right) \right)
^{2}\right. \right. \\
&&\qquad \left. \left. -\frac{\bar{A}^{2}}{2\sigma ^{2}}\left(
P_{i}^{1+\gamma }\left( t\right) K_{i}^{\alpha }\left( t\right)
-U_{2,i}\right) ^{2}+\delta \left( \tilde{X}_{i}\left( t\right)
-U_{1,i}\right) \right) \right)
\end{eqnarray*}%
where $\delta \left( \tilde{X}_{i}\left( t\right) -U_{1,i}\right) $ is the
delta function and: 
\begin{eqnarray*}
U_{1i} &=&\kappa AP_{i}\left( t\right) K_{i}^{\alpha }\left( t\right)
\sum_{j,k}\frac{P_{k}\left( t\right) }{P_{j}^{\gamma }\left( t\right) }\exp
\left( -\frac{d_{ij}\left( t\right) +d_{ik}\left( t\right) }{d}\right) \\
U_{2,i} &=&\kappa \sum_{j,k}P_{j}\left( t\right) K_{j}^{\alpha }\left(
t\right) P_{k}\left( t\right) \exp \left( -\frac{d_{ij}\left( t\right)
+d_{kj}\left( t\right) }{d}\right)
\end{eqnarray*}

The full weight of the system is thus:%
\begin{eqnarray}
&&\exp \left( -\frac{\left( \dot{K}_{i}\left( t\right) +\delta K_{i}\left(
t\right) -AP_{i}\left( t\right) K_{i}^{\alpha }\left( t\right) +\tilde{X}%
_{1}\right) ^{2}}{2\sigma ^{2}}-\frac{\bar{A}^{2}\left( P_{i}^{1+\gamma
}\left( t\right) K_{i}^{\alpha }\left( t\right) -U_{2}\right) ^{2}}{2\sigma
^{2}}+\delta \left( \tilde{X}_{1}-U_{1}\right) \right)  \label{fllwgt} \\
&&\times \exp \left( -\left( \sum_{i}\left( \frac{\left( \dot{X}_{i}\left(
t\right) \right) ^{2}}{\sigma _{X}^{2}}+V_{0}\left( X_{i}\left( t\right)
\right) \right) +\sum_{i,j}V_{1}\left( d_{ij}\left( t\right) \right)
+\sum_{i,j,k}V_{2}\left( d_{ij}\left( t\right) ,d_{ik}\left( t\right)
,d_{jk}\left( t\right) \right) \right) \right)  \notag
\end{eqnarray}

\subsection{Appendix 1}

We use the techniques defined in Gosselin, Lotz and Wambst (2017, 2018)%
\textbf{\ }to switch from the probabilistic description of the model to the
field theoretic formalism. A presentation can found in Gosselin, Lotz and
Wambst (2018),\textbf{\ }but we recall here what is needed for our purpose.

The idea is the following. For a large number of agents, the system
described by (\ref{fllwgt}), involves a large number of variables $X\left(
t\right) $\ that are difficult to handle. We consider the space $H$\ of
complex functions defined on the space of a single agent's actions. The
space $H$ describes the collective behavior of the system. Each function $%
\Psi $ of $H$ encodes a particular state of the system. Then, to\ each
function $\Psi $ of $H$, we associate a statistical weight, i.e. a
probability describing the state encoded in $\Psi $. This probability is
written $\exp \left( -S\left( \Psi \right) \right) $, where $S\left( \Psi
\right) $ is a functional, i.e. a function of the function $\Psi $. The form
of $S\left( \Psi \right) $ is derived directly from the form of (\ref{fllwgt}%
).

The weight considered in this work is a particular case of those presented
in Gosselin, Lotz and Wambst (2018), plus some adaptation. In Gosselin, Lotz
and Wambst (2018), we showed that for a{\small \ }weight describing
interaction between individual agents:%
\begin{eqnarray}
&&\sum_{i}\int_{0}^{T}\left( \frac{1}{\sigma ^{2}}\int_{0}^{T}\left( \frac{d%
}{ds}X_{i}\left( s\right) \right) ^{2}+V_{1}\left( X_{s}^{\left( i\right)
}\right) \right) ds+\frac{1}{\eta ^{2}}\sum_{i}\int_{0}^{T}\left( \frac{d}{dt%
}X_{i}\left( t\right) -H\left( X_{i}\left( t\right) \right) \right) ^{2}dt
\label{gblw3} \\
&&+\sum_{k\geqslant 2}\sum_{i_{1},...,i_{k}}\int_{0}^{T}\int_{0}^{T}\frac{%
V_{k}\left( X_{s_{1}}^{\left( i_{1}\right) },...,X_{s_{k}}^{\left(
i_{k}\right) }\right) }{\theta ^{2}}ds_{1}...ds_{k}  \notag
\end{eqnarray}%
the following field action functional contains the same information about
the system: 
\begin{eqnarray*}
S\left( \Psi \right) &=&\int \!\!\left( \Psi ^{\dag }\left( X\right)
\!\!\left( -\sigma ^{2}\nabla ^{2}+V_{1}\left( X\right) +\alpha \right) \Psi
\left( X\right) \right) dX-\sum_{i}\int \Psi ^{\dag }\left( X\right)
\!\!\left( \eta ^{2}\nabla ^{2}+\nabla .H\left( x\right) \right) \Psi \left(
X\right) dX \\
&&+\frac{1}{\theta ^{2}}\sum_{k\geqslant 2}\sum_{k\geqslant 2}\int \Psi
\left( X_{1}\right) ...\Psi \left( X_{k}\right) V_{k}\left(
X_{1}...X_{k}\right) \Psi ^{\dag }\left( X_{1}\right) ...\Psi ^{\dag }\left(
X_{k}\right) dX_{1}...dX_{k}
\end{eqnarray*}%
where $\alpha $ is the parameter arising in the Laplace transform of the
statistical weight (\ref{gblw3}), (see Gosselin, Lotz, Wambst 2017) and
where $\Psi ^{\dag }\left( x\right) $ denotes the complex conjugate of $\Psi
\left( x\right) $. The operator $\nabla $ is the gradient operator, a vector
whose $i$-th coordinate is the first derivative $\frac{\partial }{\partial
x_{i}}$: $\nabla =\left( \frac{\partial }{\partial X_{i}}\right) $. The
operator $\nabla ^{2}$ denotes the Laplacian:%
\begin{equation*}
\nabla ^{2}=\sum_{i}\frac{\partial ^{2}}{\partial X_{i}^{2}}
\end{equation*}%
where the sum runs over the coordinates $x_{i}$ of the vector $x$. Applying
this to our previous example, where $\Psi \left( X\right) =\Psi \left(
c,k\right) $, we get $\nabla ^{2}=\frac{\partial ^{2}}{\partial c^{2}}+\frac{%
\partial ^{2}}{\partial k^{2}}$.

Coming back to the model at stake here, and discarding for the moment the
weight associated to the position in the exchange space, the statistical
weight for the Price plus Capital part was (\ref{fllwgt}):

\begin{equation}
\exp \left( -\frac{1}{2\sigma ^{2}}\left( \dot{K}_{i}\left( t\right) +\delta
K_{i}\left( t\right) -AP_{i}\left( t\right) K_{i}^{\alpha }\left( t\right)
+U_{1i}\right) ^{2}-\frac{\bar{A}^{2}}{2\sigma ^{2}}\left( P_{i}^{1+\gamma
}\left( t\right) K_{i}^{\alpha }\left( t\right) -U_{2,i}\right) ^{2}\right)
\label{wgt}
\end{equation}%
with:%
\begin{eqnarray*}
U_{1i} &=&\kappa AP_{i}\left( t\right) K_{i}^{\alpha }\left( t\right)
\sum_{j,k}\frac{P_{k}\left( t\right) }{P_{j}^{\gamma }\left( t\right) }\exp
\left( -\frac{\left\vert X_{i}\left( t\right) -X_{j}\left( t\right)
\right\vert }{d}-\frac{\left\vert X_{i}\left( t\right) -X_{k}\left( t\right)
\right\vert }{d}\right) \\
U_{2i} &=&\kappa \sum_{j,k}P_{j}\left( t\right) K_{j}^{\alpha }\left(
t\right) P_{k}\left( t\right) \exp \left( -\frac{\left\vert X_{i}\left(
t\right) -X_{j}\left( t\right) \right\vert }{d}-\frac{\left\vert X_{k}\left(
t\right) -X_{j}\left( t\right) \right\vert }{d}\right)
\end{eqnarray*}%
The difference with (\ref{gblw3}) is that, here, the interactions between
different agents (terms involving sums over different agents $j,k,...$, are
local, i.e. the quantities involved in these terms are considered at the
same time, whereas it is not required in (\ref{gblw3}). As explained in
Gosselin, Lotz and Wambst (2017), the introduction of local interactions can
be done by introducing for each agent $i$, a counting variable $\Theta
_{i}\left( s\right) $ which is roughly equal to $s$ apart some random
fluctuation. A statistical weight has to be introduced for this counting
variable $\Theta _{i}\left( s\right) $ and then, one has to replace this
weight by its field counterpart:%
\begin{equation*}
\int \exp \left( -\int \frac{\left( \dot{\Theta}_{i}\left( s\right)
-1\right) ^{2}}{2\vartheta ^{2}}\right) \mathcal{D}\Theta _{i}\left(
s\right) \rightarrow \Psi ^{\dag }\left( K,P,\hat{P},X,\theta \right) \nabla
_{\theta }.\left( \vartheta ^{2}\nabla _{\theta }-2\right) \Psi \left( K,P,%
\hat{P},X,\theta \right)
\end{equation*}%
We will now show that the overall action for the Price + Capital part
associated to (\ref{fllwgt}) writes: 
\begin{eqnarray}
&&\Psi ^{\dag }\left( K,X,\theta \right) \left( -\sigma ^{2}\nabla _{K}^{2}+%
\frac{\left( \delta K-APK^{\alpha }\left( 1-\hat{U}_{1}\right) \right) ^{2}}{%
\sigma ^{2}}+\sigma _{P}^{2}\left( \frac{AK^{\alpha +1}}{\sigma ^{2}\left(
\alpha +1\right) }\right) ^{2}+\frac{\bar{A}^{2}\left( P^{1+\gamma
}K^{\alpha }+U_{2}\right) ^{2}}{\sigma ^{2}}\right.  \label{fldthvrsn} \\
&&\qquad \qquad \left. -\left( \sigma _{P}^{2}\nabla _{P}^{2}+\sigma
_{X}^{2}\nabla _{X}^{2}\right) -\vartheta ^{2}\nabla _{\theta }^{2}+\frac{1}{%
\vartheta ^{2}}+\alpha \right) \Psi \left( K,X,\theta \right)  \notag \\
&\simeq &\Psi ^{\dag }\left( K,X,\theta \right) \left( -\sigma ^{2}\nabla
_{K}^{2}+\frac{\left( \delta K-APK^{\alpha }\left( 1-\hat{U}_{1}\right)
\right) ^{2}}{\sigma ^{2}}+\left( P^{1+\gamma }K^{\alpha }+U_{2}\right)
^{2}-\sigma _{X}^{2}\nabla _{X}^{2}-\vartheta ^{2}\nabla _{\theta }^{2}+%
\frac{1}{\vartheta ^{2}}+\alpha \right) \Psi \left( K,X,\theta \right) 
\notag
\end{eqnarray}%
which is in first approximation:%
\begin{equation}
\Psi ^{\dag }\left( K,X,\theta \right) \left( -\sigma ^{2}\nabla _{K}^{2}+%
\frac{\left( \delta K-APK^{\alpha }\left( 1-\hat{U}_{1}\right) \right) ^{2}}{%
\sigma ^{2}}+\left( P^{1+\gamma }K^{\alpha }+U_{2}\right) ^{2}-\sigma
_{X}^{2}\nabla _{X}^{2}-\vartheta ^{2}\nabla _{\theta }^{2}+\frac{1}{%
\vartheta ^{2}}+\alpha \right) \Psi \left( K,X,\theta \right)  \notag
\end{equation}

\bigskip

with:%
\begin{eqnarray*}
\hat{U}_{1} &=&\kappa \int \frac{P_{3}\exp \left( -\left( \frac{D_{12}+D_{13}%
}{d}\right) \right) }{P_{2}^{\gamma }}\left\vert \bar{\Psi}\left(
Z_{2},\theta \right) \right\vert ^{2}\left\vert \bar{\Psi}\left(
Z_{3},\theta \right) \right\vert ^{2} \\
U_{2} &=&-\kappa \int P_{2}\left( K_{2}\right) ^{\alpha }P_{3}\exp \left(
-\left( \frac{D_{12}+D_{23}}{d}\right) \right) \left\vert \Psi \left(
Z_{2},\theta \right) \right\vert ^{2}\left\vert \Psi \left( Z_{3},\theta
\right) \right\vert ^{2}
\end{eqnarray*}

To prove this result, we first start with an arbitrary weight with
interactions between agents and including the counting variable $\Theta
\left( s\right) $. Such a weight includes our model as a particular case.%
\begin{equation}
\exp \left( -\int \frac{\left( \dot{X}_{i}\left( t\right) +G\left(
X_{i}\left( t\right) \right) -\sum_{j}V\left( X_{i}\left( t\right)
,X_{j}\left( t\right) \right) \right) ^{2}}{2\sigma ^{2}}\right)
\label{wgtpr}
\end{equation}%
This form describes the first part of (\ref{wgt}) if we set:%
\begin{eqnarray}
G\left( X_{i}\left( t\right) \right) &=&\delta K_{i}\left( t\right)
-AP_{i}\left( t\right) K_{i}^{\alpha }\left( t\right)  \label{GV} \\
V\left( X_{i}\left( t\right) ,X_{j}\left( t\right) \right) &=&-U_{1}  \notag
\end{eqnarray}%
The second part of (\ref{wgt}): 
\begin{equation*}
-\frac{\bar{A}^{2}}{2\sigma ^{2}}\left( P_{i}^{1+\gamma }\left( t\right)
K_{i}^{\alpha }\left( t\right) -U_{2,i}\right) ^{2}
\end{equation*}%
has a direct equivalent in term of field theory:%
\begin{equation}
\Psi ^{\dag }\left( K,X,\theta \right) \left( \left( P^{1+\gamma }K^{\alpha
}+U_{2}\right) ^{2}\right) \Psi \left( K,X,\theta \right)  \label{scndwgt}
\end{equation}%
so we skip it from the following argument and reintroduce it at the end of
the proof.

Then, as in Gosselin, Lotz and Wambst (2017), we replace in $V\left(
X_{i}\left( t\right) ,X_{j}\left( t\right) \right) $, the time by two
independent parameters $t_{i}$ and $t_{j}$, and impose the equality of the
counting variables associated to these parameters: $\theta _{j}\left(
t_{j}\right) -\theta _{i}\left( t_{i}\right) =0$. Then the weight (\ref{wgt}%
) rewrites:

\begin{equation*}
\rightarrow \exp \left( -\frac{\left( \dot{X}_{i}\left( t_{i}\right)
+G\left( X_{i}\left( t_{i}\right) \right) -\sum_{j}\int V\left( X_{i}\left(
t_{i}\right) ,X_{j}\left( t_{j}\right) \right) \delta \left( \theta
_{j}\left( t_{j}\right) -\theta _{i}\left( t_{i}\right) \right) \right) ^{2}%
}{2\sigma ^{2}}\right)
\end{equation*}%
To simplify the notations, and include possible generalizations, we will
also replace:%
\begin{eqnarray*}
\int V\left( X_{i}\left( t_{i}\right) ,X_{j}\left( t_{j}\right) \right)
\delta \left( \theta _{j}\left( t_{j}\right) -\theta _{i}\left( t_{i}\right)
\right) &\rightarrow &\int V\left( X_{i}\left( t_{i}\right) ,\theta
_{i}\left( t_{i}\right) ,X_{j}\left( t_{j}\right) ,\theta _{j}\left(
t_{j}\right) \right) \\
&=&\int V\left( Y_{i}\left( t_{i}\right) ,Y_{j}\left( t_{j}\right) \right)
\end{eqnarray*}%
where we set:%
\begin{equation*}
Y_{i}\left( t_{i}\right) =\left( X_{i}\left( t_{i}\right) ,\theta _{i}\left(
t_{i}\right) \right)
\end{equation*}%
That is, the $\delta $ factor of the potential is replaced by an arbitrary
function of the counting variables $\theta _{i}\left( t_{i}\right) $, $%
\theta _{j}\left( t_{j}\right) $. Then, we introduce an auxiliary variable $%
\tilde{X}_{i}\left( t_{i}\right) $, equal up to a random error of small
square deviation $\sigma _{2}^{2}$, to $\sum_{j}\int V\left( Y_{i}\left(
t_{i}\right) ,Y_{j}\left( t_{j}\right) \right) $ and the weight (\ref{wgtpr}%
) becomes:

\begin{eqnarray*}
&&\exp \left( -\frac{\left( \dot{X}_{i}\left( t_{i}\right) +G\left(
X_{i}\left( t_{i}\right) \right) -\sum_{j}\int V\left( Y_{i}\left(
t_{i}\right) ,Y_{j}\left( t_{j}\right) \right) \right) ^{2}}{2\sigma ^{2}}%
\right) \\
&=&\exp \left( -\frac{\left( \dot{X}_{i}\left( t\right) +G\left( X_{i}\left(
t_{i}\right) \right) -\tilde{X}_{i}\left( t_{i}\right) \right) ^{2}}{2\sigma
^{2}}-\frac{\left( \tilde{X}_{i}\left( t_{i}\right) -\sum_{j}\int V\left(
Y_{i}\left( t_{i}\right) ,Y_{j}\left( t_{j}\right) \right) \right) ^{2}}{%
2\sigma _{2}^{2}}\right) \\
&=&\exp \left( -\frac{\left( \dot{X}_{i}\left( t\right) +G\left( X_{i}\left(
t_{i}\right) \right) -\tilde{X}_{i}\left( t_{i}\right) \right) ^{2}}{2\sigma
^{2}}\right. \\
&&\left. -\frac{\left( \tilde{X}_{i}^{2}\left( t_{i}\right) \right) ^{2}-2%
\tilde{X}_{i}\left( t_{i}\right) \sum_{j}\int V\left( Y_{i}\left(
t_{i}\right) ,Y_{j}\left( t_{j}\right) \right) +\sum_{j}\int V\left(
Y_{i}\left( t_{i}\right) ,Y_{j}\left( t_{j}\right) \right) \sum_{j}\int
V\left( Y_{i}\left( t_{i}\right) ,Y_{j}\left( t_{j}\right) \right) }{2\sigma
_{2}^{2}}\right)
\end{eqnarray*}%
We also assume that:%
\begin{equation*}
\sigma _{2}^{2}<<\sigma ^{2}
\end{equation*}%
This weight has the required form to apply the techniques developped in
Gosselin, Lotz and Wambst (2017), and the corresponding field theory action
writes:%
\begin{eqnarray*}
&&\Psi ^{\dagger }\left( Y,\tilde{X}\right) \left( -\nabla _{\theta }\left(
\vartheta ^{2}\nabla _{\theta }-2\right) -\frac{1}{2}\nabla _{X}\left(
\sigma ^{2}\nabla _{X}+G\left( X\right) +\tilde{X}\right) +G^{\prime }\left(
X\right) -\frac{1}{2}\varepsilon ^{2}\nabla _{\tilde{X}}^{2}-\frac{1}{2}%
\omega \tilde{X}^{2}\right) \Psi \left( Y,\tilde{X}\right) \\
&&+\int \frac{\tilde{X}^{2}}{2\sigma _{2}^{2}}\left\vert \Psi \left( Y,%
\tilde{X}\right) \right\vert ^{2}-\int \frac{\tilde{X}V\left( \left( Y,%
\tilde{X}\right) ,\left( Y^{\prime },\tilde{X}^{\prime }\right) \right) }{%
\sigma _{2}^{2}}\left\vert \Psi \left( Y,\tilde{X}\right) \right\vert
^{2}\left\vert \Psi \left( Y^{\prime },\tilde{X}^{\prime }\right)
\right\vert ^{2} \\
&&+\int \frac{V\left( \left( Y,\tilde{X}\right) ,\left( Y^{\prime },\tilde{X}%
^{\prime }\right) \right) V\left( \left( Y,\tilde{X}\right) ,\left( Y^{"},%
\tilde{X}"\right) \right) }{2\sigma _{2}^{2}}\left\vert \Psi \left( Y,\tilde{%
X}\right) \right\vert ^{2}\left\vert \Psi \left( Y^{\prime },\tilde{X}%
^{\prime }\right) \right\vert ^{2}\left\vert \Psi \left( Y^{"},\tilde{X}%
"\right) \right\vert ^{2}
\end{eqnarray*}%
We apply the transformation on the field%
\begin{equation}
\Psi \left( Y,\tilde{X}\right) =\exp \left( -\int \left( G\left( X\right) +%
\tilde{X}\right) \right) \bar{\Psi}\left( Y,\tilde{X}\right)  \label{chngvr}
\end{equation}%
and change of notation for the sake of simplicity:%
\begin{equation*}
\bar{\Psi}\left( Y,\tilde{X}\right) \rightarrow \Psi \left( Y,\tilde{X}%
\right)
\end{equation*}%
For an action:%
\begin{eqnarray*}
&&\Psi ^{\dagger }\left( Y,\tilde{X}\right) \left( -\nabla _{\theta }\left(
\vartheta ^{2}\nabla _{\theta }-2\right) -\frac{1}{2}\nabla _{X}^{2}+\left(
G\left( X\right) +\tilde{X}\right) ^{2}-\frac{1}{2}\varepsilon ^{2}\nabla _{%
\tilde{X}}^{2}-\frac{1}{2}\omega \tilde{X}^{2}\right) \Psi \left( Y,\tilde{X}%
\right) \\
&&+\int \frac{\tilde{X}^{2}}{2\sigma _{2}^{2}}\left\vert \Psi \left( Y,%
\tilde{X}\right) \right\vert ^{2}-\int \frac{\tilde{X}V\left( \left( Y,%
\tilde{X}\right) ,\left( Y^{\prime },\tilde{X}^{\prime }\right) \right) }{%
\sigma _{2}^{2}}\left\vert \Psi \left( Y,\tilde{X}\right) \right\vert
^{2}\left\vert \Psi \left( Y^{\prime },\tilde{X}^{\prime }\right)
\right\vert ^{2} \\
&&+\int \frac{V\left( \left( Y,\tilde{X}\right) ,\left( Y^{\prime },\tilde{X}%
^{\prime }\right) \right) V\left( \left( Y,\tilde{X}\right) ,\left( Y^{"},%
\tilde{X}"\right) \right) }{2\sigma _{2}^{2}}\left\vert \Psi \left( Y,\tilde{%
X}\right) \right\vert ^{2}\left\vert \Psi \left( Y^{\prime },\tilde{X}%
^{\prime }\right) \right\vert ^{2}\left\vert \Psi \left( Y^{"},\tilde{X}%
"\right) \right\vert ^{2}
\end{eqnarray*}%
For potential satisfying: $V\left( \left( Y,\tilde{X}\right) ,\left(
Y^{\prime },\tilde{X}^{\prime }\right) \right) =V\left( Y,Y^{\prime }\right) 
$, which is the case considered in this work, it simplifies as:%
\begin{eqnarray*}
&&\Psi ^{\dagger }\left( Y,\tilde{X}\right) \left( -\nabla _{\theta }\left(
\vartheta ^{2}\nabla _{\theta }-2\right) -\frac{1}{2}\nabla _{X}^{2}+\left(
G\left( X\right) +\tilde{X}\right) ^{2}-\frac{1}{2}\varepsilon ^{2}\nabla _{%
\tilde{X}}^{2}-\frac{1}{2}\omega \right) \Psi \left( Y,\tilde{X}\right) \\
&&+\int \frac{\tilde{X}^{2}}{2\sigma _{2}^{2}}\left\vert \Psi \left( Y,%
\tilde{X}\right) \right\vert ^{2}-\int \frac{\tilde{X}V\left( Y,Y^{\prime
}\right) }{\sigma _{2}^{2}}\left\vert \Psi \left( Y,\tilde{X}\right)
\right\vert ^{2}\left\vert \Psi \left( Y^{\prime },\tilde{X}^{\prime
}\right) \right\vert ^{2} \\
&&+\int \frac{V\left( Y,Y^{\prime }\right) V\left( Y,Y^{"}\right) }{2\sigma
_{2}^{2}}\left\vert \Psi \left( Y,\tilde{X}\right) \right\vert
^{2}\left\vert \Psi \left( Y^{\prime },\tilde{X}^{\prime }\right)
\right\vert ^{2}\left\vert \Psi \left( Y^{"},\tilde{X}"\right) \right\vert
^{2}
\end{eqnarray*}%
and this is equal to:%
\begin{eqnarray}
&&\Psi ^{\dagger }\left( Y,\tilde{X}\right) \left( -\nabla _{\theta }\left(
\vartheta ^{2}\nabla _{\theta }-2\right) -\frac{1}{2}\nabla _{X}^{2}+\left(
G\left( X\right) +\tilde{X}\right) ^{2}-\frac{1}{2}\varepsilon ^{2}\nabla _{%
\tilde{X}}^{2}-\frac{1}{2}\omega \right) \Psi \left( Y,\tilde{X}\right) 
\notag \\
&&+\int \frac{\left( \tilde{X}-\int V\left( Y,Y^{\prime }\right) \left\vert
\Psi \left( Y^{\prime },\tilde{X}^{\prime }\right) \right\vert ^{2}\right)
^{2}}{2\sigma _{2}^{2}}\left\vert \Psi \left( Y,\tilde{X}\right) \right\vert
^{2}  \label{ptnt}
\end{eqnarray}%
Now, given our assumptions: $\varepsilon ^{2}<<1,\sigma _{2}^{2}<<1$, we can
consider that: $\varepsilon ^{2}\nabla _{\tilde{X}}^{2}\rightarrow 0$. As a
consequence, the condition%
\begin{equation*}
0=\int \Psi ^{\dagger }\left( Y,\tilde{X}\right) \left( \tilde{X}-\int
V\left( Y,Y^{\prime }\right) \left\vert \Psi \left( Y^{\prime },\tilde{X}%
^{\prime }\right) \right\vert ^{2}\right) ^{2}\Psi \left( Y,\tilde{X}\right)
\end{equation*}%
imposed by $\sigma _{2}^{2}$ in (\ref{ptnt}) is obtained for a function of
the type: 
\begin{equation*}
\Psi \left( Y,\tilde{X}\right) =\delta \left( \tilde{X}-\int \Psi ^{\dagger
}\left( Y^{\prime }\right) V\left( Y,Y^{\prime }\right) \Psi \left(
Y^{\prime }\right) \right) \Psi \left( Y\right)
\end{equation*}%
The $\delta \left( \tilde{X}-\int \Psi ^{\dagger }\left( Y^{\prime }\right)
V\left( Y,Y^{\prime }\right) \Psi \left( Y^{\prime }\right) \right) $ is
gaussian of norm equal to $1$ peaked around $\int \Psi ^{\dagger }\left(
Y^{\prime }\right) V\left( Y,Y^{\prime }\right) \Psi \left( Y^{\prime
}\right) $. As a consequence:

\begin{eqnarray*}
&&\int \Psi ^{\dagger }\left( Y,\tilde{X}\right) \left( G\left( X\right) +%
\tilde{X}\right) ^{2}\Psi \left( Y,\tilde{X}\right) \\
&=&\int \Psi ^{\dagger }\left( Y,\tilde{X}\right) \left( G\left( X\right)
+\int V\left( Y,Y^{\prime }\right) \left\vert \Psi \left( Y^{\prime },\tilde{%
X}^{\prime }\right) \right\vert ^{2}\right) ^{2}\Psi \left( Y,\tilde{X}%
\right) \\
&&-2\int \Psi ^{\dagger }\left( Y,\tilde{X}\right) \left( G\left( X\right)
+\int \tilde{X}V\left( Y,Y^{\prime }\right) \left\vert \Psi \left( Y^{\prime
},\tilde{X}^{\prime }\right) \right\vert ^{2}\right) \\
&&\times \left( \tilde{X}-\int V\left( Y,Y^{\prime }\right) \left\vert \Psi
\left( Y^{\prime },\tilde{X}^{\prime }\right) \right\vert ^{2}\right) \Psi
\left( Y,\tilde{X}\right) \\
&&+\int \Psi ^{\dagger }\left( Y,\tilde{X}\right) \left( \tilde{X}-\int
V\left( Y,Y^{\prime }\right) \left\vert \Psi \left( Y^{\prime },\tilde{X}%
^{\prime }\right) \right\vert ^{2}\right) ^{2}\Psi \left( Y,\tilde{X}\right)
\\
&=&\int \Psi ^{\dagger }\left( Y,\tilde{X}\right) \left( G\left( X\right)
+\int V\left( Y,Y^{\prime }\right) \left\vert \Psi \left( Y^{\prime },\tilde{%
X}^{\prime }\right) \right\vert ^{2}\right) ^{2}\Psi Y,\tilde{X} \\
&&-2\int \Psi ^{\dagger }\left( Y,\tilde{X}\right) \left( G\left( X\right)
+\int V\left( Y,Y^{\prime }\right) \left\vert \Psi \left( Y^{\prime },\tilde{%
X}^{\prime }\right) \right\vert ^{2}\right) \\
&&\times \left( \tilde{X}-\int \Psi ^{\dagger }\left( Y^{\prime },\tilde{X}%
^{\prime }\right) V\left( Y,Y^{\prime }\right) \Psi \left( Y^{\prime },%
\tilde{X}^{\prime }\right) \right) \Psi \left( Y,\tilde{X}\right)
\end{eqnarray*}%
This last quantity is negligible since:%
\begin{eqnarray}
&&\left\vert \int \Psi ^{\dagger }\left( Y,\tilde{X}\right) \left( G\left(
X\right) +\int \tilde{X}V\left( Y,Y^{\prime }\right) \left\vert \Psi \left(
Y^{\prime },\tilde{X}^{\prime }\right) \right\vert ^{2}\right) \left( \tilde{%
X}-\int \left\vert \Psi \left( Y^{\prime },\tilde{X}^{\prime }\right)
\right\vert ^{2}\tilde{X}V\left( Y,Y^{\prime }\right) \right) \Psi \left( Y,%
\tilde{X}\right) \right\vert  \notag \\
&\leqslant &\left\Vert \Psi ^{\dagger }\left( Y,\tilde{X}\right) \left(
G\left( X\right) +\int \tilde{X}V\left( Y,Y^{\prime }\right) \left\vert \Psi
\left( Y^{\prime },\tilde{X}^{\prime }\right) \right\vert ^{2}\right)
\right\Vert \left\Vert \left( \tilde{X}-\int \left\vert \Psi \left(
Y^{\prime },\tilde{X}^{\prime }\right) \right\vert ^{2}\tilde{X}V\left(
Y,Y^{\prime }\right) \right) \Psi \left( Y,\tilde{X}\right) \right\Vert
\label{Nrrm}
\end{eqnarray}%
and the norm of the last factor in (\ref{Nrm}) is close to zero, since:%
\begin{eqnarray*}
&&\left\Vert \left( \tilde{X}-\int \Psi ^{\dagger }\left( Y^{\prime },\tilde{%
X}^{\prime }\right) V\left( Y,Y^{\prime }\right) \Psi \left( Y^{\prime },%
\tilde{X}^{\prime }\right) \right) \Psi \left( Y,\tilde{X}\right)
\right\Vert ^{2} \\
&=&\int \Psi ^{\dagger }\left( Y,\tilde{X}\right) \left( \tilde{X}-\int \Psi
^{\dagger }\left( Y^{\prime },\tilde{X}^{\prime }\right) V\left( Y,X^{\prime
}\right) \Psi \left( Y^{\prime },\tilde{X}^{\prime }\right) \right) ^{2}\Psi
\left( Y,\tilde{X}\right) \\
&\rightarrow &0
\end{eqnarray*}%
Thus, the potential (\ref{ptnt}) simplifies as:%
\begin{eqnarray*}
&&\Psi ^{\dagger }\left( Y,\tilde{X}\right) \left( -\nabla _{\theta }\left(
\vartheta ^{2}\nabla _{\theta }-2\right) -\frac{1}{2}\nabla _{X}^{2}\right)
\Psi \left( Y,\tilde{X}\right) \\
&&+\int \Psi ^{\dagger }\left( Y,\tilde{X}\right) \left( G\left( X\right)
+\int \Psi ^{\dagger }\left( Y^{\prime },\tilde{X}^{\prime }\right) V\left(
Y,Y^{\prime }\right) \Psi \left( Y^{\prime },\tilde{X}^{\prime }\right)
\right) ^{2}\Psi \left( Y,\tilde{X}\right) \\
&=&\Psi ^{\dagger }\left( Y\right) \left( -\frac{\sigma ^{2}}{2}\nabla
_{X}^{2}+\left( G\left( X\right) +\int \Psi ^{\dagger }\left( Y^{\prime
}\right) V\left( Y,Y^{\prime }\right) \Psi \left( Y^{\prime }\right) \right)
^{2}\right) \Psi \left( Y\right)
\end{eqnarray*}%
We can now come back to the initial variables, by letting $Y\rightarrow
\left( X,\theta \right) $, and this yields: 
\begin{eqnarray*}
&&\Psi ^{\dagger }\left( X,\theta \right) \left( -\nabla _{\theta }\left(
\vartheta ^{2}\nabla _{\theta }-2\right) -\frac{\sigma ^{2}}{2}\nabla
_{X}^{2}+\left( G\left( X\right) +\int V\left( X,X^{\prime }\right) \delta
\left( \theta -\theta ^{\prime }\right) \left\vert \Psi \left( X^{\prime
},\theta ^{\prime }\right) \right\vert ^{2}\right) ^{2}\right) \Psi \left(
X,\theta \right) \\
&=&\Psi ^{\dagger }\left( X,\theta \right) \left( -\nabla _{\theta }\left(
\vartheta ^{2}\nabla _{\theta }-2\right) -\frac{\sigma ^{2}}{2}\nabla
_{X}^{2}+\left( G\left( X\right) +\int V\left( X,X^{\prime }\right)
\left\vert \Psi \left( X^{\prime },\theta ^{\prime }\right) \right\vert
^{2}\right) ^{2}\right) \Psi \left( X,\theta \right)
\end{eqnarray*}%
We also change of variable for the dependence in the counting variable:%
\begin{equation*}
\Psi \left( X,\theta \right) =\exp \left( \frac{\theta }{\vartheta ^{2}}%
\right) \bar{\Psi}\left( X,\theta \right)
\end{equation*}%
and reset again:%
\begin{equation*}
\bar{\Psi}\left( X,\theta \right) \rightarrow \Psi \left( X,\theta \right)
\end{equation*}%
Given our choice for $G$ and $V$ (\ref{GV}), and adding (\ref{scndwgt}), the
field theoretic version of (\ref{wgt}), one obtains (\ref{fldthvrsn}).

To close this section, we note that due to the change of variable\ \ (\ref%
{chngvr})\ \ the source terms have to be included as they participate to the
action.\ One thus has the action plus source terms:%
\begin{eqnarray*}
&&\Psi ^{\dagger }\left( X,\tilde{X}\right) \left( -\vartheta ^{2}\nabla
_{\theta }^{2}+\frac{1}{\vartheta ^{2}}-\frac{1}{2}\nabla _{X}^{2}\right)
\Psi \left( X,\tilde{X}\right) \\
&&+\int \Psi ^{\dagger }\left( X,\tilde{X}\right) \left( G\left( X\right)
+\int \Psi ^{\dagger }\left( X^{\prime },\tilde{X}^{\prime }\right) V\left(
X,X^{\prime }\right) \Psi \left( X^{\prime },\tilde{X}^{\prime }\right)
\right) ^{2}\Psi \left( X,\tilde{X}\right) \\
&&+\int J\left( X\right) \exp \left( -\int \left( G\left( X\right) +\tilde{X}%
\right) \right) \Psi \left( X,\tilde{X}\right) +J^{\dagger }\left( X\right)
\exp \left( \int \left( G\left( X\right) +\tilde{X}\right) \right) \Psi
^{\dagger }\left( X,\tilde{X}\right) \\
&=&\Psi ^{\dagger }\left( X\right) \left( -\vartheta ^{2}\nabla _{\theta
}^{2}+\frac{1}{\vartheta ^{2}}-\frac{\sigma ^{2}}{2}\nabla _{X}^{2}+\left(
G\left( X\right) +\int \Psi ^{\dagger }\left( X^{\prime }\right) V\left(
X\right) \Psi \left( X^{\prime }\right) \right) ^{2}\right) \Psi \left(
X\right) \\
&&+\int J\left( X\right) \exp \left( -\int \left( G\left( X\right) +\tilde{X}%
\right) \right) \Psi \left( X\right) +J^{\dagger }\left( X\right) \exp
\left( \int \left( G\left( X\right) +\tilde{X}\right) \right) \Psi ^{\dagger
}\left( X\right)
\end{eqnarray*}

for an action including the source terms:%
\begin{eqnarray*}
&&\Psi ^{\dagger }\left( X\right) \left( -\vartheta ^{2}\nabla _{\theta
}^{2}+\frac{1}{\vartheta ^{2}}-\frac{1}{2}\nabla _{X}^{2}\right) \Psi \left(
X\right) +\int \Psi ^{\dagger }\left( X\right) \left( G\left( X\right) +\int
\Psi ^{\dagger }\left( X^{\prime }\right) V\left( X,X^{\prime }\right) \Psi
\left( X^{\prime }\right) \right) ^{2}\Psi \left( X\right) \\
&&+\int J\left( X\right) \exp \left( -\int^{X}G\left( X\right) +X\tilde{X}%
\left( X\right) \right) \Psi \left( X\right) +\int J^{\dagger }\left(
X\right) \exp \left( \int^{X}G\left( X\right) +X\tilde{X}\left( X\right)
\right) \Psi ^{\dagger }\left( X\right)
\end{eqnarray*}%
The Green function are thus computed through the following formula:

\begin{eqnarray*}
G\left( X,Y\right) &=&\left\langle \exp \left( -\left( \int^{X}G\left(
X\right) +X\int \Psi ^{\dagger }\left( X^{\prime }\right) V\left(
X,X^{\prime }\right) \Psi \left( X^{\prime }\right) \right) \right) \right.
\\
&&\times \left. \Psi \left( X\right) \Psi ^{\dagger }\left( Y\right) \exp
\left( \left( \int^{Y}G\left( X\right) +Y\int \Psi ^{\dagger }\left(
X^{\prime }\right) V\left( Y,X^{\prime }\right) \Psi \left( X^{\prime
}\right) \right) \right) \right\rangle \\
&=&\left\langle \exp \left( -\left( \int_{Y}^{X}G\left( X\right) +\int \Psi
^{\dagger }\left( X^{\prime }\right) \left( XV\left( X,X^{\prime }\right)
-YV\left( Y,X^{\prime }\right) \right) \Psi \left( X^{\prime }\right)
\right) \right) \Psi \left( X\right) \Psi ^{\dagger }\left( Y\right)
\right\rangle
\end{eqnarray*}

\subsection{\textbf{Appendix 2}}

We inspect the case $\rho =0$.

As said in the text we look for a configuration that satisfies the market
clearing condition. That is for $\bar{A}>>A$ we replace \ 
\begin{equation*}
\Psi \left( K_{2},P_{2},X_{2},\theta \right) \rightarrow \delta \left(
P-H\left( K,X,\theta \right) \right) \Psi \left( K_{2},X_{2},\theta \right)
\end{equation*}

\bigskip As in the text we set $Z=\left( K,P,X,\theta \right) $. At zeroth
order in $\frac{A}{\bar{A}}$ the potential becomes:

\begin{equation*}
\left( \delta K-APK^{\alpha }\left( 1-\frac{\kappa }{d^{2}}\int \frac{%
P_{3}\exp \left( -\left( \frac{\left\vert X-X_{2}\right\vert }{d}+\frac{%
\left\vert X-X_{3}\right\vert }{d}\right) \right) }{P_{2}^{\gamma }}%
\left\vert \Psi \left( Z_{2},\theta \right) \right\vert ^{2}\left\vert \Psi
\left( Z_{3},\theta \right) \right\vert ^{2}\right) \right) ^{2}
\end{equation*}%
\bigskip with constraint:

\begin{equation*}
P^{1+\gamma }K^{\alpha }=\frac{\kappa }{d^{2}}\int P_{2}\left( K_{2}\right)
^{\alpha }P_{3}\exp \left( -\left( \frac{\left\vert X-X_{2}\right\vert }{d}+%
\frac{\left\vert X_{2}-X_{3}\right\vert }{d}\right) \right) \left\vert \Psi
\left( Z_{2},\theta \right) \right\vert ^{2}\left\vert \Psi \left(
Z_{3},\theta \right) \right\vert ^{2}
\end{equation*}%
The constraint is solved by looking for a field configuration (projection) 
\begin{equation*}
\Psi \left( K,P,X,\theta \right) =\delta \left( P-H\left( K,X\right) \right)
\Psi \left( K,X,\theta \right)
\end{equation*}%
for a function $P$%
\begin{equation*}
P=H\left( K,X\right) =\left( K\right) ^{-\frac{\alpha }{1+\gamma }}f\left(
X\right)
\end{equation*}%
and replacing in the constraint, leads to:%
\begin{equation}
\left( f\left( X\right) \right) ^{1+\gamma }=\kappa \int P_{2}\left(
K_{2}\right) ^{\alpha }P_{3}\exp \left( -\left( \frac{\left\vert
X-X_{2}\right\vert }{d}+\frac{\left\vert X_{2}-X_{3}\right\vert }{d}\right)
\right) \left\vert \Psi \left( K_{2},X_{2},\theta \right) \right\vert
^{2}\left\vert \Psi \left( K_{3},X_{3},\theta \right) \right\vert ^{2}
\label{qnf}
\end{equation}%
in average.

The potential for $K$ becomes:%
\begin{eqnarray*}
&&\left( \delta K-A\left( K\right) ^{\frac{\alpha \gamma }{1+\gamma }%
}f\left( X\right) \left( 1-\kappa \int \frac{\left( K_{3}\right) ^{-\frac{%
\alpha }{1+\gamma }}f\left( X_{3}\right) \exp \left( -\left( \frac{%
\left\vert X-X_{2}\right\vert +\left\vert X-X_{3}\right\vert }{d}\right)
\right) \left\vert \Psi \left( K_{2},X_{2},\theta \right) \right\vert
^{2}\left\vert \Psi \left( K_{3},X_{3},\theta \right) \right\vert ^{2}}{%
\left( \left( K_{2}\right) ^{-\frac{\alpha }{1+\gamma }}f\left( X_{2}\right)
\right) ^{\gamma }}\right) \right) ^{2} \\
&=&\left( \delta K-A\left( K\right) ^{\frac{\alpha \gamma }{1+\gamma }%
}f\left( X\right) \left( 
\begin{array}{c}
1-\kappa \int \left( K_{2}\right) ^{\frac{\alpha \gamma }{1+\gamma }}\left(
K_{3}\right) ^{-\frac{\alpha }{1+\gamma }}f\left( X_{3}\right) \left(
f\left( X_{2}\right) \right) ^{-\gamma } \\ 
\times \exp \left( -\left( \frac{\left\vert X-X_{2}\right\vert }{d}+\frac{%
\left\vert X-X_{3}\right\vert }{d}\right) \right) \left\vert \Psi \left(
K_{2},X_{2},\theta \right) \right\vert ^{2}\left\vert \Psi \left(
K_{3},X_{3},\theta \right) \right\vert ^{2}%
\end{array}%
\right) \right) ^{2}
\end{eqnarray*}

\bigskip for $\alpha \gamma <<1$, $\left( K\right) ^{\frac{\alpha \gamma }{%
1+\gamma }}\simeq \left\langle K\right\rangle _{X}^{\frac{\alpha \gamma }{%
1+\gamma }}$, and this describes, for a given $X$, an harmonic oscillator of
average $\left\langle K\right\rangle _{X}$ that satisfies:%
\begin{equation}
\delta \left\langle K\right\rangle _{X}-A\left( \left\langle K\right\rangle
_{X}\right) ^{\frac{\alpha \gamma }{1+\gamma }}f\left( X\right) \left( 
\begin{array}{c}
1-\kappa \int \left( K_{2}\right) ^{\frac{\alpha \gamma }{1+\gamma }}\left(
K_{3}\right) ^{-\frac{\alpha }{1+\gamma }}f\left( X_{3}\right) \left(
f\left( X_{2}\right) \right) ^{-\gamma }\times \\ 
\times \exp \left( -\left( \frac{\left\vert X-X_{2}\right\vert }{d}+\frac{%
\left\vert X-X_{3}\right\vert }{d}\right) \right) \left\vert \Psi \left(
K_{2},X_{2},\theta \right) \right\vert ^{2}\left\vert \Psi \left(
K_{3},X_{3},\theta \right) \right\vert ^{2}%
\end{array}%
\right) =0  \label{qtnKX}
\end{equation}

with solution:%
\begin{equation}
\left\langle K\right\rangle _{X}=\left( \frac{A}{\delta }f\left( X\right)
\right) ^{\frac{1+\gamma }{1+\gamma \left( 1-\alpha \right) }}\left( 
\begin{array}{c}
1-\kappa \int \left( K_{2}\right) ^{\frac{\alpha \gamma }{1+\gamma }}\left(
K_{3}\right) ^{-\frac{\alpha }{1+\gamma }}f\left( X_{3}\right) \left(
f\left( X_{2}\right) \right) ^{-\gamma }\times \\ 
\times \exp \left( -\left( \frac{\left\vert X-X_{2}\right\vert }{d}+\frac{%
\left\vert X-X_{3}\right\vert }{d}\right) \right) \left\vert \Psi \left(
K_{2},X_{2},\theta \right) \right\vert ^{2}\left\vert \Psi \left(
K_{3},X_{3},\theta \right) \right\vert ^{2}%
\end{array}%
\right) ^{\frac{1+\gamma }{1+\gamma \left( 1-\alpha \right) }}  \label{KX}
\end{equation}%
\bigskip This expression can thus be used in (\ref{qfn}), to identify $%
f\left( X\right) $: 
\begin{eqnarray}
\left( f\left( X\right) \right) ^{1+\gamma } &=&\kappa \int \left(
K_{2}\right) ^{\frac{\alpha \gamma }{1+\gamma }}\left( K_{3}\right) ^{-\frac{%
\alpha }{1+\gamma }}f\left( X_{3}\right) f\left( X_{2}\right)  \label{qfn2}
\\
&&\exp \left( -\left( \frac{\left\vert X-X_{2}\right\vert }{d}+\frac{%
\left\vert X_{2}-X_{3}\right\vert }{d}\right) \right) \left\vert \Psi \left(
K_{2},X_{2},\theta \right) \right\vert ^{2}\left\vert \Psi \left(
K_{3},X_{3},\theta \right) \right\vert ^{2}  \notag
\end{eqnarray}%
At the lowest order of perturbation theory (\ref{qfn2}) becomes:%
\begin{eqnarray*}
\left( f\left( X\right) \right) ^{1+\gamma } &\simeq &\kappa \int G\left(
\left( K_{3},X_{3},\theta \right) ,\left( K_{2},X_{2},\theta \right) \right)
G\left( \left( K_{2},X_{2},\theta \right) ,\left( K_{3},X_{3},\theta \right)
\right) \left( K_{2}\right) ^{\frac{\alpha \gamma }{1+\gamma }}\left(
K_{3}\right) ^{-\frac{\alpha }{1+\gamma }} \\
&&\times f\left( X_{3}\right) f\left( X_{2}\right) \exp \left( -\left( \frac{%
\left\vert X-X_{2}\right\vert }{d}+\frac{\left\vert X_{2}-X_{3}\right\vert }{%
d}\right) \right)
\end{eqnarray*}%
In Appendix 3, we will derive the Green functions. It appears that for $%
\theta ^{\prime }\simeq \theta $, The Green functions are those of harmonic
oscillators for a propagation time of order $\vartheta ^{2}$ that we will
write $G\left( \left( K_{3},X_{3}\right) ,\left( K_{2},X_{2}\right) \right) $%
. We thus have: 
\begin{equation*}
\left( f\left( X\right) \right) ^{1+\gamma }\simeq \kappa \int G\left(
\left( K_{3},X_{2}\right) ,\left( K_{2},X_{2}\right) \right) G\left( \left(
K_{2},X_{2}\right) ,\left( K_{3},X_{3}\right) \right) \left( K_{2}\right) ^{%
\frac{\alpha \gamma }{1+\gamma }}\left( K_{3}\right) ^{-\frac{\alpha }{%
1+\gamma }}\left( f\left( X_{2}\right) \right) ^{2}\exp \left( -\frac{%
\left\vert X-X_{2}\right\vert }{d}\right)
\end{equation*}%
We also assume that the dynamic for $K$ is "faster" than the dynamics for $X$%
, so that the Green functions have the form:%
\begin{equation*}
G\left( \left( K_{3},X_{3}\right) ,\left( K_{2},X_{2}\right) \right)
=G_{X_{2}}\left( K_{3},K_{2}\right) G\left( X_{2},X_{2}\right)
\end{equation*}%
As a consequence, the integrals over $K_{2}$ and $K_{3}$:%
\begin{equation*}
\int G_{X_{2}}\left( K_{3},K_{2}\right) \left( K_{2}\right) ^{\frac{\alpha
\gamma }{1+\gamma }}\left( K_{3}\right) ^{-\frac{\alpha }{1+\gamma }%
}G_{X_{3}}\left( K_{3},K_{2}\right) dK_{2}dK_{3}
\end{equation*}%
compute the average of $\left( K_{2}\right) ^{\frac{\alpha \gamma }{1+\gamma 
}}\left( K_{3}\right) ^{-\frac{\alpha }{1+\gamma }}$ for a transition from
state $X_{2}$ to state $X_{3}$. As a consequence: 
\begin{eqnarray*}
&&\int G\left( \left( K_{3},X_{3}\right) ,\left( K_{2},X_{2}\right) \right)
\left( K_{2}\right) ^{\frac{\alpha \gamma }{1+\gamma }}\left( K_{3}\right)
^{-\frac{\alpha }{1+\gamma }}G\left( \left( K_{2},X_{2}\right) ,\left(
K_{3},X_{3}\right) \right) dK_{2}dK_{3} \\
&\simeq &G\left( X_{2},X_{3}\right) G\left( X_{3},X_{2}\right) \left\langle
K\right\rangle _{X_{2}}^{\frac{\alpha \gamma }{1+\gamma }}\left\langle
K\right\rangle _{X_{3}}^{-\frac{\alpha }{1+\gamma }}
\end{eqnarray*}%
where $\left\langle K\right\rangle _{X_{2}}$ is the average level of capital
for agent with $X=X_{2}$ is computed below by identification.

As a consequence:%
\begin{equation*}
\left( f\left( X\right) \right) ^{1+\gamma }\simeq \frac{\kappa }{d^{2}}\int
G\left( X_{2},X_{3}\right) G\left( X_{3},X_{2}\right) \left\langle
K\right\rangle _{X_{2}}^{\frac{\alpha \gamma }{1+\gamma }}\left\langle
K\right\rangle _{X_{3}}^{-\frac{\alpha }{1+\gamma }}f\left( X_{3}\right)
f\left( X_{2}\right) \exp \left( -\left( \frac{\left\vert X-X_{2}\right\vert 
}{d}+\frac{\left\vert X_{2}-X_{3}\right\vert }{d}\right) \right)
\end{equation*}%
The potential for $X$ is harmonic in zeroth order in $\kappa $ with low
frequency and then we can assume that the integral is distributed around $%
X_{2}-X_{3}\simeq 0$ and replace $\exp \left( -\left( \frac{\left\vert
X_{2}-X_{3}\right\vert }{d}\right) \right) $ by $d$ times a delta function.
As a consequence: 
\begin{equation*}
\left( f\left( X\right) \right) ^{1+\gamma }\simeq \frac{\kappa }{d}\int
\left\langle K\right\rangle _{X_{2}}^{\frac{\alpha \left( \gamma -1\right) }{%
1+\gamma }}\left( f\left( X_{2}\right) \right) ^{2}G\left(
X_{2},X_{2}\right) \exp \left( -\left( \frac{\left\vert X-X_{2}\right\vert }{%
d}\right) \right)
\end{equation*}%
If the exchange positions are homogenously spread on $\left[ -1,1\right] $,
one can replace the probability $G\left( X_{2},X_{2}\right) $ by a constant
density, here $\frac{1}{2}$ in first approximation, to normalize the
probability of the interval $\left[ -1,1\right] $ to $1$.

\begin{equation*}
\left( f\left( X\right) \right) ^{1+\gamma }\simeq \frac{\kappa }{2d}\int
\left\langle K\right\rangle _{X_{2}}^{\frac{\alpha \left( \gamma -1\right) }{%
1+\gamma }}\left( f\left( X_{2}\right) \right) ^{2}\exp \left( -\frac{%
\left\vert X-X_{2}\right\vert }{d}\right)
\end{equation*}%
Thus, using (\ref{KX})\ one can express $\left\langle K\right\rangle _{X}$
as a function of $X$: 
\begin{eqnarray}
\left\langle K\right\rangle _{X} &=&\left( \frac{A}{\delta }f\left( X\right)
\right) ^{\frac{1+\gamma }{1+\gamma \left( 1-\alpha \right) }}\left( 
\begin{array}{c}
1-\frac{\kappa }{4d^{2}}\int \left( K_{2}\right) ^{\frac{\alpha \gamma }{%
1+\gamma }}\left( K_{3}\right) ^{-\frac{\alpha }{1+\gamma }}f\left(
X_{3}\right) \left( f\left( X_{2}\right) \right) ^{-\gamma }\times \\ 
\times \exp \left( -\left( \frac{\left\vert X-X_{2}\right\vert }{d}+\frac{%
\left\vert X-X_{3}\right\vert }{d}\right) \right) \left\vert \Psi \left(
K_{2},X_{2},\theta \right) \right\vert ^{2}\left\vert \Psi \left(
K_{3},X_{3},\theta \right) \right\vert ^{2}%
\end{array}%
\right) ^{\frac{1+\gamma }{1+\gamma \left( 1-\alpha \right) }}  \label{Kidtn}
\\
&\simeq &\left( \frac{A}{\delta }f\left( X\right) \right) ^{\frac{1+\gamma }{%
1+\gamma \left( 1-\alpha \right) }}\left( 1-\frac{\kappa }{4d^{2}}\int
\left\langle K\right\rangle _{X_{2}}^{\frac{\alpha \gamma }{1+\gamma }%
}\left\langle K\right\rangle _{X_{3}}^{-\frac{\alpha }{1+\gamma }}f\left(
X_{3}\right) \left( f\left( X_{2}\right) \right) ^{-\gamma }\exp \left(
-\left( \frac{\left\vert X-X_{2}\right\vert }{d}+\frac{\left\vert
X-X_{3}\right\vert }{d}\right) \right) \right)  \notag
\end{eqnarray}%
for $\gamma <<1$. The factor $\frac{1}{4}$ appears again to normalize the
probability each interval $\left[ -1,1\right] $ to $1$. We thus postulate
the following form for $\left\langle K\right\rangle _{X}$:

\begin{equation}
\left\langle K\right\rangle _{X}=\left( \frac{A}{\delta }f\left( X\right)
\left( 1-h\exp \left( -\frac{\left\vert X\right\vert }{d}\right) \left( 1-%
\frac{\cosh \frac{X}{d}}{\exp \left( \frac{1}{d}\right) }\right) \right)
\right) ^{\frac{1+\gamma }{1+\gamma \left( 1-\alpha \right) }}  \label{Kpos}
\end{equation}%
with $h$\ to be determined. In (\ref{Kidtn}), we can replace for the sake of
simplicity, $1-h\exp \left( -\frac{\left\vert X\right\vert }{d}\right)
\left( 1-\frac{\cosh \frac{X}{d}}{\exp \left( \frac{1}{d}\right) }\right) $
by its average \ which is approximatively equal to $1-h\exp \left( -\frac{1}{%
2d}\right) \left( 1-\exp \left( -\frac{1}{2d}\right) \right) $, ($h<\exp
\left( \frac{1}{2d}\right) $), so that (\ref{Kidtn}) rewrites:

\begin{eqnarray*}
\left\langle K\right\rangle _{X} &\simeq &\left( \frac{A}{\delta }f\left(
X\right) \right) ^{\frac{1+\gamma }{1+\gamma \left( 1-\alpha \right) }%
}\left( 
\begin{array}{c}
1-\frac{\kappa }{4d^{2}}\int \left( \frac{A}{\delta }f\left( X_{2}\right)
\left( 1-h\exp \left( -\frac{1}{2d}\right) \left( 1-\exp \left( -\frac{1}{2d}%
\right) \right) \right) \right) ^{\frac{\alpha \gamma }{1+\gamma \left(
1-\alpha \right) }} \\ 
\times \left( \frac{A}{\delta }f\left( X_{3}\right) \left( 1-h\right)
\right) ^{-\frac{\alpha }{1+\gamma \left( 1-\alpha \right) }}f\left(
X_{3}\right) \left( f\left( X_{2}\right) \right) ^{-\gamma }\exp \left(
-\left( \frac{\left\vert X-X_{2}\right\vert }{d}+\frac{\left\vert
X-X_{3}\right\vert }{d}\right) \right)%
\end{array}%
\right) \\
&\simeq &\left( \frac{A}{\delta }f\left( X\right) \right) ^{\frac{1+\gamma }{%
1+\gamma \left( 1-\alpha \right) }}\left( 
\begin{array}{c}
1-\frac{\kappa }{4d^{2}}\left( \frac{A}{\delta }\left( 1-h\exp \left( -\frac{%
1}{2d}\right) \left( 1-\exp \left( -\frac{1}{2d}\right) \right) \right)
\right) ^{-\frac{\alpha \left( 1-\gamma \right) }{1+\gamma \left( 1-\alpha
\right) }} \\ 
\times \int \left( f\left( X_{3}\right) \right) ^{\frac{\left( 1-\alpha
\right) \left( 1+\gamma \right) }{1+\gamma \left( 1-\alpha \right) }}\left(
f\left( X_{2}\right) \right) ^{-\gamma \frac{\left( 1-\alpha \right) \left(
1+\gamma \right) }{1+\gamma \left( 1-\alpha \right) }}\exp \left( -\left( 
\frac{\left\vert X-X_{2}\right\vert }{d}+\frac{\left\vert X-X_{3}\right\vert 
}{d}\right) \right)%
\end{array}%
\right)
\end{eqnarray*}%
and the equation for $f\left( X\right) $ becomes:%
\begin{eqnarray*}
\left( f\left( X\right) \right) ^{1+\gamma } &\simeq &\frac{\kappa }{2d}\int
\left\langle K\right\rangle _{X_{2}}^{\frac{\alpha \left( \gamma -1\right) }{%
1+\gamma }}\left( f\left( X_{2}\right) \right) ^{2}\exp \left( -\frac{%
\left\vert X-X_{2}\right\vert }{d}\right) \\
&\simeq &\frac{\kappa }{2d}\left( \frac{A}{\delta }\right) ^{\frac{\alpha
\left( \gamma -1\right) }{1+\gamma \left( 1-\alpha \right) }}\int \left(
1-h\exp \left( -\frac{\left\vert X\right\vert }{d}\right) \left( 1-\frac{%
\cosh \left( \frac{X}{d}\right) }{\exp \left( \frac{1}{d}\right) }\right)
\right) ^{\frac{\alpha \left( \gamma -1\right) }{1+\gamma \left( 1-\alpha
\right) }} \\
&&\times \left( f\left( X_{2}\right) \right) ^{2+\frac{\alpha \left( \gamma
-1\right) }{1+\gamma \left( 1-\alpha \right) }}\exp \left( -\frac{\left\vert
X-X_{2}\right\vert }{d}\right)
\end{eqnarray*}%
Again, we can replace for the sake of simplicity, $1-h\exp \left( -\frac{%
\left\vert X\right\vert }{d}\right) \left( 1-\frac{\cosh \left( \frac{X}{d}%
\right) }{\exp \left( \frac{1}{d}\right) }\right) $ by its average $\simeq
1-h\exp \left( -\frac{1}{2d}\right) \left( 1-\frac{\exp \left( -\frac{1}{2d}%
\right) }{2}\right) $, ($h<\exp \left( \frac{1}{2d}\right) $). We look for a
solution:%
\begin{equation*}
f\left( X\right) =D\exp \left( -c\frac{\left\vert X\right\vert }{d}\right)
\end{equation*}%
and the identification becomes for $\gamma <<1$ :%
\begin{eqnarray*}
\exp \left( -\left( 1+\gamma \right) c\frac{\left\vert X\right\vert }{d}%
\right) &\simeq &\frac{\kappa }{2d}\left( \frac{A}{\delta }\right) ^{\frac{%
\alpha \left( \gamma -1\right) }{1+\gamma \left( 1-\alpha \right) }}D^{\frac{%
\left( 1-\gamma ^{2}\right) \left( 1-\alpha \right) }{1+\gamma \left(
1-\alpha \right) }}\left( \left( 1-h\exp \left( -\frac{1}{2d}\right) \left(
1-\frac{\exp \left( -\frac{1}{2d}\right) }{2}\right) \right) \right) ^{\frac{%
\alpha \left( \gamma -1\right) }{1+\gamma \left( 1-\alpha \right) }} \\
&&\times \int \exp \left( -c\frac{\left( 1+\gamma \right) \allowbreak \left(
2-\alpha \right) }{1+\gamma \left( 1-\alpha \right) }\frac{\left\vert
X_{2}\right\vert }{d}\right) \exp \left( -\frac{\left\vert
X-X_{2}\right\vert }{d}\right)
\end{eqnarray*}%
The integral can be estimated as ($X>0$):

\begin{eqnarray*}
&&\int_{-1}^{0}\exp \left( \frac{2aX_{2}}{d}\right) \exp \left( -\frac{%
\left( X-X_{2}\right) }{d}\right) +\int_{0}^{X}\exp \left( -\frac{2aX_{2}}{d}%
\right) \exp \left( -\frac{\left( X-X_{2}\right) }{d}\right) \\
&&+\int_{X}^{1}\exp \left( -\frac{2aX_{2}}{d}\right) \exp \left( \frac{%
\left( X-X_{2}\right) }{d}\right) \\
&=&d\left( \frac{\exp \left( -\frac{X}{d}\right) \left( 1-\exp \left( -\frac{%
\left( 2a+1\right) }{d}\right) \right) }{2a+1}+\left( \frac{\exp \left( -%
\frac{2aX}{d}\right) -\exp \left( -\frac{X}{d}\right) }{1-2a}\right) +\frac{%
\exp \left( -\frac{2aX}{d}\right) -\exp \left( \frac{X-\left( 2a+1\right) }{d%
}\right) }{2a+1}\right) \\
&\simeq &d\frac{4a\exp \left( -\frac{X}{d}\right) -2\exp \left( -\frac{2aX}{d%
}\right) }{4a^{2}-1}
\end{eqnarray*}%
and the identification for $f\left( X\right) $ becomes:%
\begin{eqnarray*}
\exp \left( -\left( 1+\gamma \right) c\frac{\left\vert X\right\vert }{d}%
\right) &=&\frac{\kappa }{2}\left( \frac{A}{\delta }\right) ^{\frac{\alpha
\left( \gamma -1\right) }{1+\gamma \left( 1-\alpha \right) }}D^{\frac{\left(
1-\gamma ^{2}\right) \left( 1-\alpha \right) }{1+\gamma \left( 1-\alpha
\right) }}\left( \left( 1-h\exp \left( -\frac{1}{2d}\right) \left( 1-\exp
\left( -\frac{1}{2d}\right) \right) \right) \right) ^{\frac{\alpha \left(
\gamma -1\right) }{1+\gamma \left( 1-\alpha \right) }} \\
&&\times \left( \frac{4a\exp \left( -\frac{X}{d}\right) -2\exp \left( -2a%
\frac{X}{d}\right) }{4a^{2}-1}\right)
\end{eqnarray*}%
For $\gamma <1$, one has:%
\begin{eqnarray*}
c &\simeq &\frac{1}{1+\gamma } \\
2a &=&\frac{\left( 1+\gamma \right) \allowbreak \left( 2-\alpha \right) }{%
1+\gamma \left( 1-\alpha \right) }\frac{1}{1+\gamma } \\
&=&\frac{2-\alpha }{1+\gamma \left( 1-\alpha \right) }>1
\end{eqnarray*}%
That yields the identification for $D$:%
\begin{equation*}
\frac{\kappa }{2}\left( \frac{A}{\delta }\right) ^{\frac{\alpha \left(
\gamma -1\right) }{1+\gamma \left( 1-\alpha \right) }}D^{\frac{\left(
1-\gamma ^{2}\right) \left( 1-\alpha \right) }{1+\gamma \left( 1-\alpha
\right) }}\frac{\frac{2\left( 2-\alpha \right) }{1+\gamma \left( 1-\alpha
\right) }}{\left( \frac{2-\alpha }{1+\gamma \left( 1-\alpha \right) }\right)
^{2}-1}\left( \left( 1-h\exp \left( -\frac{1}{2d}\right) \left( 1-\frac{\exp
\left( -\frac{1}{2d}\right) }{2}\right) \right) \right) ^{\frac{\alpha
\left( \gamma -1\right) }{1+\gamma \left( 1-\alpha \right) }}=1
\end{equation*}%
whose solution is:%
\begin{equation*}
D\simeq \left( \frac{1+\gamma \left( 1-\alpha \right) }{\left( 2-\alpha
\right) \kappa }\left( \left( \frac{2-\alpha }{1+\gamma \left( 1-\alpha
\right) }\right) ^{2}-1\right) \left( \frac{A}{\delta }\left( 1-h\exp \left(
-\frac{1}{2d}\right) \left( 1-\frac{\exp \left( -\frac{1}{2d}\right) }{2}%
\right) \right) \right) ^{\frac{\alpha \left( 1-\gamma \right) }{1+\gamma
\left( 1-\alpha \right) }}\right) ^{\frac{1+\gamma \left( 1-\alpha \right) }{%
\left( 1-\gamma ^{2}\right) \left( 1-\alpha \right) }}
\end{equation*}%
and we are lead ultimately to:

\begin{eqnarray}
f\left( X\right) &=&D\exp \left( -\frac{\left\vert X\right\vert }{\left(
1+\gamma \right) d}\right)  \label{fctn} \\
&=&\left( \frac{A}{\delta }\left( 1-h\exp \left( -\frac{1}{2d}\right) \left(
1-\frac{\exp \left( -\frac{1}{2d}\right) }{2}\right) \right) \right) ^{\frac{%
\alpha }{\left( 1-\alpha \right) \left( \gamma +1\right) }}  \notag \\
&&\times \left( \frac{1+\gamma \left( 1-\alpha \right) }{\left( 2-\alpha
\right) \kappa }\left( \left( \frac{2-\alpha }{1+\gamma \left( 1-\alpha
\right) }\right) ^{2}-1\right) \right) ^{\frac{1+\gamma \left( 1-\alpha
\right) }{\left( 1-\gamma ^{2}\right) \left( 1-\alpha \right) }}\exp \left( -%
\frac{\left\vert X\right\vert }{d\left( 1+\gamma \right) }\right)  \notag
\end{eqnarray}%
and using (\ref{Kpos}):%
\begin{eqnarray*}
\left\langle K\right\rangle _{X} &=&\left( \frac{1+\gamma \left( 1-\alpha
\right) }{\left( 2-\alpha \right) \kappa }\left( \left( \frac{2-\alpha }{%
1+\gamma \left( 1-\alpha \right) }\right) ^{2}-1\right) \right) ^{\frac{1}{%
\left( 1-\gamma \right) \left( 1-\alpha \right) }} \\
&&\times \left( \left( 1-h\exp \left( -\frac{1}{2d}\right) \left( 1-\frac{%
\exp \left( -\frac{1}{2d}\right) }{2}\right) \right) \right) ^{\frac{\alpha 
}{\left( 1-\alpha \right) \left( 1+\gamma \left( 1-\alpha \right) \right) }%
}\left( \frac{A}{\delta }\right) ^{-\frac{1}{\left( 1-\alpha \right) }} \\
&&\times \exp \left( -\frac{\left\vert X\right\vert }{\left( 1+\gamma \left(
1-\alpha \right) \right) d}\right) \left( \left( 1-h\exp \left( -\frac{%
\left\vert X\right\vert }{d}\right) \left( 1-\frac{\cosh \left( \frac{X}{d}%
\right) }{\exp \left( \frac{1}{d}\right) }\right) \right) \right) ^{\frac{%
1+\gamma }{1+\gamma \left( 1-\alpha \right) }}
\end{eqnarray*}%
Then we can use the condition (\ref{Kidtn}) on $\left\langle K\right\rangle
_{X}$ and compute $g$ and $h$ (we consider $\gamma <<1$): 
\begin{eqnarray*}
U &=&1-\frac{\kappa }{4d^{2}}\left( \frac{A}{\delta }\left( 1-h\exp \left( -%
\frac{1}{2d}\right) \left( 1-\frac{\exp \left( -\frac{1}{2d}\right) }{2}%
\right) \right) \right) ^{-\frac{\alpha \left( 1-\gamma \right) }{1+\gamma
\left( 1-\alpha \right) }}\int \left( f\left( X_{3}\right) \right) ^{\frac{%
\left( 1-\alpha \right) \left( 1+\gamma \right) }{1+\gamma \left( 1-\alpha
\right) }}\left( f\left( X_{2}\right) \right) ^{-\gamma \frac{\left(
1-\alpha \right) \left( 1+\gamma \right) }{1+\gamma \left( 1-\alpha \right) }%
} \\
&&\times \exp \left( -\left( \frac{\left\vert X-X_{2}\right\vert }{d}+\frac{%
\left\vert X-X_{3}\right\vert }{d}\right) \right) \\
&\simeq &1-\frac{1}{4d^{2}}\left( \frac{1+\gamma \left( 1-\alpha \right) }{%
\left( 2-\alpha \right) }\left( \left( \frac{2-\alpha }{1+\gamma \left(
1-\alpha \right) }\right) ^{2}-1\right) \right) \\
&&\times \int \exp \left( -\frac{\left( 1-\alpha \right) }{1+\gamma \left(
1-\alpha \right) }\frac{\left\vert X_{3}\right\vert }{d}\right) \exp \left(
-\gamma \frac{\left( 1-\alpha \right) }{1+\gamma \left( 1-\alpha \right) }%
\frac{\left\vert X_{2}\right\vert }{d}\right) \exp \left( -\left( \frac{%
\left\vert X-X_{2}\right\vert }{d}+\frac{\left\vert X-X_{3}\right\vert }{d}%
\right) \right) \\
&\simeq &1-\frac{1}{4d^{2}}\left( \frac{1+\gamma \left( 1-\alpha \right) }{%
\left( 2-\alpha \right) }\left( \left( \frac{2-\alpha }{1+\gamma \left(
1-\alpha \right) }\right) ^{2}-1\right) \right) \\
&&\times \int \exp \left( -\frac{\left( 1-\alpha \right) }{1+\gamma \left(
1-\alpha \right) }\frac{\left\vert X_{3}\right\vert }{d}\right) \exp \left(
-\left( \frac{\left\vert X-X_{2}\right\vert }{d}+\frac{\left\vert
X-X_{3}\right\vert }{d}\right) \right) \\
&\simeq &1-\frac{1}{2d^{2}}\frac{1+\gamma \left( 1-\alpha \right) }{\left(
2-\alpha \right) }\left( \left( \frac{2-\alpha }{1+\gamma \left( 1-\alpha
\right) }\right) ^{2}-1\right) \int \exp \left( -\frac{\left( 1-\alpha
\right) }{1+\gamma \left( 1-\alpha \right) }\frac{\left\vert
X_{3}\right\vert }{d}\right) \exp \left( -\frac{\left\vert
X-X_{3}\right\vert }{d}\right) \\
&=&1-\frac{1}{2d}\frac{1+\gamma \left( 1-\alpha \right) }{\left( 2-\alpha
\right) }\left( \left( \frac{2-\alpha }{1+\gamma \left( 1-\alpha \right) }%
\right) ^{2}-1\right) \int \exp \left( -\frac{\left( 1-\alpha \right) }{%
1+\gamma \left( 1-\alpha \right) }\frac{\left\vert X_{3}\right\vert }{d}%
\right) \left( 1-\frac{\cosh \left( \frac{X}{d}\right) }{\exp \left( \frac{1%
}{d}\right) }\right)
\end{eqnarray*}%
since:%
\begin{equation*}
\int \exp \left( -\left( \frac{\left\vert X-X_{2}\right\vert }{d}\right)
\right) =2d\left( 1-\frac{\cosh \left( \frac{X}{d}\right) }{\exp \left( 
\frac{1}{d}\right) }\right)
\end{equation*}%
Then $g$ and $h$\ will be found by computing the following integral (for $%
X>0 $, the general case being obtained by replacing $X$ with $\left\vert
X\right\vert $:

\begin{eqnarray*}
&&\int \exp \left( -\frac{\left( 1-\alpha \right) }{1+\gamma \left( 1-\alpha
\right) }\frac{\left\vert X_{3}\right\vert }{d}\right) \exp \left( -\frac{%
\left\vert X-X_{3}\right\vert }{d}\right) \\
&\simeq &\int \exp \left( -\left( 1-\alpha \right) \frac{\left\vert
X_{3}\right\vert }{d}\right) \exp \left( -\frac{\left\vert
X-X_{3}\right\vert }{d}\right) dX_{3} \\
&&\int_{-1}^{0}\exp \left( \left( 1-\alpha \right) \frac{u}{d}-\frac{X-u}{d}%
\right) du+\int_{X}^{1}\exp \left( -\left( 1-\alpha \right) \frac{u}{d}+%
\frac{X-u}{d}\right) du \\
&&+\int_{0}^{X}\exp \left( -\left( 1-\alpha \right) \frac{u}{d}-\frac{X-u}{d}%
\right) du \\
&=&d\left( \frac{e^{\alpha \frac{X}{d}}-1}{\alpha }e^{-\frac{X}{d}}+\frac{e^{%
\frac{X+\alpha -2}{d}}-e^{\frac{X}{d}\left( \alpha -1\right) }}{\alpha -2}%
-e^{-\frac{X}{d}}\frac{1-e^{\frac{\alpha -2}{d}}}{\alpha -2}\right)
\end{eqnarray*}%
which can be approximated at the zeroth order in $\alpha $:

\begin{eqnarray*}
&\simeq &d\allowbreak \left( \frac{1}{2}e^{-\frac{X}{d}}-\frac{1}{2}e^{\frac{%
X-2}{d}}-e^{-\frac{X}{d}}\left( \frac{1}{2}e^{-\frac{2}{d}}-\frac{1}{2}%
\right) +\allowbreak Xe^{-\frac{X}{d}}\right) \\
&\simeq &d\left( \frac{X}{d}+1-\frac{1}{2}e^{\frac{2X-2}{d}}\right) e^{-%
\frac{X}{d}}
\end{eqnarray*}

The correction terms of order $\alpha $ are negligible in the neighborhood
of $d=1$. In average, the previous formula is equal to $d\left( \frac{3}{2d}-%
\frac{1}{2}e^{-1}\right) e^{-\frac{X}{d}}$. Restoring the absolute value,
one finds for $d$ close to $1$:%
\begin{equation*}
h\simeq \frac{1+\gamma \left( 1-\alpha \right) }{2\left( 1-\alpha \right) }%
\left( \frac{1+\gamma \left( 1-\alpha \right) }{\left( 2-\alpha \right) }%
\left( \left( \frac{2-\alpha }{1+\gamma \left( 1-\alpha \right) }\right)
^{2}-1\right) \right) \left( \frac{3}{2}-\frac{e^{-\frac{1}{d}}}{2}\right)
\end{equation*}%
As a consequence of the previous computations, the potential becomes:

\begin{equation*}
\delta ^{2}\left( K-\left\langle K\right\rangle _{X}\right) ^{2}
\end{equation*}%
which justifies the assumption of harmonic oscillations.

The price can be rewritten as a function of $\left\langle K\right\rangle
_{X} $ using (\ref{Kpos}):%
\begin{eqnarray*}
P &=&H\left( K,X\right) =\left( K\right) ^{-\frac{\alpha }{1+\gamma }%
}f\left( X\right) \\
&&\left( \frac{K}{\left\langle K\right\rangle _{X}}\right) ^{-\frac{\alpha }{%
1+\gamma }}\frac{f\left( X\right) }{\left( \frac{A}{\delta }f\left( X\right)
\left( 1-h\exp \left( -\frac{\left\vert X\right\vert }{d}\right) \left( 1-%
\frac{\cosh \frac{X}{d}}{\exp \left( \frac{1}{d}\right) }\right) \right)
\right) ^{\frac{\alpha }{1+\gamma \left( 1-\alpha \right) }}} \\
&=&\left( \frac{K}{\left\langle K\right\rangle _{X}}\right) ^{-\frac{\alpha 
}{1+\gamma }}\frac{\left( f\left( X\right) \right) ^{\frac{\left( 1+\gamma
\right) \left( 1-\alpha \right) }{1+\gamma \left( 1-\alpha \right) }}}{%
\left( \frac{A}{\delta }\left( 1-h\exp \left( -\frac{\left\vert X\right\vert 
}{d}\right) \left( 1-\frac{\cosh \frac{X}{d}}{\exp \left( \frac{1}{d}\right) 
}\right) \right) \right) ^{\frac{\alpha }{1+\gamma \left( 1-\alpha \right) }}%
}
\end{eqnarray*}%
and thus, (\ref{Kpos}): 
\begin{equation}
P=\frac{\left( \frac{1+\gamma \left( 1-\alpha \right) }{\left( 2-\alpha
\right) \kappa }\left( \left( \frac{2-\alpha }{1+\gamma \left( 1-\alpha
\right) }\right) ^{2}-1\right) \right) ^{\frac{1}{1-\gamma }}\exp \left( -%
\frac{\left( 1-\alpha \right) \left\vert X\right\vert }{d\left( 1+\gamma
\left( 1-\alpha \right) \right) }\right) }{\left( \frac{K}{\left\langle
K\right\rangle _{X}}\right) ^{\frac{\alpha }{1+\gamma }}\left( \left( \frac{%
1-h\exp \left( -\frac{\left\vert X\right\vert }{d}\right) \left( 1-\frac{%
\cosh \frac{X}{d}}{\exp \left( \frac{1}{d}\right) }\right) }{1-h\exp \left( -%
\frac{1}{2d}\right) \left( 1-\frac{\exp \left( -\frac{1}{2d}\right) }{2}%
\right) }\right) \right) ^{\frac{\alpha }{1+\gamma \left( 1-\alpha \right) }}%
}  \label{Prn}
\end{equation}%
Having found the price for $\bar{A}\rightarrow \infty $, we can now consider
the corrections due to $A$:

$\bar{A}>>A$: \ once again we consider fields of the form: $\Psi \left(
K,P,X,\theta \right) \rightarrow \delta \left( P-F\left( K,X,\theta \right)
\right) \Psi \left( K,X,\theta \right) $.

\bigskip The potential considered is now:%
\begin{eqnarray*}
&&\left( \delta K-APK^{\alpha }\left( 1-\kappa \int \frac{P_{3}\exp \left(
-\left( \left\vert X_{1}-X_{2}\right\vert +\left\vert X_{1}-X_{3}\right\vert
\right) \right) }{P_{2}^{\gamma }}\left\vert \Psi \left( K_{2},X_{2},\theta
\right) \right\vert ^{2}\left\vert \Psi \left( K_{3},X_{3},\theta \right)
\right\vert ^{2}\right) \right) ^{2} \\
&&+\bar{A}^{2}\left( P^{1+\gamma }K^{\alpha }-\kappa \int P_{2}\left(
K_{2}\right) ^{\alpha }P_{3}\exp \left( -\left( \left\vert
X_{1}-X_{2}\right\vert +\left\vert X_{2}-X_{3}\right\vert \right) \right)
\left\vert \Psi \left( K_{2},X_{2},\theta \right) \right\vert ^{2}\left\vert
\Psi \left( K_{3},X_{3},\theta \right) \right\vert ^{2}\right) ^{2} \\
&=&\left( \delta K-APK^{\alpha }U\right) ^{2}+\bar{A}^{2}\left( P^{1+\gamma
}K^{\alpha }-V\right) ^{2}
\end{eqnarray*}%
with:%
\begin{eqnarray*}
U &=&\left( 1-\kappa \int \frac{P_{3}\exp \left( -\left( \left\vert
X_{1}-X_{2}\right\vert +\left\vert X_{1}-X_{3}\right\vert \right) \right) }{%
P_{2}^{\gamma }}\left\vert \Psi \left( K_{2},X_{2},\theta \right)
\right\vert ^{2}\left\vert \Psi \left( K_{3},X_{3},\theta \right)
\right\vert ^{2}\right) \\
V &=&\kappa \int P_{2}\left( K_{2}\right) ^{\alpha }P_{3}\exp \left( -\left(
\left\vert X_{1}-X_{2}\right\vert +\left\vert X_{2}-X_{3}\right\vert \right)
\right) \left\vert \Psi \left( K_{2},X_{2},\theta \right) \right\vert
^{2}\left\vert \Psi \left( K_{3},X_{3},\theta \right) \right\vert ^{2}
\end{eqnarray*}%
The price is found through the minimization of the sum:

\begin{equation*}
\left( \delta K-APK^{\alpha }U\right) ^{2}+\bar{A}^{2}\left( P^{1+\gamma
}K^{\alpha }-V\right) ^{2}
\end{equation*}%
That yields the condition:%
\begin{equation}
-UAK^{\alpha }\left( \delta K-APK^{\alpha }U\right) +\left( 1+\gamma \right)
P^{\gamma }\bar{A}^{2}K^{\alpha }\left( P^{1+\gamma }K^{\alpha }-V\right) =0
\label{fstr}
\end{equation}%
A first order approximation in $\gamma $ of (\ref{fstr})\ is: 
\begin{equation}
-AK^{\alpha }\left( \delta K-APK^{\alpha }\right) +\left( 1+\gamma \right)
P^{\gamma }\bar{A}^{2}K^{\alpha }\left( P^{1+\gamma }K^{\alpha }-V\right)
\simeq 0  \label{fstrpr}
\end{equation}%
We we then use (\ref{fstrpr}) at first order in $\gamma $and in $\frac{A^{2}%
}{\bar{A}^{2}}$ to compute $P^{1+\gamma }K^{\alpha }$: 
\begin{eqnarray*}
P^{1+\gamma }K^{\alpha } &=&V+\frac{A\left( \delta K-APK^{\alpha }\right) }{%
\bar{A}^{2}\left( 1+\gamma \right) P^{\gamma }}\simeq V+\frac{A\left( \delta
K-APK^{\alpha }\right) }{\bar{A}^{2}} \\
&\simeq &V+\frac{AK^{\alpha }\left( \delta K-AV\right) }{\bar{A}%
^{2}K^{\alpha }}
\end{eqnarray*}

Now, we compute $P^{1+\gamma }K^{\alpha }$ at first order in $\frac{A^{2}}{%
\bar{A}^{2}}$:

\begin{eqnarray}
P^{1+\gamma }K^{\alpha }-A\frac{\delta K-AV}{A^{2}+\bar{A}^{2}} &=&V
\label{Prcr} \\
P^{1+\gamma }K^{\alpha }-\frac{\delta AK}{A^{2}+\bar{A}^{2}} &=&\frac{\bar{A}%
^{2}V}{A^{2}+\bar{A}^{2}}  \notag \\
P^{1+\gamma }K^{\alpha } &=&V\left( 1+A\frac{\delta K-AV}{V\left( A^{2}+\bar{%
A}^{2}\right) }\right)  \notag
\end{eqnarray}

For this value, the potentials become:%
\begin{eqnarray}
&&\left( \delta K-APK^{\alpha }U\right) ^{2}+\bar{A}^{2}\left( P^{1+\gamma
}K^{\alpha }-V\right) ^{2}  \label{Ptcr} \\
&\simeq &\left( \delta K-APK^{\alpha }U\right) ^{2}+\bar{A}^{2}A^{2}\left( 
\frac{\delta K-AUV}{\left( A^{2}U^{2}+\bar{A}^{2}\right) }\right) ^{2} 
\notag \\
&=&\delta ^{2}\left( 1+\frac{\bar{A}^{2}A^{2}}{\left( A^{2}U^{2}+\bar{A}%
^{2}\right) ^{2}}\right) \left( K-\frac{APK^{\alpha }U+\frac{\bar{A}^{2}A^{2}%
}{\left( A^{2}U^{2}+\bar{A}^{2}\right) ^{2}}AUV}{\delta \left( 1+\frac{\bar{A%
}^{2}A^{2}}{\left( A^{2}U^{2}+\bar{A}^{2}\right) ^{2}}\right) }\right) ^{2} 
\notag \\
&&+\frac{\bar{A}^{2}A^{2}\left( APK^{\alpha }U-AUV\right) ^{2}}{\bar{A}%
^{2}A^{2}+\left( A^{2}U^{2}+\bar{A}^{2}\right) ^{2}}  \notag
\end{eqnarray}

Equations (\ref{Prcr}) and (\ref{Ptcr}) can be simplified. Since we are
looking for first order corrections in $\frac{A^{2}}{\bar{A}^{2}}$, in (\ref%
{Prcr}) we can replace we can replace \ $\delta K-AV$ by its lowest order
approximation, that is, given that we have assumed $\gamma <<1$:%
\begin{equation*}
\delta K-AV=\delta K-AP^{1+\gamma }K^{\alpha }\simeq \delta K-APK^{\alpha }
\end{equation*}%
To this order of approximation, one can also replace $K$ by $\left\langle
K\right\rangle _{X}$ and using (\ref{qtnKX}): 
\begin{equation*}
\delta \left\langle K\right\rangle _{X}=AP\left\langle K\right\rangle
_{X}^{\alpha }U
\end{equation*}
and (\ref{Kpos}):%
\begin{eqnarray*}
\frac{\delta K-AV}{V} &\simeq &A\frac{\delta \left\langle K\right\rangle
_{X}-AP\left\langle K\right\rangle _{X}^{\alpha }}{AP\left\langle
K\right\rangle _{X}^{\alpha }} \\
&\simeq &A\frac{\delta \left\langle K\right\rangle _{X}-AP\left\langle
K\right\rangle _{X}^{\alpha }}{AP\left\langle K\right\rangle _{X}^{\alpha }}
\\
&=&A\left( U-1\right) \simeq A\left( \left\langle U\right\rangle -1\right)
\simeq -hA\left( 1-\exp \left( -\frac{1}{d}\right) \right)
\end{eqnarray*}%
and (\ref{Prcr}) becomes:%
\begin{equation*}
P^{1+\gamma }K^{\alpha }=V\left( 1+\frac{A^{2}}{\bar{A}^{2}}\left(
\left\langle U\right\rangle -1\right) \right) \simeq \left( 1-h\frac{A^{2}}{%
\bar{A}^{2}}\left( 1-\exp \left( -\frac{1}{d}\right) \right) \right) V
\end{equation*}

Similarly, in (\ref{Ptcr}) we can replace \ $APK^{\alpha }U-AUV$ by its
lowest order approximation, that is

\begin{equation*}
APK^{\alpha }U-AUV=AU\left( PK^{\alpha }-P^{1+\gamma }K^{\alpha }\right)
\end{equation*}%
The correction associated to this term is of order $\gamma \frac{A^{2}}{\bar{%
A}^{2}}$ and can be neglected. Moreover, by the same argument, up to
corrections of order $\gamma \frac{A^{2}}{\bar{A}^{2}}$, one has :%
\begin{equation*}
\frac{APK^{\alpha }U+\frac{\bar{A}^{2}A^{2}}{\left( A^{2}U^{2}+\bar{A}%
^{2}\right) ^{2}}AUV}{\delta \left( 1+\frac{\bar{A}^{2}A^{2}}{\left(
A^{2}U^{2}+\bar{A}^{2}\right) ^{2}}\right) }\simeq AU\frac{PK^{\alpha }+%
\frac{\bar{A}^{2}A^{2}}{\left( A^{2}U^{2}+\bar{A}^{2}\right) ^{2}}PK^{\alpha
}}{\delta \left( 1+\frac{\bar{A}^{2}A^{2}}{\left( A^{2}U^{2}+\bar{A}%
^{2}\right) ^{2}}\right) }=AUPK^{\alpha }
\end{equation*}%
and the potential is:%
\begin{equation*}
\delta ^{2}\left( 1+\frac{\bar{A}^{2}A^{2}}{\left( A^{2}U^{2}+\bar{A}%
^{2}\right) ^{2}}\right) \left( K-\frac{AUPK^{\alpha }}{\delta }\right) ^{2}
\end{equation*}%
In these formula, $U$ can be estimated by $1-h\exp \left( -\frac{1}{2d}%
\right) \left( 1-\frac{\exp \left( -\frac{1}{2d}\right) }{2}\right) $. As a
consequence, the equation for $P$ is the same as before with an additional
factor $U$, and the equation for $\left\langle K\right\rangle _{X}$ is
unchanged. We can then use a similar trial function: 
\begin{equation}
P\simeq \left( K\right) ^{-\frac{\alpha }{1+\gamma }}f\left( X\right)
\label{Pcrt}
\end{equation}%
and these equations become:%
\begin{equation*}
\left( f\left( X\right) \right) ^{1+\gamma }=\bar{\kappa}\int P_{2}\left(
K_{2}\right) ^{\alpha }P_{3}\exp \left( -\left( \left\vert
X-X_{2}\right\vert +\left\vert X_{2}-X_{3}\right\vert \right) \right)
\left\vert \Psi \left( K_{3},P_{3},X_{3},\theta \right) \right\vert
^{2}\left\vert \Psi \left( K_{2},P_{2},X_{2},\theta \right) \right\vert ^{2}
\end{equation*}%
with:%
\begin{equation}
\bar{\kappa}=\kappa \left( 1-h\frac{A^{2}}{\bar{A}^{2}}\left( 1-\exp \left( -%
\frac{1}{d}\right) \right) \right)  \label{kbr}
\end{equation}%
\begin{equation*}
\left\langle K\right\rangle _{X}\simeq \left( \frac{A}{\delta }f\left(
X\right) \right) ^{\frac{1+\gamma }{1+\gamma \left( 1-\alpha \right) }%
}\left( 1-\frac{\kappa }{4d^{2}}\int \left\langle K\right\rangle _{X_{2}}^{%
\frac{\alpha \gamma }{1+\gamma }}\left\langle K\right\rangle _{X_{3}}^{-%
\frac{\alpha }{1+\gamma }}f\left( X_{3}\right) \left( f\left( X_{2}\right)
\right) ^{-\gamma }\exp \left( -\left( \frac{\left\vert X-X_{2}\right\vert }{%
d}+\frac{\left\vert X-X_{3}\right\vert }{d}\right) \right) \right)
\end{equation*}%
And the resolution is identical to the case $A\rightarrow 0$. We thus find:%
\begin{equation*}
f\left( X\right) =D\exp \left( -\frac{\left\vert X\right\vert }{\left(
1+\gamma \right) d}\right)
\end{equation*}%
with:%
\begin{equation*}
D=\left( \frac{1+\gamma \left( 1-\alpha \right) }{\left( 2-\alpha \right) 
\bar{\kappa}}\left( \left( \frac{2-\alpha }{1+\gamma \left( 1-\alpha \right) 
}\right) ^{2}-1\right) \left( \frac{A}{\delta }\left( 1-h\exp \left( -\frac{1%
}{2d}\right) \left( 1-\frac{\exp \left( -\frac{1}{2d}\right) }{2}\right)
\right) \right) ^{\frac{\alpha \left( 1-\gamma \right) }{1+\gamma \left(
1-\alpha \right) }}\right) ^{\frac{1+\gamma \left( 1-\alpha \right) }{\left(
1-\gamma ^{2}\right) \left( 1-\alpha \right) }}
\end{equation*}%
and for $\left\langle K\right\rangle _{X}$:%
\begin{equation*}
\left\langle K\right\rangle _{X}=\left( \frac{A}{\delta }f\left( X\right)
\left( 1-h\exp \left( -\frac{\left\vert X\right\vert }{d}\right) \left( 1-%
\frac{\cosh \frac{X}{d}}{\exp \left( \frac{1}{d}\right) }\right) \right)
\right) ^{\frac{1+\gamma }{1+\gamma \left( 1-\alpha \right) }}
\end{equation*}%
with:%
\begin{eqnarray*}
h &=&\frac{\bar{h}}{\left( 1-\frac{A^{2}}{\bar{A}^{2}}\bar{h}\left( 1-\exp
\left( -\frac{1}{d}\right) \right) \right) } \\
\bar{h} &=&\left( 1+\gamma \left( 1-\alpha \right) \right) \left( \frac{%
1+\gamma \left( 1-\alpha \right) }{2\left( 2-\alpha \right) }\left( \left( 
\frac{2-\alpha }{1+\gamma \left( 1-\alpha \right) }\right) ^{2}-1\right)
\right) \left( \frac{3}{2}-\frac{e^{-\frac{1}{d}}}{2}\right)
\end{eqnarray*}%
to the first order in $\frac{A^{2}}{\bar{A}^{2}}$. Formula (\ref{Prn}) for
the price becomes:%
\begin{equation*}
P=\frac{\left( \frac{1+\gamma \left( 1-\alpha \right) }{\left( 2-\alpha
\right) \bar{\kappa}}\left( \left( \frac{2-\alpha }{1+\gamma \left( 1-\alpha
\right) }\right) ^{2}-1\right) \right) ^{\frac{1}{1-\gamma }}\exp \left( -%
\frac{\left( 1-\alpha \right) \left\vert X\right\vert }{d\left( 1+\gamma
\left( 1-\alpha \right) \right) }\right) }{\left( \frac{K}{\left\langle
K\right\rangle _{X}}\right) ^{\frac{\alpha }{1+\gamma }}\left( \left( \frac{%
1-h\exp \left( -\frac{\left\vert X\right\vert }{d}\right) \left( 1-\frac{%
\cosh \frac{X}{d}}{\exp \left( \frac{1}{d}\right) }\right) }{1-h\exp \left( -%
\frac{1}{2d}\right) \left( 1-\frac{\exp \left( -\frac{1}{2d}\right) }{2}%
\right) }\right) \right) ^{\frac{\alpha }{1+\gamma \left( 1-\alpha \right) }}%
}
\end{equation*}

\subsection{\textbf{Appendix 3}}

For $\rho >0$, the contribution of the mean cnfgrtn added to the
fluctuations. Previous computations are valid, but average values have to be
computed in states including the fundamental's contribution in the state $%
\Psi _{0}+\Psi $. We first consider the state $\Psi _{0}\left( K,P,X,\theta
\right) $ only, and in this state, we replace $P$, find the equations for $%
f\left( X\right) $ and $\left\langle K\right\rangle _{X}$. Then, we compute
the form of fundamental $\Psi _{0}$ and the value of $\rho $. We then
compute the correction due to $\Psi \left( K,P,X,\theta \right) $ and find
the effective action.

\subsubsection{\protect\bigskip Equations for $P$, $f\left( X\right) $ and $%
\left\langle K\right\rangle _{X}$ in the state $\Psi _{0}$}

In the state $\Psi _{0}\left( K,P,X,\theta \right) $, the previous equations
defining the potential, $P$ and $\left\langle K\right\rangle _{X}$ are the
same as before. We will include directly the first order corrections in $%
\frac{A^{2}}{\bar{A}^{2}}$ since it amounts, as in Appendix $2$, to replace $%
\kappa $ by $\bar{\kappa}$ (see \ref{kbr}) in the equation for $f\left(
X\right) $.

Now, to compute $f\left( X\right) $ we first find the average of $K$,
denoted $\left\langle K\right\rangle _{X}$ computed in a state $\rho \Psi
_{0}\left( K,X\right) $, which amounts to replace now, given our order of
approximations, for $\Psi _{0}\left( K,X\right) $ 
\begin{equation*}
\kappa \rightarrow \kappa \rho ^{4}
\end{equation*}%
One sets again:%
\begin{equation*}
P=\left( K\right) ^{-\frac{\alpha }{1+\gamma }}f\left( X\right)
\end{equation*}%
The equation for $f\left( X\right) $ is the same as before 
\begin{equation*}
f^{1+\gamma }\left( X\right) =\bar{\kappa}\rho ^{4}\int P_{2}\left(
K_{2}\right) ^{\alpha }P_{3}\exp \left( -\left( \left\vert
X-X_{2}\right\vert +\left\vert X_{2}-X_{3}\right\vert \right) \right)
\left\vert \Psi _{0}\left( K_{2},P_{2},X_{2},\theta \right) \right\vert
^{2}\left\vert \Psi _{0}\left( K_{3},P_{3},X_{3},\theta \right) \right\vert
^{2}
\end{equation*}%
where: 
\begin{equation*}
\bar{\kappa}=\kappa \left( 1-h\frac{A^{2}}{\bar{A}^{2}}\left( 1-\exp \left( -%
\frac{1}{d}\right) \right) \right)
\end{equation*}%
includes the first order corrections in $\frac{A^{2}}{\bar{A}^{2}}$. The
equation for $\left\langle K\right\rangle _{X}$ is still valid in first
approximation:%
\begin{equation*}
\left\langle K\right\rangle _{X}=\left( \frac{A}{\delta }f\left( X\right)
\right) ^{\frac{1+\gamma }{1+\gamma \left( 1-\alpha \right) }}\left( 
\begin{array}{c}
1-\kappa \rho ^{4}\int \left( K_{2}\right) ^{\frac{\alpha \gamma }{1+\gamma }%
}\left( K_{3}\right) ^{-\frac{\alpha }{1+\gamma }}f\left( X_{3}\right)
\left( f\left( X_{2}\right) \right) ^{-\gamma }\times \\ 
\times \exp \left( -\left( \frac{\left\vert X-X_{2}\right\vert }{d}+\frac{%
\left\vert X-X_{3}\right\vert }{d}\right) \right) \left\vert \Psi _{0}\left(
K_{2},P_{2},X_{2},\theta \right) \right\vert ^{2}\left\vert \Psi _{0}\left(
K_{3},P_{3},X_{3},\theta \right) \right\vert ^{2}%
\end{array}%
\right) ^{\frac{1+\gamma }{1+\gamma \left( 1-\alpha \right) }}
\end{equation*}%
For the same reasons as in Appendix 2, the state $\Psi _{0}^{\dag }\left(
K,X\right) $ is centered around $\left\langle K\right\rangle _{X}$. Thus,
the equation for $f\left( X\right) $ is thus: 
\begin{equation*}
\left( f\left( X\right) \right) ^{1+\gamma }\simeq \bar{\kappa}\rho ^{4}\int
\left\langle K\right\rangle _{X_{2}}^{\frac{\alpha \left( \gamma -1\right) }{%
1+\gamma }}\left( f\left( X_{2}\right) \right) ^{2}G\left(
X_{2},X_{2}\right) \exp \left( -\left( \frac{\left\vert X-X_{2}\right\vert }{%
d}\right) \right)
\end{equation*}

Moreover, in first approximation, the defining equation for the $X$\ part of 
$\Psi _{0}\left( K,X\right) $ is an oscillator with $\kappa _{0}<<1$. The
distribution for $X$ can thus be considered as uniform. The equations for $%
f\left( X\right) $ and for $K$ are thus similar to that obtained in Appendix
2, apart from a $\rho ^{4}$ factor and write:%
\begin{equation*}
\left( f\left( X\right) \right) ^{1+\gamma }\simeq \frac{\bar{\kappa}\rho
^{4}}{2}\int \left\langle K\right\rangle _{X_{2}}^{\frac{\alpha \left(
\gamma -1\right) }{1+\gamma }}\left( f\left( X_{2}\right) \right) ^{2}\exp
\left( -\frac{\left\vert X-X_{2}\right\vert }{d}\right)
\end{equation*}%
\begin{equation*}
\left\langle K\right\rangle _{X}\simeq \left( \frac{A}{\delta }f\left(
X\right) \right) ^{\frac{1+\gamma }{1+\gamma \left( 1-\alpha \right) }%
}\left( 1-\frac{\kappa \rho ^{4}}{4}\int \left\langle K\right\rangle
_{X_{2}}^{\frac{\alpha \gamma }{1+\gamma }}\left\langle K\right\rangle
_{X_{3}}^{-\frac{\alpha }{1+\gamma }}f\left( X_{3}\right) \left( f\left(
X_{2}\right) \right) ^{-\gamma }\exp \left( -\left( \frac{\left\vert
X-X_{2}\right\vert }{d}+\frac{\left\vert X-X_{3}\right\vert }{d}\right)
\right) \right)
\end{equation*}%
Again, the form for $\left\langle K\right\rangle _{X}$ is postulated: 
\begin{equation*}
\left\langle K\right\rangle _{X}=\left( \frac{A}{\delta }f\left( X\right)
\left( 1-h\exp \left( -g\frac{\left\vert X\right\vert }{d}\right) \left( 1-%
\frac{\cosh \frac{X}{d}}{\exp \left( \frac{1}{d}\right) }\right) \right)
\right) ^{\frac{1+\gamma }{1+\gamma \left( 1-\alpha \right) }}
\end{equation*}%
the resolution is thus similar to Appendix 2 and one finds:

\begin{eqnarray*}
f\left( X\right) &=&D\exp \left( -\frac{\left\vert X\right\vert }{1+\gamma }%
\right) \\
&=&\left( \frac{A}{\delta }\left( 1-h\exp \left( -\frac{1}{2d}\right) \left(
1-\frac{\exp \left( -\frac{1}{2d}\right) }{2}\right) \right) \right) ^{\frac{%
\alpha }{\left( 1-\alpha \right) \left( \gamma +1\right) }} \\
&&\times \left( \frac{1+\gamma \left( 1-\alpha \right) }{\left( 2-\alpha
\right) \bar{\kappa}\rho ^{4}}\left( \left( \frac{2-\alpha }{1+\gamma \left(
1-\alpha \right) }\right) ^{2}-1\right) \right) ^{\frac{1+\gamma \left(
1-\alpha \right) }{\left( 1-\gamma ^{2}\right) \left( 1-\alpha \right) }%
}\exp \left( -\frac{\left\vert X\right\vert }{d\left( 1+\gamma \right) }%
\right)
\end{eqnarray*}

\bigskip where:%
\begin{eqnarray*}
h &=&\frac{\bar{h}}{\left( 1-\frac{A^{2}}{\bar{A}^{2}}\bar{h}\left( 1-\exp
\left( -\frac{1}{d}\right) \right) \right) } \\
\bar{h} &=&\left( 1+\gamma \left( 1-\alpha \right) \right) \left( \frac{%
1+\gamma \left( 1-\alpha \right) }{2\left( 2-\alpha \right) }\left( \left( 
\frac{2-\alpha }{1+\gamma \left( 1-\alpha \right) }\right) ^{2}-1\right)
\right) \left( \frac{3}{2}-\frac{e^{-\frac{1}{d}}}{2}\right)
\end{eqnarray*}%
And for the price:%
\begin{equation*}
P=\frac{\left( \frac{1+\gamma \left( 1-\alpha \right) }{\left( 2-\alpha
\right) \bar{\kappa}\rho ^{4}}\left( \left( \frac{2-\alpha }{1+\gamma \left(
1-\alpha \right) }\right) ^{2}-1\right) \right) ^{\frac{1}{1-\gamma }}\exp
\left( -\frac{\left( 1-\alpha \right) \left\vert X\right\vert }{d\left(
1+\gamma \left( 1-\alpha \right) \right) }\right) }{\left( \frac{K}{%
\left\langle K\right\rangle _{X}}\right) ^{\frac{\alpha }{1+\gamma }}\left(
\left( \frac{1-h\exp \left( -\frac{\left\vert X\right\vert }{d}\right)
\left( 1-\frac{\cosh \frac{X}{d}}{\exp \left( \frac{1}{d}\right) }\right) }{%
1-h\exp \left( -\frac{1}{2d}\right) \left( 1-\frac{\exp \left( -\frac{1}{2d}%
\right) }{2}\right) }\right) \right) ^{\frac{\alpha }{1+\gamma \left(
1-\alpha \right) }}}
\end{equation*}

\subsubsection{Computation of $\protect\rho $ and $\Psi _{0}$}

\bigskip The value of $\rho $ can now be computed. As said before, this
phase is possible approximatively for: 
\begin{equation*}
\kappa _{1}^{2}-2\kappa _{0}^{\frac{1}{2}}\kappa _{2}>0
\end{equation*}%
In this phase, taking into account only the contribution in $X$, then $\rho $
can be found approximatively by: 
\begin{equation*}
\rho \simeq \frac{\kappa _{1}+\sqrt{\kappa _{1}^{2}-2\kappa _{0}^{\frac{1}{2}%
}\kappa _{2}}}{2\kappa _{2}}
\end{equation*}%
with $\frac{\partial ^{2}S}{\partial \rho ^{2}}>0$. The value of $\rho $ is
found more precisely by writing the equation for the state $\Psi _{0}\left(
K,X\right) $. For the part in $X$:

\begin{eqnarray*}
&&\sigma _{X}^{2}\left( \nabla \Psi \left( x\right) \right) ^{2}+\frac{%
\kappa _{0}}{\sigma _{X}^{2}}\Psi ^{\dag }\left( x\right) \left(
x-\left\langle x\right\rangle \right) ^{2}\Psi \left( x\right) +\left( \Psi
^{\dag }\left( x\right) \Psi \left( x\right) \right) V_{1}\left( \left\vert
x-y\right\vert \right) \left( \Psi ^{\dag }\left( y\right) \Psi \left(
y\right) \right) \\
&&+\left( \Psi ^{\dag }\left( x\right) \Psi \left( x\right) \right)
V_{2}\left( \left\vert x-y\right\vert ,\left\vert x-z\right\vert ,\left\vert
y-z\right\vert \right) \left( \Psi ^{\dag }\left( y\right) \Psi \left(
y\right) \right) \left( \Psi ^{\dag }\left( z\right) \Psi \left( z\right)
\right)
\end{eqnarray*}%
where:%
\begin{eqnarray*}
V_{1}\left( \left\vert x-y\right\vert \right) &=&-\kappa _{1}\frac{%
KK^{\prime }\exp \left( -\chi _{1}\left\vert x-y\right\vert \right) }{%
\left\langle K\right\rangle _{x}\left\langle K\right\rangle _{y}} \\
V_{2}\left( \left\vert x-y\right\vert ,\left\vert x-z\right\vert ,\left\vert
y-z\right\vert \right) &=&\kappa _{2}\exp \left( -\chi _{2}\left\vert
x-y\right\vert -\chi _{2}\left\vert x-z\right\vert -\chi _{2}\left\vert
y-z\right\vert \right)
\end{eqnarray*}%
Considering the $X$ part of $\frac{\delta }{\delta \Psi \left( x\right) }$: 
\begin{equation*}
-\sigma _{X}^{2}\nabla ^{2}\Psi \left( x\right) +\frac{\kappa _{0}}{\sigma
_{X}^{2}}\left( x-\left\langle x\right\rangle \right) ^{2}\Psi \left(
x\right) +V_{1}\left( \left\vert x-y\right\vert \right) \left\vert \Psi
\left( y\right) \right\vert ^{2}\Psi \left( x\right) +V_{2}\left( \left\vert
x-y\right\vert ,\left\vert x-z\right\vert ,\left\vert y-z\right\vert \right)
\left\vert \Psi \left( y\right) \right\vert ^{2}\left\vert \Psi \left(
z\right) \right\vert ^{2}\Psi \left( x\right)
\end{equation*}

\bigskip It can be approximated by computing the potentials $V_{1}\left(
\left\vert x-y\right\vert \right) $ and $V_{2}\left( \left\vert
x-y\right\vert ,\left\vert x-z\right\vert ,\left\vert y-z\right\vert \right) 
$ in the fundamental:%
\begin{eqnarray*}
&&\int \kappa _{1}\frac{KK^{\prime }\exp \left( -\chi _{1}\left\vert
x-y\right\vert \right) }{\left\langle K\right\rangle _{x}\left\langle
K\right\rangle _{y}}\left\vert \Psi \left( K^{\prime },y\right) \right\vert
^{2}dy \\
&\rightarrow &\int \kappa _{1}\frac{KK^{\prime }\exp \left( -\chi
_{1}\left\vert x-y\right\vert \right) }{\left\langle K\right\rangle
_{x}\left\langle K\right\rangle _{y}}\left\vert \Psi _{0}\left( K^{\prime
},y\right) \right\vert ^{2}dy \\
&\simeq &\kappa _{1}\exp \left( -\chi _{1}\left\vert x\right\vert \right) 
\frac{K}{\left\langle K\right\rangle _{x}}\rho ^{2} \\
&\simeq &\kappa _{1}\exp \left( -\chi _{1}\left\vert x\right\vert \right)
\rho ^{2}
\end{eqnarray*}

Similarly, the second part of the potential is replaced by: 
\begin{eqnarray*}
&&\kappa _{2}\exp \left( -\chi _{2}\left\vert x-y\right\vert -\chi
_{2}\left\vert x-z\right\vert -\chi _{2}\left\vert y-z\right\vert \right)
\left\vert \Psi \left( K^{\prime },y\right) \right\vert ^{2}\left\vert \Psi
\left( K",z\right) \right\vert ^{2} \\
&\rightarrow &\int \kappa _{2}\exp \left( -\chi _{2}\left\vert
x-y\right\vert -\chi _{2}\left\vert x-z\right\vert -\chi _{2}\left\vert
y-z\right\vert \right) \left\vert \Psi _{0}\left( K,y\right) \right\vert
^{2}\left\vert \Psi _{0}\left( K,z\right) \right\vert ^{2}dydz \\
&\simeq &\kappa _{2}\exp \left( -2\chi _{2}\left\vert x\right\vert \right)
\rho ^{4}
\end{eqnarray*}%
Moreover in first approximation, we can find the fundamental by taking the
expectation values of the variable $x$:%
\begin{equation*}
\left\langle x\right\rangle =0
\end{equation*}%
for an overall expression:

\begin{equation*}
-\sigma _{X}^{2}\nabla ^{2}\Psi \left( x\right) +\kappa _{0}x^{2}\Psi \left(
x\right) -\kappa _{1}\frac{K}{\left\langle K\right\rangle _{x}}\exp \left(
-\chi _{1}\left\vert x\right\vert \right) \Psi \left( K,x\right) \rho
^{2}+\kappa _{2}\exp \left( -2\chi _{2}\left\vert x\right\vert \right) \Psi
\left( K,x\right) \rho ^{4}+\alpha \Psi \left( x\right)
\end{equation*}%
Adding the $K$ a part of the action, and considering $\chi _{1}<<1$, $\chi
_{2}<<1$, at the lowest order in $\chi _{1}$ and $\chi _{2}$, the
fundamental equation is:%
\begin{eqnarray*}
&&0=\left( -\sigma _{X}^{2}\nabla _{X}^{2}-\sigma ^{2}\nabla
_{K}^{2}-\vartheta ^{2}\nabla _{\theta }^{2}\right) \Psi _{0}\left(
K,X,\theta \right) \\
&&+\left( \left( \delta ^{2}+\frac{\bar{A}^{2}A^{2}}{\left( A^{2}U^{2}+\bar{A%
}^{2}\right) ^{2}}\right) \left( K-\left\langle K\right\rangle _{X}\right)
^{2}+\kappa _{0}x^{2}+\left( \rho ^{4}\kappa _{2}-\rho ^{2}\kappa _{1}+\frac{%
1}{\vartheta ^{2}}+\alpha \right) \right) \Psi _{0}\left( K,X,\theta \right)
\end{eqnarray*}%
The fourier transform in $\theta $ shows that the fundamental does not
depend on $\theta $. Eigenstates of the operator leads to look for a
fundamental of the form $N\Psi _{0}\left( X\right) \Psi _{0}\left( K\right) $%
. We assume $\theta <\Theta $ with $\Theta >>1$, so that the integral over $%
\theta $ exists.

The normalization factor $N$ ensures that $N\Psi _{0}\left( X\right) \Psi
_{0}\left( K\right) $ has norm $\rho ^{2}$. Then $\Psi _{0}\left( X\right) $
and $\Psi _{0}\left( K\right) $ can be searched as fundamental states of
oscillators with eigenstates 
\begin{equation*}
\kappa _{0}^{\frac{1}{2}}\text{ and }\sigma \sqrt{\delta ^{2}+\frac{\bar{A}%
^{2}A^{2}}{\left( A^{2}U^{2}+\bar{A}^{2}\right) ^{2}}}
\end{equation*}%
and this leads to the relation:%
\begin{equation}
0=\alpha +\frac{1}{\vartheta ^{2}}+\frac{1}{2}\kappa _{0}^{\frac{1}{2}%
}-\kappa _{1}\rho ^{2}+\kappa _{2}\rho ^{4}+\frac{\sigma \sqrt{\delta ^{2}+%
\frac{\bar{A}^{2}A^{2}}{\left( A^{2}U^{2}+\bar{A}^{2}\right) ^{2}}}}{2} 
\notag
\end{equation}%
with:%
\begin{eqnarray*}
U &=&\left( 1-D^{1+\gamma }\int \exp \left( -\frac{\left\vert
X_{3}\right\vert }{\left( 1+\gamma \right) }\right) \exp \left( \frac{\gamma
\left\vert X_{2}\right\vert }{\left( 1+\gamma \right) }\right) \exp \left(
-\left( \left\vert X-X_{2}\right\vert +\left\vert X-X_{3}\right\vert \right)
\right) \right) \\
&\simeq &1-h\left( 1-\exp \left( -\frac{1}{d}\right) \right)
\end{eqnarray*}%
whose solution, with with $\frac{\partial ^{2}}{\partial \rho ^{2}}>0$, is:%
\begin{equation}
\rho ^{2}=\frac{\kappa _{1}+\sqrt{\kappa _{1}^{2}-2\kappa _{2}\left( 2\alpha
+\frac{2}{\vartheta ^{2}}+\sqrt{\kappa _{0}}+\sigma \sqrt{\delta ^{2}+\frac{%
\bar{A}^{2}A^{2}}{\left( A^{2}U^{2}+\bar{A}^{2}\right) ^{2}}}\right) }}{%
2\kappa _{2}}  \label{rhosq}
\end{equation}%
and the eigenstate:%
\begin{equation}
\Psi _{0}\left( K,X\right) =N\exp \left( -\frac{\kappa _{0}^{-\frac{1}{2}%
}X^{2}}{2}\right) \exp \left( -\frac{\sqrt{\delta ^{2}+\frac{\bar{A}^{2}A^{2}%
}{\left( A^{2}U^{2}+\bar{A}^{2}\right) ^{2}}}\left( K-\left\langle
K\right\rangle _{X}\right) ^{2}}{2\sigma }\right)
\end{equation}%
where $N$ is the normalisation factor:%
\begin{equation}
N=\frac{\rho }{2\pi \sqrt{\Theta \kappa _{0}^{-\frac{1}{2}}\sigma \sqrt{%
\delta ^{2}+\frac{\bar{A}^{2}A^{2}}{\left( A^{2}U^{2}+\bar{A}^{2}\right) ^{2}%
}}}}
\end{equation}%
which completes the computations for $\allowbreak P$, $\left\langle
K\right\rangle $, and the potential.

\bigskip For later purpose, we also derive the corrections to these results
at the second order in $\chi _{1}$ and $\chi _{2}$. At this order, expanding
the exponentials in $\chi _{1}$ and $\chi _{2}$\ yields the potential of the 
$X$ part of the action:

\begin{eqnarray}
&&\frac{1}{2}\kappa _{0}x^{2}+\kappa _{1}\chi _{1}\left\vert x\right\vert 
\frac{K}{\left\langle K\right\rangle _{x}}\rho ^{2}-2\kappa _{2}\chi
_{2}\left\vert x\right\vert \rho ^{4}-\frac{\kappa _{1}}{2}\chi
_{1}^{2}x^{2}\left( \frac{K}{\left\langle K\right\rangle _{x}}\rho
^{2}\right) ^{2}+2\kappa _{2}\chi _{2}^{2}x^{2}\rho ^{8}  \notag \\
&\rightarrow &\frac{1}{2}\kappa _{0}\left( x+\frac{sgn\left( x\right) }{%
\kappa _{0}}\left( \chi _{1}\kappa _{1}\frac{K}{2\left\langle K\right\rangle
_{x}}\rho ^{2}-2\chi _{2}\kappa _{2}\rho ^{4}\right) \right) ^{2}-\frac{1}{%
2\kappa _{0}}\left( \chi _{1}\kappa _{1}\frac{K}{2\left\langle
K\right\rangle _{x}}\rho ^{2}-2\chi _{2}\kappa _{2}\rho ^{4}\right) ^{2} 
\notag \\
&&+\left( -\frac{\kappa _{1}\chi _{1}^{2}\rho ^{2}}{2}+2\chi _{2}^{2}\kappa
_{2}\rho ^{4}\right) x^{2}  \notag \\
&\simeq &\frac{1}{2}\left( \kappa _{0}+\left( -\frac{\kappa _{1}\chi
_{1}^{2}\rho ^{2}}{2}+2\chi _{2}^{2}\kappa _{2}\rho ^{4}\right) \right)
\left( x+\frac{sgn\left( x\right) }{\kappa _{0}}\left( \chi _{1}\kappa _{1}%
\frac{K}{2\left\langle K\right\rangle _{x}}\rho ^{2}-2\chi _{2}\kappa
_{2}\rho ^{4}\right) \right) ^{2}  \label{Ptscd} \\
&&-\frac{1}{2\kappa _{0}}\left( \chi _{1}\kappa _{1}\frac{K}{2\left\langle
K\right\rangle _{x}}\rho ^{2}-2\chi _{2}\kappa _{2}\rho ^{4}\right) ^{2} 
\notag
\end{eqnarray}%
so that, the fundamental equation becomes:%
\begin{eqnarray*}
&&0=\left( -\sigma _{X}^{2}\nabla _{X}^{2}-\sigma ^{2}\nabla
_{K}^{2}-\vartheta ^{2}\nabla _{\theta }^{2}\right) \Psi _{0}\left(
K,X,\theta \right) \\
&&+\kappa _{0}\left( 1-\frac{\chi _{1}^{2}\kappa _{1}\rho ^{2}}{2\kappa _{0}}%
+2\chi _{2}\frac{\kappa _{2}}{\kappa _{0}}\rho ^{4}\right) \left(
x+sgn\left( x\right) \left( \chi _{1}\kappa _{1}\frac{K}{2\left\langle
K\right\rangle _{x}}\rho ^{2}-\chi _{2}^{2}\kappa _{2}\rho ^{4}\right)
\right) ^{2}\Psi _{0}\left( K,X,\theta \right) \\
&&+\left( \left( \delta ^{2}+\frac{\bar{A}^{2}A^{2}}{\left( A^{2}U^{2}+\bar{A%
}^{2}\right) ^{2}}\right) \left( K-\left\langle K\right\rangle _{X}\right)
^{2}\right) \Psi _{0}\left( K,X,\theta \right) \\
&&+\left( \rho ^{4}\kappa _{2}-\rho ^{2}\kappa _{1}-\left( \chi _{1}\kappa
_{1}\frac{K}{2\left\langle K\right\rangle _{x}}\rho ^{2}-\chi _{2}\kappa
_{2}\rho ^{4}\right) ^{2}+\frac{1}{\vartheta ^{2}}+\alpha \right) \Psi
_{0}\left( K,X,\theta \right)
\end{eqnarray*}

Again, we look for a fundamental of the form $N\Psi _{0}\left( X\right) \Psi
_{0}\left( K\right) $. The operators in $X$ and $K$ are harmonic oscillators
with frequencies:%
\begin{equation*}
\kappa _{0}^{\frac{1}{2}}\sqrt{1-\frac{\chi _{1}^{2}\kappa _{1}\rho ^{2}}{%
2\kappa _{0}}+2\chi _{2}^{2}\frac{\kappa _{2}}{\kappa _{0}}\rho ^{4}}\text{
and }\sigma \sqrt{\delta ^{2}+\frac{\bar{A}^{2}A^{2}}{\left( A^{2}U^{2}+\bar{%
A}^{2}\right) ^{2}}}
\end{equation*}

\bigskip 
\begin{eqnarray}
\Psi _{0}\left( X\right) &=&\exp \left( -\frac{\kappa _{0}^{-\frac{1}{2}%
}\left( X+\left( \chi _{1}\kappa _{1}\frac{K}{2\left\langle K\right\rangle
_{X}}\rho ^{2}-\chi _{2}\kappa _{2}\rho ^{4}\right) \right) ^{2}}{2}\right)
H\left( x\right)  \label{psi0X} \\
&&+\exp \left( -\frac{\kappa _{0}^{-\frac{1}{2}}\left( X-\left( \chi
_{1}\kappa _{1}\frac{K}{2\left\langle K\right\rangle _{X}}\rho ^{2}-\chi
_{2}\kappa _{2}\rho ^{4}\right) \right) ^{2}}{2}\right) H\left( -x\right) 
\notag \\
&\equiv &\bar{\Psi}_{0}\left( X+\left( \chi _{1}\kappa _{1}\frac{K}{%
2\left\langle K\right\rangle _{X}}\rho ^{2}-\chi _{2}\kappa _{2}\rho
^{4}\right) \right) H\left( X\right) +\bar{\Psi}_{0}\left( X-\left( \chi
_{1}\kappa _{1}\frac{K}{2\left\langle K\right\rangle _{X}}\rho ^{2}-\chi
_{2}\kappa _{2}\rho ^{4}\right) \right) H\left( -X\right)  \notag
\end{eqnarray}%
\begin{equation}
\Psi _{0}\left( K\right) =\exp \left( -\frac{\sqrt{\delta ^{2}+\frac{\bar{A}%
^{2}A^{2}}{\left( A^{2}U^{2}+\bar{A}^{2}\right) ^{2}}}\left( K-\left\langle
K\right\rangle _{X}\right) ^{2}}{2\sigma }\right)  \label{psi0K}
\end{equation}%
and the fundamental state writes:%
\begin{eqnarray}
\Psi _{0}\left( K,X\right) &=&\rho N\left[ \bar{\Psi}_{0}\left( X+\delta
X\right) H\left( X\right) +\bar{\Psi}_{0}\left( X-\delta X\right) H\left(
-X\right) \right]  \label{mnml} \\
&&\times \exp \left( -\frac{\sqrt{\delta ^{2}+\frac{\bar{A}^{2}A^{2}}{\left(
A^{2}U^{2}+\bar{A}^{2}\right) ^{2}}}\left( K-\left\langle K\right\rangle
_{X}\right) ^{2}}{2\sigma }\right)  \notag
\end{eqnarray}%
where $N$ is a normalisation factor:%
\begin{equation}
N=\frac{1}{2\pi \sqrt{\kappa _{0}^{-\frac{1}{2}}\sigma \sqrt{\delta ^{2}+%
\frac{\bar{A}^{2}A^{2}}{\left( A^{2}U^{2}+\bar{A}^{2}\right) ^{2}}}}}
\label{Nrm}
\end{equation}%
and:%
\begin{equation}
\delta X=\left( \chi _{1}\kappa _{1}\frac{K}{2\left\langle K\right\rangle }%
\rho ^{2}-\chi _{2}\kappa _{2}\rho ^{4}\right)  \label{dltX}
\end{equation}%
with $\rho $ satisfying the condition:%
\begin{eqnarray}
0 &=&\alpha +\frac{1}{\vartheta ^{2}}+\frac{1}{2}\kappa _{0}^{\frac{1}{2}}%
\sqrt{1-\frac{\chi _{1}^{2}\kappa _{1}\rho ^{2}}{2}+2\chi _{2}^{2}\kappa
_{2}\rho ^{4}}  \notag \\
&&-\kappa _{1}\rho ^{2}+\kappa _{2}\rho ^{4}-\left( \chi _{1}\kappa _{1}%
\frac{K}{2\left\langle K\right\rangle _{X}}\rho ^{2}-\chi _{2}\kappa
_{2}\rho ^{4}\right) ^{2}+\frac{\sigma \sqrt{\delta ^{2}+\frac{\bar{A}%
^{2}A^{2}}{\left( A^{2}U^{2}+\bar{A}^{2}\right) ^{2}}}}{2}  \label{rhqn}
\end{eqnarray}

\subsubsection{Contribution of $\Psi $}

Now, the replacement $\Psi \left( K_{3},P_{3},X_{3},\theta \right)
\rightarrow \Psi _{0}\left( K_{3},P_{3},X_{3},\theta \right) +\delta \Psi
\left( K_{3},P_{3},X_{3},\theta \right) $ for $V$ can be performed in the
following way. For any quantity, $\left( K\right) ^{-\frac{\alpha }{1+\gamma 
}}$, $f\left( X\right) $ the expectations:%
\begin{equation*}
\left( \Psi _{0}^{\dag }\left( K_{3},P_{3},X_{3},\theta \right) +\delta \Psi
^{\dag }\left( K_{3},P_{3},X_{3},\theta \right) \right) \left( 
\begin{array}{c}
\left( K\right) ^{-\frac{\alpha }{1+\gamma }} \\ 
f\left( X\right)%
\end{array}%
\right) \left( \Psi _{0}\left( K_{3},P_{3},X_{3},\theta \right) +\delta \Psi
\left( K_{3},P_{3},X_{3},\theta \right) \right)
\end{equation*}%
are approximatively:%
\begin{equation*}
\Psi _{0}^{\dag }\left( K_{3},P_{3},X_{3},\theta \right) \left( 
\begin{array}{c}
\left( K\right) ^{-\frac{\alpha }{1+\gamma }} \\ 
f\left( X\right)%
\end{array}%
\right) \Psi _{0}\left( K_{3},P_{3},X_{3},\theta \right) +\delta \Psi ^{\dag
}\left( K_{3},P_{3},X_{3},\theta \right) \left( 
\begin{array}{c}
\left( K\right) ^{-\frac{\alpha }{1+\gamma }} \\ 
f\left( X\right)%
\end{array}%
\right) \delta \Psi \left( K_{3},P_{3},X_{3},\theta \right)
\end{equation*}%
since, given their form, $\left( K\right) ^{-\frac{\alpha }{1+\gamma }}$ and 
$f\left( X\right) $ can be considered as close to their average $%
\left\langle K\right\rangle ^{-\frac{\alpha }{1+\gamma }}$ and $f\left(
\left\langle X\right\rangle \right) $ and thus%
\begin{equation*}
\left\langle \Psi _{0}^{\dag }\left( K_{3},P_{3},X_{3},\theta \right) \left( 
\begin{array}{c}
\left( K\right) ^{-\frac{\alpha }{1+\gamma }} \\ 
f\left( X\right)%
\end{array}%
\right) ,\delta \Psi \left( K_{3},P_{3},X_{3},\theta \right) \right\rangle
\simeq 0
\end{equation*}%
for a perturbation $\delta \Psi \left( K,X\right) $ orthogonal to $\Psi
_{0}\left( K,X\right) $. As a consequence the second order development of
the potential terms are:%
\begin{eqnarray*}
&&\Psi _{0}^{\dag }\left( K,X\right) \left( f^{1+\gamma }\left( X\right)
-\kappa \int P_{2}\left( K_{2}\right) ^{\alpha }P_{3}\exp \left( -\frac{%
\left\vert X-X_{2}\right\vert +\left\vert X_{2}-X_{3}\right\vert }{d}\right)
\left\vert \Psi _{0}\left( K_{2},X_{2}\right) \right\vert ^{2}\left\vert
\Psi _{0}\left( K_{3},X_{3}\right) \right\vert ^{2}\right) ^{2} \\
&&+\delta \Psi ^{\dag }\left( K,X\right) \left( f^{1+\gamma }\left( X\right)
-\kappa \int P_{2}\left( K_{2}\right) ^{\alpha }P_{3}\exp \left( -\frac{%
\left\vert X-X_{2}\right\vert +\left\vert X_{2}-X_{3}\right\vert }{d}\right)
\left\vert \Psi _{0}\left( K_{2},X_{2}\right) \right\vert ^{2}\left\vert
\Psi _{0}\left( K_{3},X_{3}\right) \right\vert ^{2}\right) \delta \Psi
\left( K,X\right) \\
&&-2\kappa \Psi _{0}^{\dag }\left( K,X\right) \left( \int \Psi _{0}^{\dag
}\left( K_{3},X_{3}\right) \delta \Psi ^{\dag }\left( K_{2},X_{2}\right)
P_{2}\left( K_{2}\right) ^{\alpha }P_{3}\exp \left( -\frac{\left\vert
X-X_{2}\right\vert +\left\vert X_{2}-X_{3}\right\vert }{d}\right) \delta
\Psi \left( K_{2},X_{2}\right) \Psi _{0}\left( K_{3},X_{3}\right) \right. \\
&&+\left. \int \Psi _{0}^{\dag }\left( K_{2},X_{2}\right) \delta \Psi ^{\dag
}\left( K_{3},X_{3}\right) P_{2}\left( K_{2}\right) ^{\alpha }P_{3}\exp
\left( -\frac{\left\vert X-X_{2}\right\vert +\left\vert
X_{2}-X_{3}\right\vert }{d}\right) \delta \Psi \left( K_{3},X_{3}\right)
\Psi _{0}\left( K_{2},X_{2}\right) \right) \\
&&\times \left( f^{1+\gamma }\left( X\right) -\kappa \int P_{2}\left(
K_{2}\right) ^{\alpha }P_{3}\exp \left( -\frac{\left\vert X-X_{2}\right\vert
+\left\vert X_{2}-X_{3}\right\vert }{d}\right) \left\vert \Psi _{0}\left(
K_{2},X_{2}\right) \right\vert ^{2}\left\vert \Psi _{0}\left(
K_{3},X_{3}\right) \right\vert ^{2}\right) \Psi _{0}\left( K,X\right)
\end{eqnarray*}%
where, as before, we define $f\left( X\right) =PK^{-\frac{\alpha }{1+\gamma }%
}$. The first term is the action part for $\Psi _{0}\left( K,X\right) $, it
can be discarded (and in fact is equal to zero given the constraint).

Given that both for $\Psi _{0}$ and $\delta \Psi $, $K$ is centered around $%
\left\langle K\right\rangle _{X,0}$ and $\left\langle K\right\rangle _{X}$,
and that for both state we may consider the $X$ ditribution as uniform and,
as a first approximation, we set:%
\begin{eqnarray*}
G_{0}\left( X_{i},X_{i}\right) &=&\rho ^{2}\Psi _{0}^{\dag }\left(
\left\langle K\right\rangle _{X_{i},0},X_{i}\right) \Psi _{0}\left(
\left\langle K\right\rangle _{X_{i},0},X_{i}\right) \\
G\left( X_{i},X_{i}\right) &=&G\left( \left( \left\langle K\right\rangle
_{X_{i}},X_{i}\right) ,\left( \left\langle K\right\rangle
_{X_{j}},X_{j}\right) \right)
\end{eqnarray*}%
and:%
\begin{eqnarray*}
f_{0}\left( X\right) &=&\int PK^{-\frac{\alpha }{1+\gamma }}\left\vert \Psi
_{0}\left( K,X\right) \right\vert ^{2}dK\simeq PK^{-\frac{\alpha }{1+\gamma }%
}\left\vert \Psi _{0}\left( \left\langle K\right\rangle _{X,0},X\right)
\right\vert ^{2} \\
f_{0}^{1+\gamma }\left( X\right) &=&\int PK^{-\frac{\alpha }{1+\gamma }%
}\left\vert \Psi _{0}\left( K,X\right) \right\vert ^{2}dK\simeq P^{1+\gamma
}K^{-\alpha }\left\vert \Psi _{0}\left( \left\langle K\right\rangle
_{X,0},X\right) \right\vert ^{2}
\end{eqnarray*}%
so that the last term rewrites:%
\begin{eqnarray*}
&&-\kappa \int G_{0}\left( X_{2},X_{2}\right) G\left( X_{3},X_{3}\right)
\left\langle K\right\rangle _{X_{2},0}^{\frac{\alpha \gamma }{1+\gamma }%
}\left\langle K\right\rangle _{X_{3}}^{-\frac{\alpha }{1+\gamma }}f\left(
X_{3}\right) f_{0}\left( X_{2}\right) \exp \left( -\frac{\left\vert
X-X_{2}\right\vert +\left\vert X_{2}-X_{3}\right\vert }{d}\right) \\
&&\times \left( \left( f_{0}\left( X\right) \right) ^{1+\gamma }-\kappa \int
G_{0}\left( X_{2},X_{2}\right) G_{0}\left( X_{3},X_{3}\right) \left\langle
K\right\rangle _{X_{2},0}^{\frac{\alpha \gamma }{1+\gamma }}\left\langle
K\right\rangle _{X_{3},0}^{-\frac{\alpha }{1+\gamma }}f_{0}\left(
X_{3}\right) f_{0}\left( X_{2}\right) \exp \left( -\frac{\left\vert
X-X_{2}\right\vert +\left\vert X_{2}-X_{3}\right\vert }{d}\right) \right) \\
&&\times G_{0}\left( X,X\right) \\
&&-\kappa \int G\left( X_{2},X_{2}\right) G_{0}\left( X_{3},X_{3}\right)
\left\langle K\right\rangle _{X_{2}}^{\frac{\alpha \gamma }{1+\gamma }%
}\left\langle K\right\rangle _{X_{3},0}^{-\frac{\alpha }{1+\gamma }%
}f_{0}\left( X_{3}\right) f\left( X_{2}\right) \exp \left( -\frac{\left\vert
X-X_{2}\right\vert +\left\vert X_{2}-X_{3}\right\vert }{d}\right) \\
&&\times \left( \left( f_{0}\left( X\right) \right) ^{1+\gamma }-\kappa \int
G_{0}\left( X_{2},X_{2}\right) G_{0}\left( X_{3},X_{3}\right) \left\langle
K\right\rangle _{X_{2},0}^{\frac{\alpha \gamma }{1+\gamma }}\left\langle
K\right\rangle _{X_{3},0}^{-\frac{\alpha }{1+\gamma }}f_{0}\left(
X_{3}\right) f_{0}\left( X_{2}\right) \exp \left( -\frac{\left\vert
X-X_{2}\right\vert +\left\vert X_{2}-X_{3}\right\vert }{d}\right) \right) \\
&&\times G_{0}\left( X,X\right)
\end{eqnarray*}%
As in Appendix 2, we can approximate the integrals by their estimations on
the diagonal, and consider in first approximation a uniform distribution for 
$X$ so that this term rewrites: 
\begin{eqnarray*}
&&-\kappa \int \left( \left\langle K\right\rangle _{X_{2},0}^{\frac{\alpha
\gamma }{1+\gamma }}\left\langle K\right\rangle _{X_{2}}^{-\frac{\alpha }{%
1+\gamma }}+\left\langle K\right\rangle _{X_{2}}^{\frac{\alpha \gamma }{%
1+\gamma }}\left\langle K\right\rangle _{X_{2},0}^{-\frac{\alpha }{1+\gamma }%
}\right) f\left( X_{2}\right) f_{0}\left( X_{2}\right) \\
&&\times \int \exp \left( -\left( \frac{\left\vert X-X_{2}\right\vert }{d}%
\right) \right) \left( \left( f_{0}\left( X\right) \right) ^{1+\gamma
}-\kappa \int \left\langle K\right\rangle _{X_{2},0}^{\frac{\alpha \left(
\gamma -1\right) }{1+\gamma }}\left( f_{0}\left( X_{2}\right) \right)
^{2}\exp \left( -\left( \frac{\left\vert X-X_{2}\right\vert }{d}\right)
\right) \right)
\end{eqnarray*}%
where we used that the $K$ part of $\Psi _{0}$ is peaked around $%
\left\langle K\right\rangle _{X,0}$, and that the $X$ part is assumed
distributed uniformly.

\begin{eqnarray*}
&\simeq &-\kappa \int \left( \left\langle K\right\rangle _{X,0}^{\frac{%
\alpha \gamma }{1+\gamma }}\left\langle K\right\rangle _{X}^{-\frac{\alpha }{%
1+\gamma }}+\left\langle K\right\rangle _{X}^{\frac{\alpha \gamma }{1+\gamma 
}}\left\langle K\right\rangle _{X,0}^{-\frac{\alpha }{1+\gamma }}\right)
f\left( X\right) f_{0}\left( X\right) \\
&&\times \left( \left( f_{0}\left( X\right) \right) ^{1+\gamma }-\kappa \int
\left\langle K\right\rangle _{X_{2},0}^{\frac{\alpha \left( \gamma -1\right) 
}{1+\gamma }}\left( f_{0}\left( X_{2}\right) \right) ^{2}\exp \left( -\left( 
\frac{\left\vert X-X_{2}\right\vert }{d}\right) \right) \right) \\
&\simeq &0
\end{eqnarray*}%
since the equality%
\begin{equation*}
\left( f_{0}\left( X\right) \right) ^{1+\gamma }-\kappa \int \left\langle
K\right\rangle _{X_{2},0}^{\frac{\alpha \left( \gamma -1\right) }{1+\gamma }%
}\left( f_{0}\left( X_{2}\right) \right) ^{2}\exp \left( -\left( \frac{%
\left\vert X-X_{2}\right\vert }{d}\right) \right) =0
\end{equation*}%
holds on $X$. Thus the constraint becomes to the second order: 
\begin{equation}
\delta \Psi ^{\dag }\left( P^{1+\gamma }K^{\alpha }-\kappa \int G_{0}\left(
X_{2},X_{2}\right) G_{0}\left( X_{3},X_{3}\right) \left\langle
K\right\rangle _{X_{2},0}^{\frac{\alpha \gamma }{1+\gamma }}\left\langle
K\right\rangle _{X_{3},0}^{-\frac{\alpha }{1+\gamma }}f_{0}\left(
X_{3}\right) f_{0}\left( X_{2}\right) \exp \left( -\frac{\left\vert
X-X_{2}\right\vert +\left\vert X_{2}-X_{3}\right\vert }{d}\right) \right)
^{2}\delta \Psi  \label{Pcd}
\end{equation}%
The identification for $\left\langle K\right\rangle _{X}$:%
\begin{equation*}
\Psi ^{\dag }\left( \delta K-APK^{\alpha }\left( 1-\kappa \int \frac{%
P_{3}\exp \left( -\frac{\left\vert X_{1}-X_{2}\right\vert +\left\vert
X_{1}-X_{3}\right\vert }{d}\right) }{P_{2}^{\gamma }}\left\vert \Psi
_{0}\left( K_{2},X_{2}\right) \right\vert ^{2}\left\vert \Psi _{0}\left(
K_{3},X_{3}\right) \right\vert ^{2}\right) \right) ^{2}\Psi
\end{equation*}%
becomes at the second order:

\begin{eqnarray*}
&&\Psi _{0}^{\dag }\left( K,X\right) \left( \delta K-APK^{\alpha }\left(
1-\kappa \int \frac{P_{3}\exp \left( -\frac{\left\vert X-X_{2}\right\vert
+\left\vert X_{2}-X_{3}\right\vert }{d}\right) }{P_{2}^{\gamma }}\left\vert
\Psi _{0}\left( K_{2},X_{2}\right) \right\vert ^{2}\left\vert \Psi
_{0}\left( K_{3},X_{3}\right) \right\vert ^{2}\right) \right) ^{2}\Psi
_{0}\left( K,X\right) \\
&&+\delta \Psi ^{\dag }\left( K,X\right) \left( \delta K-APK^{\alpha }\left(
1-\kappa \int \frac{P_{3}\exp \left( -\frac{\left\vert X-X_{2}\right\vert
+\left\vert X_{2}-X_{3}\right\vert }{d}\right) }{P_{2}^{\gamma }}\left\vert
\Psi _{0}\left( K_{2},X_{2}\right) \right\vert ^{2}\left\vert \Psi
_{0}\left( K_{3},X_{3}\right) \right\vert ^{2}\right) \right) ^{2}\delta
\Psi \left( K,X\right) \\
&&+\Psi _{0}^{\dag }\left( K,X\right) \left( \int \frac{\exp \left( -\frac{%
\left\vert X-X_{2}\right\vert +\left\vert X_{2}-X_{3}\right\vert }{d}\right) 
}{P_{2}^{\gamma }}P_{3}\left\vert \delta \Psi \left( K_{2},X_{2}\right)
\right\vert ^{2}\left\vert \Psi _{0}\left( K_{3},X_{3}\right) \right\vert
^{2}\right. \\
&&\left. +\int \frac{\exp \left( -\frac{\left\vert X-X_{2}\right\vert
+\left\vert X_{2}-X_{3}\right\vert }{d}\right) }{P_{2}^{\gamma }}%
P_{3}\left\vert \delta \Psi \left( K_{3},X_{3}\right) \right\vert
^{2}\left\vert \Psi _{0}\left( K_{2},X_{2}\right) \right\vert ^{2}\right) \\
&&\times \left( \delta K-APK^{\alpha }\left( 1-\kappa \int \frac{P_{3}\exp
\left( -\frac{\left\vert X-X_{2}\right\vert +\left\vert
X_{2}-X_{3}\right\vert }{d}\right) }{P_{2}^{\gamma }}\left\vert \Psi
_{0}\left( K_{2},X_{2}\right) \right\vert ^{2}\left\vert \Psi _{0}\left(
K_{3},X_{3}\right) \right\vert ^{2}\right) \right) \Psi _{0}\left( K,X\right)
\end{eqnarray*}

\bigskip As before, the first term is the potential in state $\Psi _{0}$ and
can be discarded. Under our asumptions, the last term rewrites:%
\begin{eqnarray*}
&&+\int \left( \int G_{0}\left( X_{3},X_{3}\right) G\left(
X_{2},X_{2}\right) \left( \left\langle K\right\rangle _{X}\right) ^{\frac{%
\alpha \gamma }{1+\gamma }}\left( \left\langle K\right\rangle _{X,0}\right)
^{-\frac{\alpha }{1+\gamma }}f\left( X_{3}\right) \left( f\left(
X_{2}\right) \right) ^{-\gamma }\exp \left( -\frac{\left\vert
X-X_{2}\right\vert +\left\vert X_{2}-X_{3}\right\vert }{d}\right) \right. \\
&&\left. +\int G_{0}\left( X_{2},X_{2}\right) G\left( X_{3},X_{3}\right)
\left( \left\langle K\right\rangle _{X,0}\right) ^{\frac{\alpha \gamma }{%
1+\gamma }}\left( \left\langle K\right\rangle _{X}\right) ^{-\frac{\alpha }{%
1+\gamma }}\exp \left( -\frac{\left\vert X-X_{2}\right\vert +\left\vert
X_{2}-X_{3}\right\vert }{d}\right) \right) \\
&&\times \left( \delta \left\langle K\right\rangle _{X,0}-AP\left\langle
K\right\rangle _{X,0}^{\alpha }\left( 1-\kappa \int \left\langle
K\right\rangle _{X_{2},0}^{\frac{\alpha \gamma }{1+\gamma }}\left\langle
K\right\rangle _{X_{3},0}^{-\frac{\alpha }{1+\gamma }}f\left( X_{3}\right)
\left( f\left( X_{2}\right) \right) ^{-\gamma }\exp \left( -\frac{\left\vert
X-X_{2}\right\vert +\left\vert X_{2}-X_{3}\right\vert }{d}\right) \right)
\right) \\
&&\times G_{0}\left( X,X\right)
\end{eqnarray*}%
and is nul in first approximation since:%
\begin{equation}
0=\delta \left\langle K\right\rangle _{X,0}-AP\left\langle K\right\rangle
_{X,0}^{\alpha }\left( 1-\kappa \int \left\langle K\right\rangle _{X_{2},0}^{%
\frac{\alpha \gamma }{1+\gamma }}\left\langle K\right\rangle _{X_{3},0}^{-%
\frac{\alpha }{1+\gamma }}f\left( X_{3}\right) \left( f\left( X_{2}\right)
\right) ^{-\gamma }\exp \left( -\frac{\left\vert X-X_{2}\right\vert
+\left\vert X_{2}-X_{3}\right\vert }{d}\right) \right)  \label{UK0}
\end{equation}%
holds for all $X$. Thus, the second order expansion becomes:%
\begin{equation}
\delta \Psi ^{\dag }\left( K,X\right) \left( \delta K-APK^{\alpha }\left(
1-\kappa \int \left\langle K\right\rangle _{X_{2},0}^{\frac{\alpha \gamma }{%
1+\gamma }}\left\langle K\right\rangle _{X_{3},0}^{-\frac{\alpha }{1+\gamma }%
}f\left( X_{3}\right) \left( f\left( X_{2}\right) \right) ^{-\gamma }\exp
\left( -\frac{\left\vert X-X_{2}\right\vert +\left\vert
X_{2}-X_{3}\right\vert }{d}\right) \right) \right) ^{2}\delta \Psi \left(
K,X\right)  \label{Kscd}
\end{equation}

$\bigskip $Along with the constraint, we deduce that $f\left( X\right)
\simeq f_{0}\left( X\right) $ and $\left\langle K\right\rangle _{X}\simeq
\left\langle K\right\rangle _{X,0}$, the values computed in state $\Psi _{0}$
previously in this Appendix. Thus, the potential becomes: 
\begin{eqnarray*}
&&\delta \Psi ^{\dag }\left( \delta K-APK^{\alpha }\left( 1-\kappa \int
\left\langle K\right\rangle _{X_{2},0}^{\frac{\alpha \gamma }{1+\gamma }%
}\left\langle K\right\rangle _{X_{3},0}^{-\frac{\alpha }{1+\gamma }}f\left(
X_{3}\right) \left( f\left( X_{2}\right) \right) ^{-\gamma }\exp \left( -%
\frac{\left\vert X-X_{2}\right\vert +\left\vert X_{2}-X_{3}\right\vert }{d}%
\right) \right) \right) ^{2}\delta \Psi \\
&=&\delta \Psi ^{\dag }\delta ^{2}\left( K-\left\langle K\right\rangle
_{X,0}\right) ^{2}\delta \Psi
\end{eqnarray*}%
as a consequence of (\ref{UK0}).

As in the case of the first phase, we can include the first order
corrections in $\frac{A^{2}}{\bar{A}^{2}}$. Since the terms appearing in the
second order expansion are (\ref{Pcd}) and (\ref{Kscd}), the potential
becomes:%
\begin{equation*}
\delta ^{2}\left( 1+\frac{\bar{A}^{2}A^{2}}{\left( A^{2}U^{2}+\bar{A}%
^{2}\right) ^{2}}\right) \delta \Psi ^{\dag }\left( \delta K-\left\langle
K\right\rangle _{X,0}\right) ^{2}\delta \Psi
\end{equation*}

\subsection{\textbf{Appendix 4}}

\bigskip In this section, we compute the Green functions in both phases.

\subsubsection{Case $\protect\rho =0$}

For $\rho =0$, without the $X$ contribution:%
\begin{equation*}
\Psi ^{\dag }\left( K,P,X,\theta \right) \left( -\sigma ^{2}\nabla
_{K}^{2}-\vartheta ^{2}\nabla _{\theta }^{2}+\left( \delta ^{2}+\frac{\bar{A}%
^{2}A^{2}}{\left( A^{2}U^{2}+\bar{A}^{2}\right) ^{2}}\right) \left(
K-\left\langle K\right\rangle \right) ^{2}+\frac{1}{\vartheta ^{2}}+\alpha
\right) \Psi \left( K,P,X,\theta \right)
\end{equation*}%
whose Green function is given by:%
\begin{eqnarray*}
G_{0}\left( K,K^{\prime },\theta ,\theta ^{\prime },t\right) &=&\sqrt{\frac{%
\omega }{2\pi \sigma ^{2}\sinh \left( \omega t\right) }}\exp \left( \left( -%
\frac{\omega }{2\sigma ^{2}\sinh \left( \omega t\right) }\right) \left(
\left( K^{2}+\left( K^{\prime }\right) ^{2}\right) \cosh \left( \omega
t\right) -2KK^{\prime }\right) \right) \\
&&\times \sqrt{\frac{1}{2\pi \vartheta ^{2}t}}\exp \left( -\frac{\left(
\theta -\theta ^{\prime }\right) ^{2}}{\vartheta ^{2}t}\right)
\end{eqnarray*}%
with:%
\begin{equation*}
\omega =\delta ^{2}+\frac{\bar{A}^{2}A^{2}}{\left( A^{2}U^{2}+\bar{A}%
^{2}\right) ^{2}}
\end{equation*}%
To include the $\Psi \left( x\right) $ part of the action: 
\begin{eqnarray*}
&&\Psi ^{\dag }\left( x\right) \left( -\sigma _{X}^{2}\nabla _{X}^{2}+\kappa
_{0}x^{2}\right) \Psi \left( x\right) -\frac{\kappa _{1}}{2}\left( \Psi
^{\dag }\left( K,x\right) \Psi \left( K,x\right) \right) \frac{KK^{\prime
}\exp \left( -\chi _{1}\left\vert x-y\right\vert \right) }{\left\langle
K\right\rangle _{x}\left\langle K\right\rangle _{y}}\left( \Psi ^{\dag
}\left( K^{\prime },y\right) \Psi \left( K^{\prime },y\right) \right) \\
&&+\frac{\kappa _{2}}{3}\left( \Psi ^{\dag }\left( K,x\right) \Psi \left(
K,x\right) \right) \exp \left( -\chi _{2}\left\vert x-y\right\vert -\chi
_{2}\left\vert x-z\right\vert -\chi _{2}\left\vert y-z\right\vert \right)
\left( \Psi ^{\dag }\left( K^{\prime },y\right) \Psi \left( K^{\prime
},y\right) \right) \left( \Psi ^{\dag }\left( K",z\right) \Psi \left(
K",z\right) \right)
\end{eqnarray*}%
and its contribution to the Green function, we can replace the interaction
potential by its average in variables $y$ and $z$:

\begin{eqnarray*}
&&-\frac{\kappa _{1}}{2}\left\vert \Psi \left( K,x\right) \right\vert ^{2}%
\frac{KK^{\prime }\exp \left( -\chi _{1}\left\vert x-y\right\vert \right) }{%
\left\langle K\right\rangle _{x}\left\langle K\right\rangle _{y}}\left\vert
\Psi \left( K^{\prime },y\right) \right\vert ^{2} \\
&\rightarrow &-\kappa _{1}\left\langle \frac{KK^{\prime }\exp \left( -\chi
_{1}\left\vert x-y\right\vert \right) }{\left\langle K\right\rangle
_{x}\left\langle K\right\rangle _{y}}\left\vert \Psi \left( K^{\prime
},y\right) \right\vert ^{2}\right\rangle \left\vert \Psi \left( K,x\right)
\right\vert ^{2} \\
&&\frac{\kappa _{2}}{3}\left\vert \Psi \left( K,x\right) \right\vert
^{2}\exp \left( -\chi _{2}\left\vert x-y\right\vert -\chi _{2}\left\vert
x-z\right\vert -\chi _{2}\left\vert y-z\right\vert \right) \left\vert \Psi
\left( K^{\prime },y\right) \right\vert ^{2}\left\vert \Psi \left(
K",z\right) \right\vert ^{2} \\
&\rightarrow &\kappa _{2}\left\vert \Psi \left( K,x\right) \right\vert
^{2}\left\langle \exp \left( -\chi _{2}\left\vert x-y\right\vert -\chi
_{2}\left\vert x-z\right\vert -\chi _{2}\left\vert y-z\right\vert \right)
\left\vert \Psi \left( K^{\prime },y\right) \right\vert ^{2}\left\vert \Psi
\left( K",z\right) \right\vert ^{2}\right\rangle
\end{eqnarray*}%
Given that at the lowest order:%
\begin{equation*}
\left\langle \frac{KK^{\prime }\exp \left( -\chi _{1}\left\vert
x-y\right\vert \right) }{\left\langle K\right\rangle _{x}\left\langle
K\right\rangle _{y}}\left\vert \Psi \left( K^{\prime },y\right) \right\vert
^{2}\right\rangle \simeq \int \frac{KK^{\prime }\exp \left( -\chi
_{1}\left\vert x-y\right\vert \right) }{\left\langle K\right\rangle
_{x}\left\langle K\right\rangle _{y}}G\left( \left( K,x\right) ,\left(
K^{\prime },y\right) \right) G\left( \left( K^{\prime },y\right) ,\left(
K,x\right) \right) dy
\end{equation*}%
Under the same hypothesis as in appendix 2, that is $K^{\prime }$ is spread
around $\left\langle K\right\rangle _{y}$ and the $X$ part of the Green
function is approximatively uniformly distributed on the interval $\left[
-1,1\right] $, we are left with:%
\begin{equation*}
\left\langle \frac{KK^{\prime }\exp \left( -\chi _{1}\left\vert
x-y\right\vert \right) }{\left\langle K\right\rangle _{x}\left\langle
K\right\rangle _{y}}\left\vert \Psi \left( K^{\prime },y\right) \right\vert
^{2}\right\rangle \simeq \frac{K}{2\left\langle K\right\rangle _{x}}\int
\exp \left( -\chi _{1}\left\vert x-y\right\vert \right) dy
\end{equation*}%
This last integral is:%
\begin{eqnarray*}
&&\int_{-1}^{1}\exp \left( -\left( \chi _{1}\right) \left\vert
x-y\right\vert \right) dy \\
&=&\int_{-1-x}^{1-x}\exp \left( -\chi _{1}\left\vert u\right\vert \right)
du=\int_{-1-x}^{0}\frac{\exp \left( \chi _{1}u\right) }{\sqrt{\alpha }}%
du+\int_{0}^{1-x}\frac{\exp \left( -\chi _{1}u\right) }{\sqrt{\alpha }}du \\
&=&\frac{\left( 2-\exp \left( -\chi _{1}\left( 1+x\right) \right) -\exp
\left( -\chi _{1}\left( 1-x\right) \right) \right) }{\chi _{1}} \\
&=&\frac{2\left( 1-\exp \left( -\chi _{1}\right) \cosh \left( \chi
_{1}x\right) \right) }{\chi _{1}} \\
&=&\frac{2\left( 1-\exp \left( -\chi _{1}\right) \right) }{\chi _{1}}+\frac{%
2\left( \exp \left( -\chi _{1}\right) \right) \left( 1-\cosh \left( \chi
_{1}x\right) \right) }{\chi _{1}} \\
&\simeq &\frac{2\left( 1-\exp \left( -\chi _{1}\right) \right) }{\chi _{1}}%
-\exp \left( -\chi _{1}\right) \chi _{1}x^{2}
\end{eqnarray*}%
By the same token, we find the evaluation of the second part of the
potential:%
\begin{eqnarray*}
&&\int \kappa _{2}\left\langle \exp \left( -\chi _{2}\left\vert
x-y\right\vert -\chi _{2}\left\vert x-z\right\vert -\chi _{2}\left\vert
y-z\right\vert \right) \left\vert \Psi \left( x\right) \right\vert
^{2}\left\vert \Psi \left( y\right) \right\vert ^{2}\left\vert \Psi \left(
z\right) \right\vert ^{2}\right\rangle \\
&\rightarrow &\int \kappa _{2}\exp \left( -\chi _{2}\left\vert
x-y\right\vert -\chi _{2}\left\vert x-z\right\vert -\chi _{2}\left\vert
y-z\right\vert \right) G\left( x,y\right) G\left( y,z\right) G\left(
z,x\right) dydz \\
&\simeq &\int \kappa _{2}\exp \left( -\chi _{2}\left( \left\vert
x-y\right\vert +\left\vert x-z\right\vert +\left\vert y-z\right\vert \right)
\right) dydz \\
&\simeq &\int \kappa _{2}\exp \left( -\chi _{2}\left( \left\vert
x-y\right\vert +\left\vert y-x\right\vert \right) \right) dy\simeq \kappa
_{2}\frac{2\left( 1-\exp \left( -\chi _{2}\right) \right) }{\chi _{2}}%
-\kappa _{2}\exp \left( -\chi _{2}\right) \chi _{2}x^{2}
\end{eqnarray*}%
Under these approximations, the $X$ part of the action becomes:%
\begin{equation*}
\left( \nabla \Psi \left( x\right) \right) ^{2}+\omega _{X}\Psi ^{\dag
}\left( x\right) x^{2}\Psi \left( x\right) +\alpha _{X}\Psi ^{\dag }\left(
x\right) \Psi \left( x\right)
\end{equation*}%
where:%
\begin{equation*}
\omega _{X}=\kappa _{0}+\kappa _{1}\frac{K}{\left\langle K\right\rangle _{X}}%
\exp \left( -\chi _{1}\right) \chi _{1}-\kappa _{2}\exp \left( -\chi
_{2}\right) \chi _{2}
\end{equation*}%
and:%
\begin{equation*}
\alpha _{X}=\alpha -\kappa _{1}\frac{K}{\left\langle K\right\rangle _{X}}%
\frac{2\left( 1-\exp \left( -\chi _{1}\right) \right) }{\chi _{1}}+\kappa
_{2}\frac{2\left( 1-\exp \left( -\chi _{2}\right) \right) }{\chi _{2}}
\end{equation*}%
As a consequence, the overall second order action becomes:%
\begin{equation}
\Psi ^{\dag }\left( K,X,\theta \right) \left( -\sigma ^{2}\nabla
_{K}^{2}-\vartheta ^{2}\nabla _{\theta }^{2}-\sigma _{X}^{2}\nabla
_{X}^{2}+\left( \delta ^{2}+\frac{\bar{A}^{2}A^{2}}{\left( A^{2}U^{2}+\bar{A}%
^{2}\right) ^{2}}\right) \left( K-\left\langle K\right\rangle \right)
^{2}+\omega _{X}x^{2}+\frac{1}{\vartheta ^{2}}+\alpha _{X}\right) \Psi
\left( K,X,\theta \right)  \label{ctnfp}
\end{equation}%
Ultimately, the Green function are modified by the change of variable (\ref%
{chngvr}) including a factor:%
\begin{equation*}
\exp \left( -\int \left( \frac{\left( \delta K-APK^{\alpha }+V_{1}\right) }{%
\sigma ^{2}}\right) \right) \Psi \left( K,P,X,\theta \right)
\end{equation*}%
with: 
\begin{equation*}
V_{1}=\kappa APK^{\alpha }\int \frac{P_{3}\exp \left( -\left( \left\vert
X_{1}-X_{2}\right\vert +\left\vert X_{1}-X_{3}\right\vert \right) \right) }{%
P_{2}^{\gamma }}\left\vert \Psi \left( K_{2},P_{2},X_{2},\theta \right)
\right\vert ^{2}\left\vert \Psi \left( K_{3},P_{3},X_{3},\theta \right)
\right\vert ^{2}
\end{equation*}%
Given that:%
\begin{eqnarray*}
&&\int APK^{\alpha }-V_{1} \\
&\simeq &\int APK^{\alpha }\left( 1-\kappa \left\langle \int \frac{P_{3}\exp
\left( -\left( \left\vert X_{1}-X_{2}\right\vert +\left\vert
X_{1}-X_{3}\right\vert \right) \right) }{P_{2}^{\gamma }}\left\vert \Psi
\left( K_{2},P_{2},X_{2},\theta \right) \right\vert ^{2}\left\vert \Psi
\left( K_{3},P_{3},X_{3},\theta \right) \right\vert ^{2}\right\rangle \right)
\\
&=&\int \kappa APK^{\alpha }U
\end{eqnarray*}%
the exponential factor rewrites:%
\begin{eqnarray*}
\exp \left( -\int \left( \frac{\left( \delta K-APK^{\alpha }+V_{1}\right) }{%
\sigma ^{2}}\right) \right) &=&\exp \left( -\int \left( \frac{\left( \delta
K-APK^{\alpha }U\right) }{\sigma ^{2}}\right) \right) \\
&=&\exp \left( -\int \left( \frac{\left( \delta K-A\left( K\right) ^{\frac{%
\alpha \gamma }{1+\gamma }}f\left( X\right) U\right) }{\sigma ^{2}}\right)
\right)
\end{eqnarray*}%
We have seen in Appendix 2 that:%
\begin{equation*}
\delta \left\langle K\right\rangle _{X}-A\left( \left\langle K\right\rangle
_{X}\right) ^{\frac{\alpha \gamma }{1+\gamma }}f\left( X\right) U=0
\end{equation*}%
so that we can rewrite the term in the exponential as:%
\begin{eqnarray*}
\frac{\delta }{\sigma ^{2}}\int \left( K-\left( \frac{K}{\left\langle
K\right\rangle _{X}}\right) ^{\frac{\alpha \gamma }{1+\gamma }}\left\langle
K\right\rangle _{X}\right) &\simeq &\frac{\delta }{\sigma ^{2}}\int \left(
K-\left\langle K\right\rangle _{X}\right) \\
&=&\frac{\left( K-\left\langle K\right\rangle _{X}\right) ^{2}}{2\sigma ^{2}}
\end{eqnarray*}%
for $\gamma <<1$. The action (\ref{ctnfp}) is now quadratic, but the
variables $K$ and $X$ are entangled through $\omega $ and $\omega _{X}$. To
find the Green function between $K,X$ and $K^{\prime },X^{\prime }$ one can
simplify the problem by replacing $K,X$\ in $\omega $ and $\omega _{X}$ by
their average trajectory values. In first approximation this means to
replace $K,X$ with $\frac{\left\langle K\right\rangle _{X}+\left\langle
K\right\rangle _{X^{\prime }}}{2},\frac{X+X^{\prime }}{2}$. Then we set:%
\begin{equation*}
\left\langle K\right\rangle =\frac{\left\langle K\right\rangle
_{X}+\left\langle K\right\rangle _{X^{\prime }}}{2}
\end{equation*}%
\begin{equation*}
\bar{\omega}_{X}=\kappa _{0}+\frac{\kappa _{1}}{2}\left( \frac{K}{%
\left\langle K\right\rangle _{X}}+\frac{K^{\prime }}{\left\langle
K\right\rangle _{Y}}\right) \exp \left( -\chi _{1}\right) \chi _{1}-\kappa
_{2}\exp \left( -\chi _{2}\right) \chi _{2}
\end{equation*}%
\begin{equation*}
\bar{\alpha}_{X}=\alpha +\frac{1}{\vartheta ^{2}}-\frac{\kappa _{1}}{2}%
\left( \frac{K}{\left\langle K\right\rangle _{X}}+\frac{K^{\prime }}{%
\left\langle K\right\rangle _{Y}}\right) \frac{2\left( 1-\exp \left( -\chi
_{1}\right) \right) }{\chi _{1}}+\kappa _{2}\frac{2\left( 1-\exp \left(
-\chi _{2}\right) \right) }{\chi _{2}}
\end{equation*}%
\begin{eqnarray*}
\bar{U} &=&\left( 1-\allowbreak \frac{\left( \gamma +1\right) ^{2}}{\left(
\gamma +2\right) \left( 2\gamma +1\right) }\left( \frac{1+\gamma }{4}\left(
\left( \frac{2}{1+\gamma }\right) ^{2}-1\right) \right) ^{\frac{1+\gamma }{%
1-\gamma }}\exp \left( -\left\vert \frac{X+X^{\prime }}{2}\right\vert
\right) \right) \\
\bar{\omega} &=&\delta ^{2}+\frac{\bar{A}^{2}A^{2}}{\left( A^{2}\bar{U}^{2}+%
\bar{A}^{2}\right) ^{2}}
\end{eqnarray*}%
and the Green function of the action (\ref{ctnfp}) is the Laplace transform
with parameter $\bar{\alpha}_{X}\simeq \alpha $ for small coupling
parameters, of the following temporal transition function : 
\begin{eqnarray*}
&&G\left( K,K^{\prime },P,P^{\prime },X,X^{\prime },\theta ,\theta ^{\prime
},t\right) \\
&=&\exp \left( -\left[ \frac{\left( K-\left\langle K\right\rangle
_{X}\right) ^{2}}{2\sigma ^{2}}\right] _{\left( K,X\right) }^{\left(
K^{\prime },X^{\prime }\right) }\right) \\
&&\times \sqrt{\frac{\bar{\omega}}{2\pi \sigma ^{2}\sinh \left( \bar{\omega}%
t\right) }}\exp \left( -\frac{\bar{\omega}\left( \left( \left(
K-\left\langle K\right\rangle \right) ^{2}+\left( K^{\prime }-\left\langle
K\right\rangle \right) ^{2}\right) \cosh \left( \bar{\omega}t\right)
-2\left( K-\left\langle K\right\rangle \right) \left( K^{\prime
}-\left\langle K\right\rangle \right) \right) }{2\sigma ^{2}\sinh \left( 
\bar{\omega}t\right) }\right) \\
&&\times \sqrt{\frac{\bar{\omega}_{X}}{2\pi \sinh \left( \bar{\omega}%
_{X}t\right) }}\exp \left( -\frac{\bar{\omega}_{X}\left( \left( X^{2}+\left(
X^{\prime }\right) ^{2}\right) \cosh \left( \bar{\omega}_{X}t\right)
-2XX^{\prime }\right) }{2\sinh \left( \bar{\omega}_{X}t\right) }\right) \\
&&\times \sqrt{\frac{1}{2\pi \vartheta ^{2}t}}\exp \left( -\frac{\left(
\theta -\theta ^{\prime }-t\right) ^{2}}{2\vartheta ^{2}t}\right) \times
\delta \left( P-\frac{D\exp \left( -\frac{\left\vert X\right\vert }{1+\gamma 
}\right) }{\left( K\right) ^{\frac{\alpha }{1+\gamma }}}\right) \delta
\left( P^{\prime }-\frac{D\exp \left( -\frac{\left\vert X^{\prime
}\right\vert }{1+\gamma }\right) }{\left( K^{\prime }\right) ^{\frac{\alpha 
}{1+\gamma }}}\right)
\end{eqnarray*}%
$\bigskip $For $\vartheta ^{2}<<1$, The variable $t$ can be replaced by $%
\theta -\theta ^{\prime }$, for $\theta >\theta ^{\prime }$. Actually, due
to the term:%
\begin{equation}
\exp \left( -\frac{\left( \theta -\theta ^{\prime }-t\right) ^{2}}{%
2\vartheta ^{2}t}\right)  \label{dmn}
\end{equation}%
the Green function is non nul for values of $\theta $ and $\theta ^{\prime }$
such that $\theta -\theta ^{\prime }-t=0$. Since $t>0$, this implies that
the replacement is only valid for $\theta >\theta ^{\prime }$, otherwise the
Green function is equal to $0$. As a consequence, we can remove the time
dependence in the Green function and we obtain: 
\begin{eqnarray*}
&&G\left( K,K^{\prime },P,P^{\prime },X,X^{\prime },\theta ,\theta ^{\prime
}\right) \\
&=&\exp \left( -\left[ \frac{\left( K-\left\langle K\right\rangle
_{X}\right) ^{2}}{2\sigma ^{2}}\right] _{\left( K,X\right) }^{\left(
K^{\prime },X^{\prime }\right) }\right) \\
&&\times \sqrt{\frac{\bar{\omega}}{2\pi \sigma ^{2}\sinh \left( \bar{\omega}%
\left( \theta -\theta ^{\prime }\right) \right) }}\exp \left( -\frac{\bar{%
\omega}\left( \left( \left( K-\left\langle K\right\rangle \right)
^{2}+\left( K^{\prime }-\left\langle K\right\rangle \right) ^{2}\right)
\cosh \left( \bar{\omega}\left( \theta -\theta ^{\prime }\right) \right)
-2\left( K-\left\langle K\right\rangle \right) \left( K^{\prime
}-\left\langle K\right\rangle \right) \right) }{2\sigma ^{2}\sinh \left( 
\bar{\omega}\left( \theta -\theta ^{\prime }\right) \right) }\right) \\
&&\times \sqrt{\frac{\bar{\omega}_{X}}{2\pi \sinh \left( \bar{\omega}%
_{X}\left( \theta -\theta ^{\prime }\right) \right) }}\exp \left( -\frac{%
\bar{\omega}_{X}\left( \left( X^{2}+\left( X^{\prime }\right) ^{2}\right)
\cosh \left( \bar{\omega}_{X}\left( \theta -\theta ^{\prime }\right) \right)
-2XX^{\prime }\right) }{2\sinh \left( \bar{\omega}_{X}\left( \theta -\theta
^{\prime }\right) \right) }\right) \\
&&\times \delta \left( P-\frac{D\exp \left( -\frac{\left\vert X\right\vert }{%
1+\gamma }\right) }{\left( K\right) ^{\frac{\alpha }{1+\gamma }}}\right)
\delta \left( P^{\prime }-\frac{D\exp \left( -\frac{\left\vert X^{\prime
}\right\vert }{1+\gamma }\right) }{\left( K^{\prime }\right) ^{\frac{\alpha 
}{1+\gamma }}}\right) H\left( \theta -\theta ^{\prime }\right)
\end{eqnarray*}

\bigskip where $H\left( \theta -\theta ^{\prime }\right) $ is the Heaviside
function.

Note that for $\theta -\theta ^{\prime }=0$, in (\ref{dmn}), the dominant
part becomes $\frac{t}{2\vartheta ^{2}}$, which means that in average one
can replace $t$ by $\vartheta ^{2}$, which describes the description in
terms of harmonic oscillators used in Appendix 2.

\subsubsection{Case $\protect\rho \neq 0$}

For $\rho \neq 0$, the $X$ part \bigskip of the action is similar to the
previous case, in 
\begin{eqnarray*}
&&\int \Psi ^{\dag }\left( K,X\right) \left( -\nabla _{X}^{2}+\kappa
_{0}X^{2}\right) \Psi \left( K,X\right) -\frac{\kappa _{1}}{2}\left\vert
\Psi \left( K,X\right) \right\vert ^{2}\frac{KK^{\prime }\exp \left( -\chi
_{1}\left\vert X-Y\right\vert \right) }{\left\langle K\right\rangle ^{2}}%
\left\vert \Psi \left( K^{\prime },Y\right) \right\vert ^{2} \\
&&+\int \frac{\kappa _{2}}{3}\exp \left( -\chi _{2}\left\vert X-Y\right\vert
-\chi _{2}\left\vert X-Z\right\vert -\chi _{2}\left\vert Y-Z\right\vert
\right) \left\vert \Psi \left( K,X\right) \right\vert ^{2}\left\vert \Psi
\left( K^{\prime },Y\right) \right\vert ^{2}\left\vert \Psi \left(
K^{"},Z\right) \right\vert ^{2}
\end{eqnarray*}%
we will replace the interaction potential by its average in variables $y$
and $z$, but now, this average is computed in the fundamental state $\Psi
_{0}$. \ We found that the norm of $\Psi _{0}$ satisfies at the zeroth order
in $\chi _{1}$ and $\chi _{2}$: 
\begin{equation}
\rho ^{2}=\frac{\kappa _{1}+\sqrt{\kappa _{1}^{2}-2\kappa _{2}\left( 2\alpha
+\frac{2}{\vartheta ^{2}}+\sqrt{\kappa _{0}}+\sigma \delta \right) }}{%
2\kappa _{2}}  \label{rhosqp}
\end{equation}%
for $\bar{A}^{2}>>A^{2}$. \ It will be useful in the sequel to find a more
precise form for $\rho ^{2}$ at the second order in $\chi _{1}$ and $\chi
_{2}$. At this order, the equation for the fundamental $\Psi _{0}$ can be
found in the following way. We developp the potential terms to second order
in $\Psi \left( K,x\right) $ around $\Psi _{0}$, that is:%
\begin{eqnarray*}
&&\int \frac{\kappa _{1}}{2}\left\vert \Psi \left( K,X\right) \right\vert
^{2}\frac{KK^{\prime }\exp \left( -\chi _{1}\left\vert X-Y\right\vert
\right) }{\left\langle K\right\rangle ^{2}}\left\vert \Psi \left( K^{\prime
},Y\right) \right\vert ^{2}dY \\
&\rightarrow &\int \kappa _{1}\left\vert \Psi \left( K,X\right) \right\vert
^{2}\frac{KK^{\prime }\exp \left( -\chi _{1}\left\vert X-Y\right\vert
\right) }{\left\langle K\right\rangle ^{2}}\left\vert \Psi _{0}\left(
K^{\prime },Y\right) \right\vert ^{2}dY \\
&\simeq &\kappa _{1}\exp \left( -\chi _{1}\left\vert X\right\vert \right)
\left\vert \Psi \left( K,X\right) \right\vert ^{2}\frac{K}{\left\langle
K\right\rangle }\rho ^{2} \\
&\simeq &\kappa _{1}\left( \frac{K}{\left\langle K\right\rangle }\exp \left(
-\chi _{1}\left\vert X\right\vert \right) \left\vert \Psi \left( K,X\right)
\right\vert ^{2}\right) \rho ^{2}
\end{eqnarray*}

\bigskip The parameter $\rho ^{2}$ has been computed before. Similarly, the
second part of the potential is replaced by: 
\begin{eqnarray*}
&&\frac{\kappa _{2}}{3}\left\vert \Psi _{0}\left( K,x\right) \right\vert
^{2}\exp \left( -\chi _{2}\left\vert x-y\right\vert -\chi _{2}\left\vert
x-z\right\vert -\chi _{2}\left\vert y-z\right\vert \right) \left\vert \Psi
\left( K^{\prime },y\right) \right\vert ^{2}\left\vert \Psi ^{\dag }\left(
K",z\right) \right\vert ^{2} \\
&\rightarrow &\int \kappa _{2}\exp \left( -\chi _{2}\left\vert
x-y\right\vert -\chi _{2}\left\vert x-z\right\vert -\chi _{2}\left\vert
y-z\right\vert \right) \left\vert \Psi _{0}\left( K,x\right) \right\vert
^{2}\left\vert \Psi _{0}\left( K^{\prime },y\right) \right\vert
^{2}\left\vert \Psi _{0}\left( K",z\right) \right\vert ^{2}dK^{\prime
}dK"dydz \\
&\simeq &\kappa _{2}\exp \left( -2\chi _{2}\left\vert x\right\vert \right)
\left\vert \Psi _{0}\left( K,x\right) \right\vert ^{2}\rho ^{4}
\end{eqnarray*}

\begin{equation*}
-\nabla ^{2}\Psi \left( x\right) +\frac{\kappa _{0}}{2}x^{2}\Psi \left(
x\right) -\kappa _{1}\left( \frac{K}{\left\langle K\right\rangle }\exp
\left( -\chi _{1}\left\vert x\right\vert \right) \left\vert \Psi \left(
K,x\right) \right\vert ^{2}\right) \rho ^{2}+\kappa _{2}\exp \left( -2\chi
_{2}\left\vert x\right\vert \right) \left\vert \Psi \left( K,x\right)
\right\vert ^{2}\rho ^{4}+\alpha \Psi \left( x\right)
\end{equation*}

expanding the exponentials in $\chi _{1}$ and $\chi _{2}$\ yields thus the
first approximation to the potential of the $X$ (\ref{Ptscd}), and thus a
second order action in $\chi _{1}$ and $\chi _{2}$:%
\begin{eqnarray*}
&&\Psi ^{\dag }\left( x\right) \left( -\sigma _{X}^{2}\nabla _{X}^{2}+\frac{%
\kappa _{0}}{2}\left( 1-\frac{\chi _{1}^{2}\kappa _{1}\rho ^{2}}{2\kappa _{0}%
}+2\chi _{2}\frac{\kappa _{2}}{\kappa _{0}}\rho ^{4}\right) \left( x+\frac{%
sgn\left( x\right) }{\kappa _{0}}\left( \chi _{1}\kappa _{1}\frac{K}{%
2\left\langle K\right\rangle }\rho ^{2}-\chi _{2}\kappa _{2}\rho ^{4}\right)
\right) ^{2}\right) \Psi \left( x\right) \\
&&+\left( \rho ^{4}\kappa _{2}-\rho ^{2}\kappa _{1}-\frac{1}{2\kappa _{0}}%
\left( \chi _{1}\kappa _{1}\frac{K}{2\left\langle K\right\rangle }\rho
^{2}-\chi _{2}\kappa _{2}\rho ^{4}\right) ^{2}\right) \left( \Psi ^{\dag
}\left( x\right) \Psi \left( x\right) \right)
\end{eqnarray*}%
The complete action is then:%
\begin{eqnarray*}
&&\int \Psi ^{\dag }\left( K,X,\theta \right) \left( -\sigma ^{2}\nabla
_{K}^{2}-\vartheta ^{2}\nabla _{\theta }^{2}-\sigma _{X}^{2}\nabla
_{X}^{2}+\left( \delta ^{2}+\frac{\bar{A}^{2}A^{2}}{\left( A^{2}U^{2}+\bar{A}%
^{2}\right) ^{2}}\right) \left( K-\left\langle K\right\rangle \right)
^{2}\right) \Psi \left( K,X,\theta \right) \\
&&+\frac{\kappa _{0}}{2}\left( 1-\frac{\chi _{1}^{2}\kappa _{1}\rho ^{2}}{%
2\kappa _{0}}+2\chi _{2}\frac{\kappa _{2}}{\kappa _{0}}\rho ^{4}\right) \int
\left( \left( x+\frac{sgn\left( x\right) }{\kappa _{0}}\left( \chi
_{1}\kappa _{1}\frac{K}{2\left\langle K\right\rangle }\rho ^{2}-\chi
_{2}\kappa _{2}\rho ^{4}\right) \right) ^{2}x^{2}+\alpha _{X}\right)
\left\vert \Psi \left( K,X,\theta \right) \right\vert ^{2} \\
&&+\left( 2\rho ^{2}\kappa _{2}-\kappa _{1}\right) \left\vert \int \Psi
^{\dag }\left( K,X,\theta \right) \Psi _{0}\left( K,X,\theta \right)
\right\vert ^{2}
\end{eqnarray*}%
where%
\begin{equation*}
\alpha _{X}=\alpha +\frac{1}{\vartheta ^{2}}+\left( -\kappa _{1}\rho
^{2}+\kappa _{2}\rho ^{4}-\frac{1}{2\kappa _{0}}\left( \chi _{1}\kappa _{1}%
\frac{K}{2\left\langle K\right\rangle }\rho ^{2}-\chi _{2}\kappa _{2}\rho
^{4}\right) ^{2}\right)
\end{equation*}%
since (\ref{rhqn}) implies that: 
\begin{eqnarray*}
0 &=&\alpha +\frac{1}{\vartheta ^{2}}+\frac{1}{2}\kappa _{0}^{\frac{1}{2}}%
\sqrt{1-\frac{\chi _{1}^{2}\kappa _{1}\rho ^{2}}{2}+2\chi _{2}^{2}\kappa
_{2}\rho ^{4}} \\
&&-\kappa _{1}\rho ^{2}+\kappa _{2}\rho ^{4}-\frac{1}{2\kappa _{0}}\left(
\chi _{1}\kappa _{1}\frac{K}{2\left\langle K\right\rangle }\rho ^{2}-\chi
_{2}\kappa _{2}\rho ^{4}\right) ^{2}+\frac{\sigma \sqrt{\delta ^{2}+\frac{%
\bar{A}^{2}A^{2}}{\left( A^{2}U^{2}+\bar{A}^{2}\right) ^{2}}}}{2}
\end{eqnarray*}%
one has:%
\begin{equation*}
\alpha _{X}=-\frac{1}{2}\kappa _{0}^{\frac{1}{2}}\sqrt{1-\frac{\chi
_{1}^{2}\kappa _{1}\rho ^{2}}{2}+2\chi _{2}^{2}\kappa _{2}\rho ^{4}}-\frac{%
\sigma \sqrt{\delta ^{2}+\frac{\bar{A}^{2}A^{2}}{\left( A^{2}U^{2}+\bar{A}%
^{2}\right) ^{2}}}}{2}
\end{equation*}%
and the second order action rewrites:%
\begin{eqnarray}
&&\int \Psi ^{\dag }\left( K,X,\theta \right) \left( -\sigma ^{2}\nabla
_{K}^{2}-\vartheta ^{2}\nabla _{\theta }^{2}-\sigma _{X}^{2}\nabla
_{X}^{2}+\left( \delta ^{2}+\frac{\bar{A}^{2}A^{2}}{\left( A^{2}U^{2}+\bar{A}%
^{2}\right) ^{2}}\right) \left( K-\left\langle K\right\rangle \right)
^{2}\right) \Psi \left( K,X,\theta \right)  \label{ctns} \\
&&+\frac{\kappa _{0}}{2}\left( 1-\frac{\chi _{1}^{2}\kappa _{1}\rho ^{2}}{%
2\kappa _{0}}+2\chi _{2}\frac{\kappa _{2}}{\kappa _{0}}\rho ^{4}\right) \int
\Psi ^{\dag }\left( K,X,\theta \right) \left( \left( x+\frac{sgn\left(
x\right) }{\kappa _{0}}\left( \chi _{1}\kappa _{1}\frac{K}{2\left\langle
K\right\rangle }\rho ^{2}-\chi _{2}\kappa _{2}\rho ^{4}\right) \right)
^{2}x^{2}\right) \Psi \left( K,X,\theta \right)  \notag \\
&&-\int \Psi ^{\dag }\left( K,X,\theta \right) \left( \frac{1}{2}\kappa
_{0}^{\frac{1}{2}}\sqrt{1-\frac{\chi _{1}^{2}\kappa _{1}\rho ^{2}}{2}+2\chi
_{2}^{2}\kappa _{2}\rho ^{4}}+\frac{\sigma \sqrt{\delta ^{2}+\frac{\bar{A}%
^{2}A^{2}}{\left( A^{2}U^{2}+\bar{A}^{2}\right) ^{2}}}}{2}\right) \Psi
\left( K,X,\theta \right)  \notag \\
&&+\left( 2\rho ^{2}\kappa _{2}-\kappa _{1}\right) \left\vert \int \Psi
^{\dag }\left( K,P,X,\theta \right) \Psi _{0}\left( K,P,X,\theta \right)
\right\vert ^{2}  \notag
\end{eqnarray}

\bigskip As said before, the fundamental level of the $X$ part of the action
has the form:%
\begin{eqnarray*}
\Psi _{0}\left( X\right) &=&N_{1}\left( \exp \left( -\frac{\kappa _{0}^{-%
\frac{1}{2}}\left( X+\delta X\right) ^{2}}{2}\right) H\left( x\right) +\exp
\left( -\frac{\kappa _{0}^{-\frac{1}{2}}\left( X-\delta X\right) ^{2}}{2}%
\right) H\left( -x\right) \right) \\
&\equiv &\bar{\Psi}_{0}\left( X+\delta X\right) H\left( X\right) +\bar{\Psi}%
_{0}\left( X-\delta X\right) H\left( -X\right)
\end{eqnarray*}%
where:%
\begin{equation*}
\delta X=\frac{1}{\kappa _{0}}\left( \chi _{1}\kappa _{1}\frac{K}{%
2\left\langle K\right\rangle }\rho ^{2}-\chi _{2}\kappa _{2}\rho ^{4}\right)
\end{equation*}%
For a given value of $K$, The Green function can be obtained by its
expansion in function of all the eigenstates $\bar{\Psi}_{n}$ of the system.%
\begin{eqnarray*}
G_{K}\left( x,y\right) &=&\sum_{n}\left( \bar{\Psi}_{n}\left( x+\delta
X\right) H\left( x\right) +\bar{\Psi}_{n}\left( x-\delta X\right) H\left(
-x\right) \right) \times \left( \bar{\Psi}_{n}\left( y+\delta X\right)
H\left( y\right) +\bar{\Psi}_{n}\left( y-\delta X\right) H\left( -y\right)
\right) \\
&=&\bar{G}\left( x+\delta X,y+\delta X\right) H\left( x\right) H\left(
y\right) +\bar{G}\left( x-\delta X,y-\delta X\right) H\left( -x\right)
H\left( -y\right) \\
&&+\bar{G}\left( x+\delta X,y-\delta X\right) H\left( x\right) H\left(
-y\right) +\bar{G}\left( x-\delta X,y+\delta X\right) H\left( -x\right)
H\left( y\right)
\end{eqnarray*}%
where we have defined:%
\begin{eqnarray*}
\bar{G}\left( x,y,t\right) &=&\sqrt{\frac{\kappa _{0}}{2\pi \sigma
_{X}^{2}\sinh \left( \kappa _{0}t\right) }}\exp \left( \left( -\frac{\kappa
_{0}}{2\sigma _{X}^{2}\sinh \left( \kappa _{0}t\right) }\right) \left(
\left( x^{2}+\left( x^{\prime }\right) ^{2}\right) \cosh \left( \kappa
_{0}t\right) -2xx^{\prime }\right) \right) \\
&=&\sqrt{\frac{\kappa _{0}}{2\pi \sigma _{X}^{2}\sinh \left( \kappa
_{0}t\right) }}\exp \left( \left( -\frac{\kappa _{0}}{2\sigma _{X}^{2}\sinh
\left( \kappa _{0}t\right) }\right) \left( \left( x-x^{\prime }\right)
^{2}+\left( \cosh \left( \kappa _{0}t\right) -1\right) \left( x^{2}+\left(
x^{\prime }\right) ^{2}\right) \right) \right)
\end{eqnarray*}%
The term associated to the change of variable can be found similarly to the
first phase:, but now it has to evaluated in the state $\Psi _{0}\left(
K,X,\theta \right) +\Psi \left( K,X,\theta \right) $. As explained in
Appendix 3, in first approximation, this is equivalent to evaluate it in the
state $\Psi _{0}\left( K,X,\theta \right) $. We find again an exponential
factor:%
\begin{equation*}
\exp \left( -\left[ \frac{\left( K-\left\langle K\right\rangle _{X}\right)
^{2}}{2\sigma ^{2}}_{\left( K,X\right) }^{\left( K^{\prime },X^{\prime
}\right) }\right] _{\left( K,X\right) }^{\left( K^{\prime },X^{\prime
}\right) }\right)
\end{equation*}%
with $\left\langle K\right\rangle _{X}$ computed in the state $\Psi
_{0}\left( K,X,\theta \right) $ as in Appendix 3.

\bigskip Ultimately, as in the first phase, to (\ref{ctns}) one can
associate a Green function that is the Laplace transform with parameter $%
\alpha _{X}$ of a temporal Green function:

\begin{eqnarray*}
&&G\left( K,K^{\prime },P,P^{\prime },X,X^{\prime },\theta ,\theta ^{\prime
},t\right) \\
&=&\exp \left( -\left[ \frac{\left( K-\left\langle K\right\rangle
_{X}\right) ^{2}}{2\sigma ^{2}}_{\left( K,X\right) }^{\left( K^{\prime
},X^{\prime }\right) }\right] _{\left( K,X\right) }^{\left( K^{\prime
},X^{\prime }\right) }\right) \\
&&\times \sqrt{\frac{\bar{\omega}}{2\pi \sigma ^{2}\sinh \left( \bar{\omega}%
t\right) }}\exp \left( -\frac{\bar{\omega}\left( \left( \left(
K-\left\langle K\right\rangle \right) ^{2}+\left( K^{\prime }-\left\langle
K\right\rangle \right) ^{2}\right) \cosh \left( \bar{\omega}t\right)
-2\left( K-\left\langle K\right\rangle \right) \left( K^{\prime
}-\left\langle K\right\rangle \right) \right) }{2\sigma ^{2}\sinh \left( 
\bar{\omega}t\right) }\right) \\
&&\times G_{\frac{K+K^{\prime }}{2}}\left( X,Y,t\right) \\
&&\times \sqrt{\frac{1}{2\pi \vartheta ^{2}t}}\exp \left( -\frac{\left(
\theta -\theta ^{\prime }\right) ^{2}}{2\vartheta ^{2}t}+\frac{\theta
-\theta ^{\prime }}{\vartheta ^{2}}\right) \times \delta \left( P-\frac{%
D\exp \left( -\frac{\left\vert X\right\vert }{1+\gamma }\right) }{\left( 
\frac{K}{\left\langle K\right\rangle }\right) ^{\frac{\alpha }{1+\gamma }}}%
\right) \times \delta \left( P^{\prime }-\frac{D\exp \left( -\frac{%
\left\vert X^{\prime }\right\vert }{1+\gamma }\right) }{\left( \frac{%
K^{\prime }}{\left\langle K\right\rangle }\right) ^{\frac{\alpha }{1+\gamma }%
}}\right)
\end{eqnarray*}

Eventhough $\alpha _{X}<0$, the Laplace transform is well defined since $%
-\alpha _{X}$ is the lower bound of the terms in the exponential ($-\alpha
_{X}$ is the lowest eigenvalue associated to the evolution operator whose $G$
is the Green function).

As in phase 1, this Green function is centered around $t=\theta -\theta
^{\prime }$, so that we can replace $t=\theta -\theta ^{\prime }$ in the
Green function, leads to:

\begin{eqnarray*}
&&G\left( K,K^{\prime },P,P^{\prime },X,X^{\prime },\theta ,\theta ^{\prime
}\right) \\
&=&\exp \left( -\left[ \frac{\left( K-\left\langle K\right\rangle
_{X}\right) ^{2}}{2\sigma ^{2}}_{\left( K,X\right) }^{\left( K^{\prime
},X^{\prime }\right) }\right] _{\left( K,X\right) }^{\left( K^{\prime
},X^{\prime }\right) }\right) \\
&&\times \sqrt{\frac{\bar{\omega}}{2\pi \sigma ^{2}\sinh \left( \bar{\omega}%
\left( \theta -\theta ^{\prime }\right) \right) }}\exp \left( -\frac{\bar{%
\omega}\left( \left( \left( K-\left\langle K\right\rangle \right)
^{2}+\left( K^{\prime }-\left\langle K\right\rangle \right) ^{2}\right)
\cosh \left( \bar{\omega}\left( \theta -\theta ^{\prime }\right) \right)
-2\left( K-\left\langle K\right\rangle \right) \left( K^{\prime
}-\left\langle K\right\rangle \right) \right) }{2\sigma ^{2}\sinh \left( 
\bar{\omega}\left( \theta -\theta ^{\prime }\right) \right) }\right) \\
&&\times G_{\frac{K+K^{\prime }}{2}}\left( X,Y\right) \times \delta \left( P-%
\frac{D\exp \left( -\frac{\left\vert X\right\vert }{1+\gamma }\right) }{%
\left( \frac{K}{\left\langle K\right\rangle }\right) ^{\frac{\alpha }{%
1+\gamma }}}\right) \times \delta \left( P^{\prime }-\frac{D\exp \left( -%
\frac{\left\vert X^{\prime }\right\vert }{1+\gamma }\right) }{\left( \frac{%
K^{\prime }}{\left\langle K\right\rangle }\right) ^{\frac{\alpha }{1+\gamma }%
}}\right) H\left( \theta -\theta ^{\prime }\right)
\end{eqnarray*}

\subsection{\protect\bigskip \textbf{Appendix 5}}

We consider a general model:

\begin{equation*}
S\left( \Psi \right) =\int \Psi ^{\dag }\left( X\right) \left( -\sigma
_{X}^{2}\nabla _{X}^{2}+V\right) \Psi \left( X\right) +\int \frac{1}{2}\Psi
^{\dag }\left( X\right) \Psi ^{\dag }\left( Y\right) W\left( X,Y\right) \Psi
\left( X\right) \Psi \left( Y\right)
\end{equation*}%
that encompasses the model studied in this work. The field $\Psi \left(
X\right) $ depends on an arbitrary number of variables $X$ belonging to some
configuration space, and $W\left( X,Y\right) =W\left( Y,X\right) $. We have
chosen a fourth order interaction term, but a more general choice, such a s
a sum of powers, would not change the result. We assume that there is a non
trivial minimum to the action $S\left( \Psi \right) $, so that the equation: 
\begin{equation*}
\left( -\sigma _{X}^{2}\nabla _{X}^{2}+V\right) \Psi \left( X\right) +\left(
\int \Psi ^{\dag }\left( Y\right) W\left( X,Y\right) \Psi \left( Y\right)
\right) \Psi \left( X\right) =0
\end{equation*}%
has a solution $\rho \Psi _{0}\left( X\right) \neq 0$, and $\Psi _{0}$ of
norm equal to $1$. We want to show that the non trivial vacuum implies to
separate the system into two systems defined on two half-space of the
configuration space.

Given the field $\Psi _{0}$ minimizing the action, the second order
variation of $S\left( \Psi \right) $ is: 
\begin{eqnarray*}
&&\delta \Psi ^{\dag }\left( X\right) \left( -\sigma _{X}^{2}\nabla
_{X}^{2}+V\right) \delta \Psi \left( X\right) +\delta \Psi ^{\dag }\left(
X\right) \left( \int \Psi _{0}^{\dag }\left( Y\right) W\left( X,Y\right)
\Psi _{0}\left( Y\right) \right) \delta \Psi \left( X\right) \\
&&+2\func{Re}\left( \delta \Psi ^{\dag }\left( X\right) \left( \Psi
_{0}^{\dag }\left( Y\right) W\left( X,Y\right) \Psi _{0}\left( X\right)
\right) \delta \Psi \left( Y\right) \right) \\
&=&\delta \Psi ^{\dag }\left( X\right) \left( -\sigma _{X}^{2}\nabla
_{X}^{2}+V\left( X\right) \right) \delta \Psi \left( X\right) +\delta \Psi
^{\dag }\left( X\right) V_{0}\left( X\right) \delta \Psi \left( X\right)
+\delta \Psi ^{\dag }\left( X\right) W_{0}\left( X,Y\right) \delta \Psi
\left( Y\right)
\end{eqnarray*}%
where:%
\begin{eqnarray*}
W_{0}\left( Y,X\right) &=&\Psi _{0}^{\dag }\left( Y\right) W\left(
X,Y\right) \Psi _{0}\left( X\right) \\
W_{0}^{\dag }\left( X,Y\right) &=&W_{0}\left( Y,X\right)
\end{eqnarray*}%
and:%
\begin{equation*}
V_{0}\left( X\right) =\int \Psi _{0}^{\dag }\left( Y\right) W\left(
X,Y\right) \Psi _{0}\left( Y\right)
\end{equation*}%
Now, assume that $\Psi _{0}\left( X\right) $ is peaked around some $X_{0}$,
which is the case in the text for the $K$ part of $\Psi _{0}$, then:%
\begin{eqnarray*}
V_{0}\left( X\right) &\simeq &\int \Psi _{0}^{\dag }\left( Y\right) W\left(
X,X_{0}\right) \Psi _{0}\left( Y\right) =W\left( X,X_{0}\right) \rho ^{2} \\
W_{0}\left( Y,X\right) &\simeq &\rho ^{2}\int \Psi _{0}^{\dag }\left(
X_{0}\right) W\left( X_{0},X_{0}\right) \Psi _{0}\left( X\right) \delta
\left( X-X_{0}\right) \delta \left( Y-X_{0}\right)
\end{eqnarray*}%
and the following contributions of the second order variation becomes:%
\begin{eqnarray*}
\delta \Psi ^{\dag }\left( X\right) V_{0}\left( X\right) \delta \Psi \left(
X\right) &\rightarrow &\rho ^{2}\delta \Psi ^{\dag }\left( X\right) W\left(
X,X_{0}\right) \delta \Psi \left( X\right) \\
\delta \Psi ^{\dag }\left( X\right) W_{0}\left( X,Y\right) \delta \Psi
\left( Y\right) &\rightarrow &\rho ^{2}\left\vert \Psi _{0}\left( X\right)
\right\vert ^{2}W\left( X_{0},X_{0}\right) \delta \Psi ^{\dag }\left(
X_{0}\right) \delta \Psi \left( X_{0}\right)
\end{eqnarray*}%
For $\rho ^{2}W\left( X_{0},X_{0}\right) >>1$, the contributions for fields $%
\delta \Psi $ such that $\delta \Psi \left( X_{0}\right) \neq 0$ are
suppressed. Then the integrals over the $\delta \Psi $ can be limited to
contributions such that $\delta \Psi \left( X_{0}\right) =0$.

This means that $\delta \Psi $ can be decomposed in two parts: 
\begin{equation*}
\delta \Psi \left( X\right) =\delta \Psi _{+}\left( X\right) +\delta \Psi
_{-}\left( X\right)
\end{equation*}%
and $\delta \Psi _{+}\left( X\right) $ and $\delta \Psi _{-}\left( X\right) $
are independent and defined on two half space $X_{\pm }$ respectively and
satisfy:%
\begin{equation*}
\delta \Psi _{\pm }\left( X_{\mp }\right) =0
\end{equation*}%
And the second order action for $\delta \Psi _{\pm }$ becomes: 
\begin{eqnarray*}
&&\delta \Psi ^{\dag }\left( X\right) \left( -\sigma _{X}^{2}\nabla
_{X}^{2}+V\left( X\right) \right) \delta \Psi \left( X\right) +\delta \Psi
^{\dag }\left( X\right) V_{0}\left( X\right) \delta \Psi \left( X\right)
+\delta \Psi ^{\dag }\left( X\right) W_{0}\left( X,Y\right) \delta \Psi
\left( Y\right) \\
&=&\delta \Psi _{\pm }^{\dag }\left( X\right) \left( -\sigma _{X}^{2}\nabla
_{X}^{2}+V\left( X\right) \right) \delta \Psi _{\pm }\left( X\right) +\rho
^{2}\delta \Psi _{\pm }^{\dag }\left( X\right) W\left( X,X_{0}\right) \delta
\Psi _{\pm }\left( X\right)
\end{eqnarray*}%
This action models also two independent fields with constraint $\delta \bar{%
\Psi}_{\pm }\left( X_{\mp }\right) =0$ and defined on the all space, but
subject to a wall potential $H_{\pm }\left( X\right) $, the wall being
defined on the space $X_{\mp }$:%
\begin{equation*}
\delta \bar{\Psi}_{\pm }^{\dag }\left( X\right) \left( -\sigma
_{X}^{2}\nabla _{X}^{2}+V\left( X\right) +H_{\pm }\left( X\right) \right)
\delta \bar{\Psi}_{\pm }\left( X\right) +\rho ^{2}\delta \bar{\Psi}_{\pm
}^{\dag }\left( X\right) W\left( X,X_{0}\right) \delta \bar{\Psi}_{\pm
}\left( X\right)
\end{equation*}%
and this models two sets of different agents, evolving on $X_{\pm }$ and
subject to a "wall" potential. Applied to our case, this result are the
following. Recall that we found the fundamental state (\ref{mnml}):%
\begin{eqnarray*}
\Psi _{0}\left( K,X\right) &=&\rho N\left[ \bar{\Psi}_{0}\left( X+\delta
X\right) H\left( X\right) +\bar{\Psi}_{0}\left( X-\chi \delta X\right)
H\left( -X\right) \right] \\
&&\times \exp \left( -\frac{\sqrt{\delta ^{2}+\frac{\bar{A}^{2}A^{2}}{\left(
A^{2}U^{2}+\bar{A}^{2}\right) ^{2}}}\left( K-\left\langle K\right\rangle
_{X}\right) ^{2}}{2\sigma }\right)
\end{eqnarray*}

where $N$ and $\rho $ are given by (\ref{Nrm}) and (\ref{rhosq}) and (\ref%
{dltX}) respectively. If we assume, as in Appendix 2 that:%
\begin{equation*}
\frac{\sqrt{\delta ^{2}+\frac{\bar{A}^{2}A^{2}}{\left( A^{2}U^{2}+\bar{A}%
^{2}\right) ^{2}}}}{\sigma }>\kappa _{0}^{\frac{1}{2}}
\end{equation*}%
i.e. the $X$ variable is more spread than $K$, then $\Psi _{0}\left(
K,X\right) $ is peaked on the hypersurface $K=\left\langle K\right\rangle
_{X}$. \ As a consequence, the space $\left( K,X\right) $ is divided into
two subspaces $S_{+}$ (defined by $K>\left\langle K\right\rangle _{X}$) and $%
S_{-}$(defined by $K<\left\langle K\right\rangle _{X}$). To these halfspaces
correspond two systems, independent in first approximation. An agent
starting in $S_{+}$ ($S_{-}$ respectively) will remain in $S_{+}$ ($S_{-}$
respectively).\newpage

\section*{References}


\begin{description}
\item Abergel F, Chakraborti A, Muni Toke I and Patriarca M (2011)
Econophysics review: I. Empirical facts, Quantitative Finance, Vol. 11, No.
7, 991-1012.

\item Abergel F, Chakraborti A, Muni Toke I and Patriarca M (2011)
Econophysics review: II. Agent-based models, Quantitative Finance, Vol. 11,
No. 7, 1013-1041.

\item Barro RJ, Sala-i-Martin X (1995) Economic Growth. McGraw-Hill,
New-York.

\item Ciarli T, Lorentz A, Savona M, Valente M (2010) The effect of
consumption and production structure on growth and distribution. A micro to
macro model. Metroeconomica, 61(1):180.

\item Dawid H, Gemkow S, Harting P, van der Hoog S, Neugart M (2011) An
agent-based macroeconomic model for economic policy analysis. Technical
report, Working paper. Universitat Bielefeld.

\item Dosi G, Nelson RR (2010) Technical change and industrial dynamics as
evolutionary processes. Handbook of the Economics of Innovation - Vol-I, p.
51-128.

\item Dosi G, Fagiolo G, Roventini A. Schumpeter meeting Keynes, A
policy-friendly model of endogenous growth and business cycles. Journal of
Economic Dynamics and Control, 34(9):1748 1767, September 2010.

\item Dosi G, Fagiolo G, Napoletano M, Roventini A, Treibich T (2015) Fiscal
and monetary policies in complex evolving economies. Journal of Economic
Dynamics and Control, 52:166.

\item Fujita M, Thisse J-F, Economics of Agglomeration, Cambridge, Cambridge
University Press 2002.

\item Gaffard J-L and Napoletano M Editors: Agent-based models and economic
policy. Ofce 2012.

\item Gualdi S, Mandel A (2016) Endogenous Growth in Production Networks.
Documents de travail du Centre d'Economie de la Sorbonne 2016.54 - ISSN :
1955-611X.

\item Gosselin P, Lotz A and Wambst M (2017) A Path Integral Approach to
Interacting Economic Systems with Multiple Heterogeneous Agents. IF\_PREPUB.
2017. hal-01549586v2.

\item Gosselin P, Lotz A and Wambst M (2018) A Path Integral Approach to
Business Cycle Models with Large Number of Agents arXiv:1810.07178 [econ.GN].

\item Handbook of Economic Growth, Volume 1, Part A, Pages 1-1060, I1-I46
(2005). Edited by Philippe Aghion and Steven N. Durlauf.

\item Jackson M (2010) Social and Economic Networks, Princeton University
Press 2010.

\item Kleinert H (1989) Gauge fields in condensed matter Vol. I , Superflow
and vortex lines, Disorder Fields, Phase Transitions, Vol. II, Stresses and
defects, Differential Geometry, Crystal Melting, World Scientific, Singapore
1989.

\item Kleinert H (2009) Path Integrals in Quantum Mechanics, Statistics,
Polymer Physics, and Financial Markets 5th edition, World Scientific,
Singapore 2009.

\item Krugman P (1991) Increasing Returns and Economic Geography. Journal of
Political Economy, 99(3), 483-499.

\item Lucas, Robert (1976) "Econometric Policy Evaluation: A Critique". In
Brunner, K.; Meltzer, A. The Phillips Curve and Labor Markets.
Carnegie-Rochester Conference Series on Public Policy. 1. New York: American
Elsevier. pp. 19--46. ISBN 0-444-11007-0.

\item Mandel A, Jaeger C, F\"{u}rst S, Lass W, Lincke D, Meissner F,
Pablo-Marti F, Wolf S. Agent-based dynamics in disaggregated growth models.
Documents de travail du Centre d'Economie de la Sorbonne 10077, Universit%
\'{e} Pantheon-Sorbonne (Paris 1), Centre d'Economie de la Sorbonne,
September 2010.

\item Mandel A (2012) Agent-based dynamics in the general equilibrium model.
Complexity Economics 1, 105--121.

\item Solow RM (1957) Technical Change and the Aggregate Production
Function. The Review of Economics and Statistics, Vol. 39, No. 3, pp.
312-320, August.

\item Wolf S, F\"{u}rst S, Mandel A, Lass W, Lincke D, Pablo-Marti F, Jaeger
C (2013) A multi-agent model of several economic regions. Environmental
Modelling and Software, Elsevier 44, pp.25-43.
\end{description}


\end{document}